\documentclass[12pt,leqno]{report}

\usepackage{graphicx}
\usepackage{indentfirst,csquotes}
\usepackage{amssymb,amsthm,amsmath}
\usepackage{xcolor,paralist,hyperref,titlesec,fancyhdr,etoolbox}
\usepackage{lipsum}
\usepackage[margin=2.5cm]{geometry}

\titlespacing*{\section}{0pt}{0ex}{0ex}

\hypersetup{colorlinks=true, linkcolor=black, filecolor=black, urlcolor=black}

\begin{document}

\title{Inference of the degree of dissociation of weakly collisional hydrogen plasmas using collisional-radiative models}

\author{Bin Ahn\\
Department of Nuclear and Quantum Engineering\\
Korea Advanced Institute of Science and Technology\\
Daejeon, South Korea\\
}

\maketitle

\renewcommand{\thefootnote}{}
\footnotetext{This paper is a revised version of the master's thesis by Bin Ahn.}
\renewcommand{\thefootnote}{\arabic{footnote}}

\begin{abstract}
    A new analysis technique for the inference of the degree of dissociation of weakly collisional hydrogen plasmas was developed and tested with an experiment.
    There exist neutrals, namely hydrogen atoms and molecules, in low temperature hydrogen plasmas.
    These neutrals are critical to a wide range of plasma-based technologies, and thus, it is important to investigate the neutral properties in such plasmas.
    The most important physical parameter for neutrals in hydrogen plasmas is the degree of dissociation, and the technique for its inference was developed in this work.
    To improve the accuracy of the analysis, the collisional-radiative models for hydrogen atom and molecule were constructed and modified to handle bi-Maxwellian electron energy distribution and radiation trapping effect.
    An additional analysis of Fulcher-alpha transitions was conducted to obtain gas temperature and ground vibrational temperature from rotational-vibrational distribution of excited molecules.
    The technique involves multiple steps to generate calculated excited state distributions from experimental data, which are then compared to measured distributions to infer the most probable degree of dissociation.
    To test and verify the analysis technique, experiments were conducted.
    Hydrogen plasmas were generated in a large cylindrical chamber named MAXIMUS, by DC discharge with a hot cathode, under the gas pressure ranging from $3 \,-\, 6 \mathrm{mTorr}$.
    Various diagnostics including the optical emission and absorption spectroscopy, and the Langmuir probe measurement were performed to obtain spectra and electron parameters.
    Based on the measured data and the analysis technique, the degrees of dissociation for the generated plasmas were inferred to be around $1\%$, increasing with the gas pressure.
    Some problems were identified in the analysis technique, due to insufficiently accurate atomic data, incomplete inclusion of contributing channels in the models, and imperfect optical measurements.
	The possible causes and implications were discussed and improvement strategies were outlined.
	Overall, the analysis technique developed in this work is expected to play a valuable role in future diagnostic and analytical studies of hydrogen plasmas.
\end{abstract}

\tableofcontents

\chapter{Introduction}
\noindent
Hydrogen plasmas are of widespread interest in various technological fields \cite{Rousseau1994, Pigarov2003, Otorbaev1994, Iordanova2011}.
Neutral particles, hydrogen atoms and molecules, are present along with electrons and ions in hydrogen plasmas, and they play crucial roles in these fields.
Hydrogen neutrals react with materials to reduce oxides, deposit films, and etch surfaces, which is an important process in surface modification \cite{Rousseau1994}.
Hydrogen neutrals are also dominantly present in divertor regions of tokamaks for nuclear fusion, and their physics drastically influence the operation of these devices \cite{Pigarov2003, Otorbaev1994}.
Recombination processes involving hydrogen neutrals in the region affect heat fluxes on the divertor, and plasma-wall interaction alters the nature of inner wall of fusion reactors.
In addition, the production of negative hydrogen ions for neutral beam injection (NBI) for heating fusion plasmas is governed by the densities of hydrogen neutrals \cite{Iordanova2011}.

Information of neutrals in hydrogen plasmas is therefore the key to successfully control and utilize hydrogen plasmas in these field.
The degree of dissociation(DOD) is the most important physical parameter about hydrogen neutrals, since it tells us the ratio between densities of hydrogen molecules and atoms.
The DOD is defined as $DOD=(2n_{H_{2}}/n_{H}+1)^{-1}$, where $n_{H_{2}}$, $n_{H}$ are number densities of hydrogen molecules and atoms, respectively.
Since they affect the physical property of the hydrogen plasmas that need to be utilized as mentioned above, researchers attempted to measure their densities, thus the DOD, using various techniques in the past \cite{Iordanova2011, Gathen1996, AbdelRahman2006, Lavrov2006, Fantz2006a}.

Measurement of densities of neutrals is quite difficult, since there is no simple and passive way to directly detect them.
They can however be measured indirectly with fractionally populated excited states.
Excited states decay down to lower states with finite rates, and emit photons with the corresponding amounts of energy in the process.
A spontaneous emission rate is $-dn_{u}/dt=dn_{l}/dt=A_{ul}n_{u}$, where $n_{u}$, $n_{l}$ are the densities of upper and lower states and $A_{ul}$ is the Einstein A coefficient of the corresponding transition.
This rate is equivalent to the production rate of the photon, which can be measured with a spectrometer, and so a upper state density can be calculated.
Once, the ratio of measured upper state densities are calculated, the densities of neutrals of interest can be inferred with the analysis technique involving the population models.

The measurement method described above is the optical emission spectroscopy(OES), and it has been performed in a number of different ways to obtain hydrogen atom densities.
A fraction of upper states is typically calculated with certain equilibrium assumptions.
The excited state is assumed to be populated only by electron impact excitation from the ground state and depopulated only by spontaneous emission transitions to all lower states.
Then, the excited state density is calculated to be $n_{u}=\tau^{life}_{u}C_{1\rightarrow u}n_{e}n_{1}$, where $\tau^{life}_{u}$ is the lifetime due to the excited state spontaneous decays, $C_{1\rightarrow u}$ is the electron impact excitation rate coefficient, $n_{e}$ is the electron density, and $n_{u}$, $n_{1}$ are the densities of excited and ground states, respectively.
This is so-called the corona model(CoronaM).
As mentioned above photon counts measured by a spectrometer are proportional to the density of excited states.
Therefore, the ratio between ground state densities of different neutrals can be calculated, with fraction information of excited states from the CoronaM, plus the density ratio of excited states from measured spectra.

One way to achieve this is the actinometry method \cite{Iordanova2011, Gathen1996, AbdelRahman2006, Lavrov2006, Fantz2006a}.
A foreign species is introduced to the plasma, and densities of excited state of hydrogen atom and the actinometer are measured simultaneously with a spectrometer.
Since the amount of actinometer added is known, the density of hydrogen atom is deduced in the same manner as described above.
The actinometry with argon was performed numerously in the past \cite{Iordanova2011, Gathen1996}.
Krypton and neon gas was also used for the same purpose \cite{AbdelRahman2006}.
There is also a proposed way to perform a similar actinometry without foreign gas.
Since there exists excited states of hydrogen molecules that emit photons, hydrogen atom density can be inferred from the known hydrogen molecule density \cite{Gathen1996, Lavrov2006, Fantz2006a}.
In addition, more rigorous calculations involving the collisional-radiative model(CRM) was used instead of the CoronaM \cite{Iordanova2011, Lavrov2006, Fantz2006a}.
The CRM is the model that calculates excited state densities considering more significant transitions, such as stepwise excitation and radiative cascade, and therefore, gives more accurate results than CoronaM.
All these methods would require information of electron energy distribution function(EEDF), since electron impact transitions are considered.
EEDF can be directly measured using a Langmuir probe.
Effective electron temperature was also deduced from measured line ratios \cite{Gathen1996, Lavrov2006}, and an approximation to neglect EEDF was proposed as well \cite{Fantz2006a}.

Another method to measure neutrals is the optical absorption spectroscopy(OAS) \cite{Otorbaev1994}.
Hydrogen atoms, for example, are capable of absorbing certain photons, e.g. the Lyman photons.
The absorption probability of the said photons is proportional to the density of the neutral, which can therefore be calculated with measured absorption.
The OAS does not need additional foreign gas or EEDF information for the calculation, but it is disadvantageous in that it requires external light sources and it may affect the measured plasma.

In this work, A newly developed analysis technique to infer the DOD of weakly collisional DC hydrogen plasmas is introduced.
For this technique, CRMs for hydrogen atom and molecule are newly developed to improve the calculation.
In order to more accurately use EEDF information to calculate state transition rate, the models are modified so that they are able to handle bi-Maxwellian EEDF, with additional parameters.
The increased production of hydrogen atom excited states from photon absorption is also considered with a newly modeled optical escape factors.
In addition, the Fulcher-$\alpha$ transition is analyzed to obtain gas temperature and ground vibrational temperature, which are also utilized to improve the DOD inference process.
The philosophy of this technique is to utilize all available information from measurement to infer the DOD most precisely.
To test this analysis technique, hydrogen plasmas of several conditions are generated in a cylindrical chamber by DC discharge, and their emission spectra and EEDF are measured.
The inferred hydrogen atom density using this technique is compared with that measured by the OAS.
The subsequent chapters present the experimental setup and measurements, outline the theoretical background, detail the developed analysis technique and experimental results, and conclude with a comprehensive discussion.

\chapter[Experiment \& Measurement]{Experiment \& Measurement}
An experiment is conducted to test and verify the analysis technique developed in this work, which are to be introduced in the following chapters.
$H$ plasmas are generated and various measurements are performed.
The details of the experiment and the measurement results are shown in this chapter first in order to help the reader understand example data used to explain the theory and the technique presented in this work.

\section{Experiment}

\subsection{Apparatus}

\begin{figure}[htbp]
	\centering
	\includegraphics[width=0.8\textwidth]{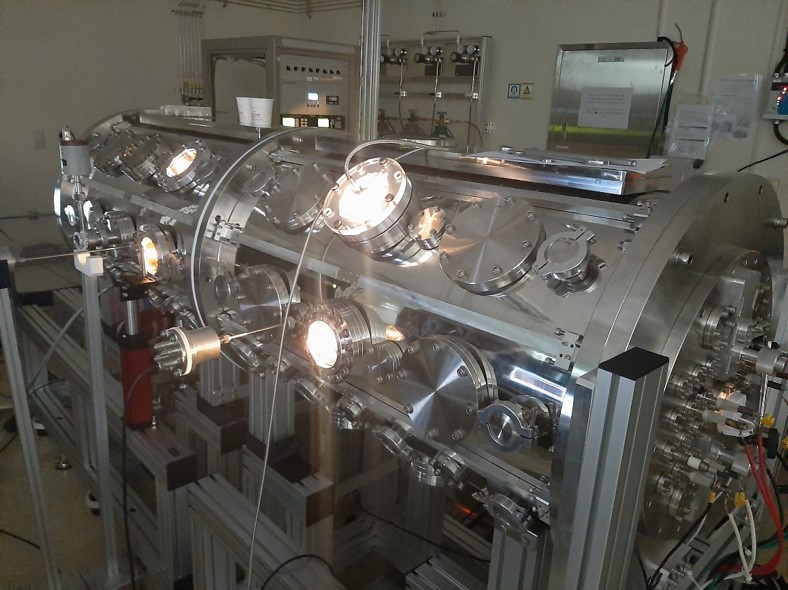}
	\caption[Picture of MAXIMUS]{The picture of the MAXIMUS experimental device.}
	\label{fig:maximus1}
\end{figure}

\begin{figure}[htbp]
	\centering
	\includegraphics[width=0.8\textwidth]{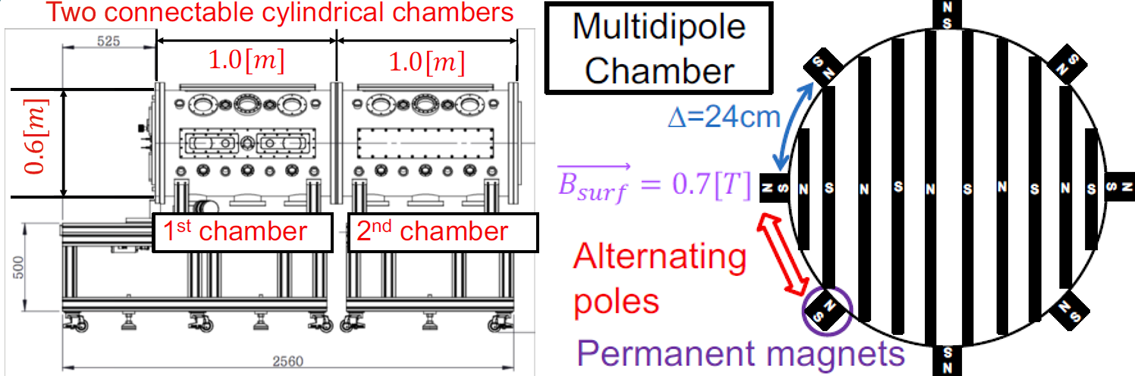}
	\caption[Layout and magnets of MAXIMUS]{The layout of the MAXIMUS device (left) and the permanent magnets attached (right).}
	\label{fig:layout_of_maximus}
\end{figure}

\begin{figure}[htbp]
	\centering
	\includegraphics[width=0.6\textwidth]{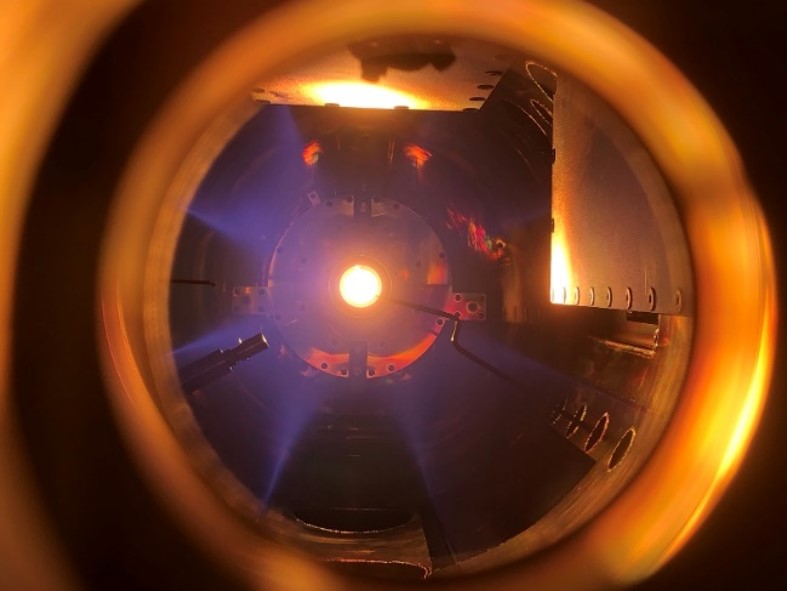}
	\caption[$\mathrm{H}$ plasma from $\mathrm{LaB}_{6}$ cathode]{Hydrogen plasma generated by the $\mathrm{LaB}_{6}$ crystal cathode, with a Langmuir probe located at the radial center, viewed from an end port.}
	\label{fig:LaB6_plasma}
\end{figure}

Plasmas are generated in a large cylindrical chamber named MAXIMUS (Figure \ref{fig:maximus1}).
The name is the acronym for the MAgnetic X-point sIMUlation System.
The MAXIMUS is a multi-dipole chamber that is surrounded by alternating poles of permanent magnets, which are installed on the side and the both ends to improve the confinement of the plasmas generated (Figure \ref{fig:layout_of_maximus}).
Gas is filled inside the chamber and its pressure is controlled by mass flow meters and pump speeds.
Species of gas that have been tested in the chamber are $Ar, He$ and $H$.
Working pressures are $P_{working}=0.1 \,-\, 50 mTorr$, while base pressure is around $P_{base}=10^{-3} mTorr$.
Plasmas are generated by heating and biasing voltages of cathodes located at the end of the chamber.
Thoriated tungsten filaments($W$) and a lanthanum hexaboride($LaB_{6}$) crystal are used as the cathodes, and the latter one is used for the experiment conducted (Figure \ref{fig:LaB6_plasma}).

\begin{figure}[htbp]
	\centering
	\includegraphics[width=0.7\textwidth]{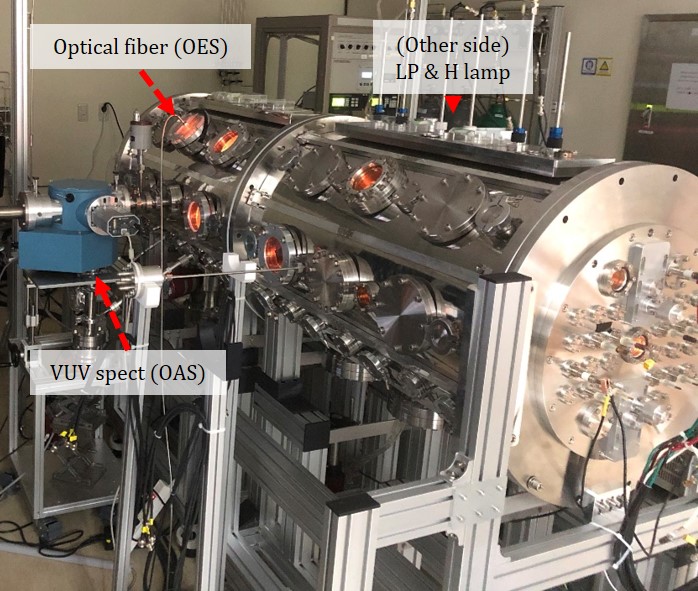}
	\caption[MAXIMUS with VUV system]{The MAXIMUS device, equipped with the VUV spectroscopy system.}
	\label{fig:maximus2}
\end{figure}

Electron parameters are measured mainly by Langmuir probes(LP) inserted to the MAXIMUS (Figure \ref{fig:maximus2}).
Probe voltage is swept several tens of times to obtain number of $I-V$ curves that are ensemble-averaged and evaluated for electron temperatures and electron densities.
Energies of electrons of the plasmas generated in the MAXIMUS typically do not reach a Maxwellian distribution, but rather, bi-Maxwellian distribution, which means the electrons reach two separate thermodynamic equilibria.
The cause of this phenomenon is speculated to be a low collision frequency of electrons.
In this case, there are cold electron temperature and density, and hot electron temperature and density.
Therefore, the LP measures these four electron parameters.

There are two optical measurement systems.
One is for the optical emission spectroscopy(OES), and the other is for the optical absorption spectroscopy(OAS).
The OES is performed with a visible(VIS) spectrometer(IsoPlane-320) with an EMCCD camera(PIXIS), an intensity calibration source(LSVN0407), and a fiber bundle (all from Princeton Instruments).
The groove density of the most used grating of the spectrometer is $d_{gr}=1800gr/mm$.
The optical fiber connected to the entrance slit of the VIS spectrometer is mounted on a quartz window of the MAXIMUS by means of a custom holder with a collimator (Figure \ref{fig:maximus2}).
The VIS spectrometer measures the emission spectra on the line-of-sight of a plasma.
A vacuum-ultraviolet(VUV) spectrometer(from McPherson) with a $H$ lamp(from Hamamatsu Photonics) is used for the OAS.
Radiations from the $H$ lamp are measured by the VUV spectrometer, with and without generated plasmas, so that the absorption of Lyman photons can be calculated.

\subsection{Experimental condition \& Measurement}

The plasmas generated for test and verification of the technique are $H$ plasmas discharged by the $LaB_{6}$ crystal cathode (Figure \ref{fig:LaB6_plasma}).
The heating current and voltage are $V_{heating}=164 V \;/\; I_{heating}=25 A$.
The discharge voltage is kept at $100 V$.
The gas pressure are varied to generate plasmas with different conditions, and the discharge current changed with the pressure (Table \ref{tb:pressure_Idischarge}).

\begin{table}[htbp]
	\caption[The gas pressures and the discharge currents of the $H$ plasmas]{The gas pressures and the discharge currents of the $H$ plasmas
	}
	\label{tb:pressure_Idischarge}
	\begin{center}
		\begin{tabular} {ccccccccccc}
			\hline\hline
			& Case & 1 & 2 & 3 & 4 & 5 & 6 &\\
			\hline
			& $P_{g} \quad [mTorr]$ & 3.0 & 3.5 & 4.0 & 4.5 & 5.0 & 5.5 &\\
			& $I_{discharge} \quad [A]$ & 0.38 & 0.65 & 0.93 & 1.25 & 1.69 & 2.26 &\\
			\hline\hline
		\end{tabular}
	\end{center}
\end{table}

The OES is performed to measure the emission spectra 20 times for each plasma condition.
The wavelength range is $\lambda=420 \,-\, 490, 590 \,-\, 660 nm$, the resolution is $\Delta\lambda\cong0.02nm$, and the exposure time is $t_{exposure}=300 ms$.
The Langmuir probe located at the center next to the line-of-sight of the OES measures the electron parameters of the plasmas with 30 sweeps.
The OAS is performed on the same axial position simultaneously.\footnote{Collaborating reseachers(Y. Lim from NQE, KAIST \& D. cho from Jeonbuk National Univ.) analyze the LP and the OES data to provide the electron parameters and the $H$ atom densities.}

\section{Measurement result}

\begin{figure}[htbp]
	\centering
    \includegraphics[width=15cm]{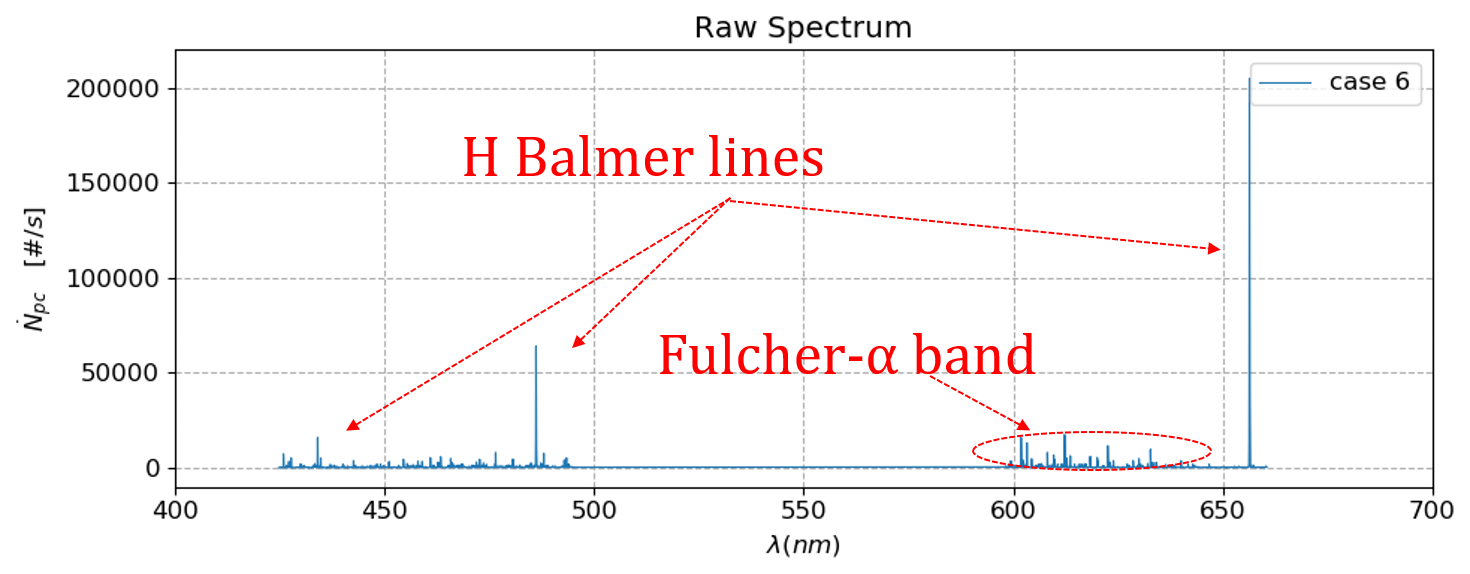}
	\caption[Measured raw spectrum (Case 6)
	]{Measured raw spectrum (Case 6)
	} \label{fig:measured_raw_spectrum}
\end{figure}

The emission spectra are successfully measured by the VIS spectrometer.
Figure \ref{fig:measured_raw_spectrum} is an example spectrum of Case 6.
Among many present peaks, there are the $H$ Balmer lines and the Fulcher-$\alpha$ band, which contain important information of $H$ neutrals.
These are the lines of interest.
The three shown Balmer peaks correspond to $H_{\alpha}, H_{\beta}, H_{\gamma}$, which are from excited states of $H$ atoms.
The Fulcher-$\alpha$ band is from a excited state of $H_{2}$ molecules.
The detailed origin of the radiations are explained in Chapter 3.

\begin{figure}[htbp]
	\centering
    \includegraphics[width=14cm]{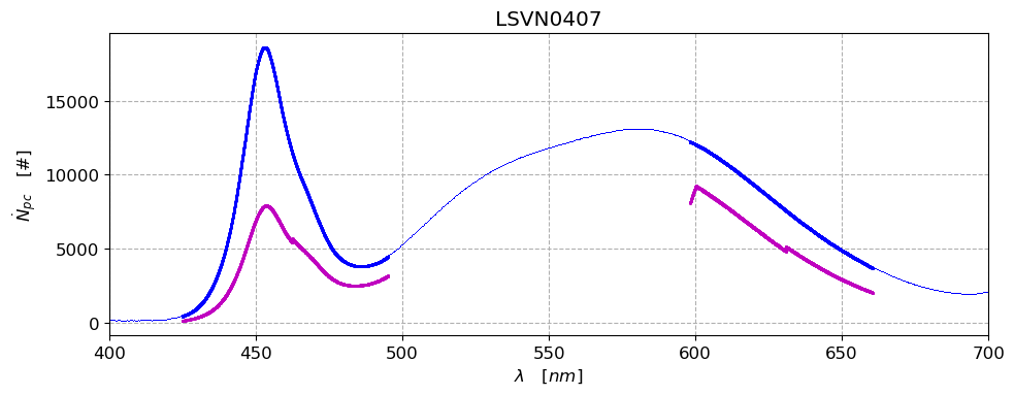}
	\caption[The reference spectrum(blue) and the measured spectrum(magenta) of the LSVN0407 lamp
	]{The reference spectrum(blue) and the measured spectrum(magenta) of the LSVN0407 lamp. The intensity scale is relative.
	} \label{fig:The reference spectrum(blue) and the measured spectrum(magenta) of the LSVN0407 lamp}
\end{figure}

\begin{figure}[htbp]
	\centering
    \includegraphics[width=14cm]{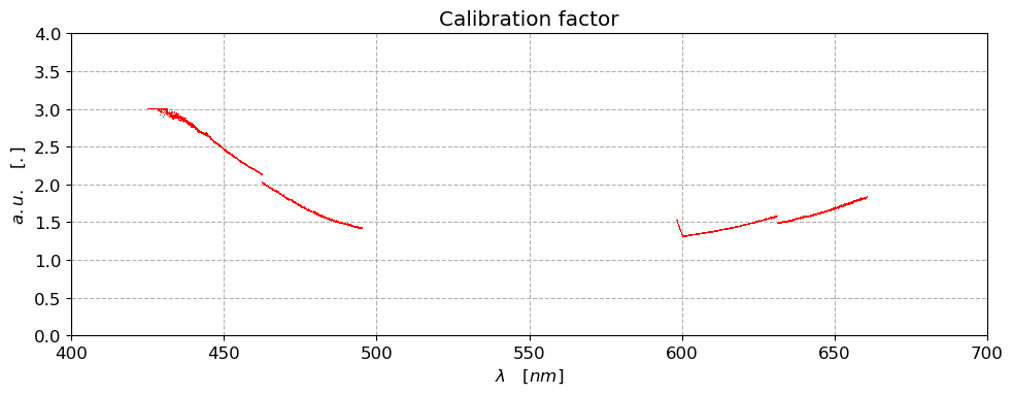}
	\caption[The calibration factor calculated
	]{The calibration factor calculated. It is multiplied to measured plasma spectra to calibrate their intensity.
	} \label{fig:The calibration factor calculated}
\end{figure}

\begin{figure}[htbp]
	\centering
    \includegraphics[width=14cm]{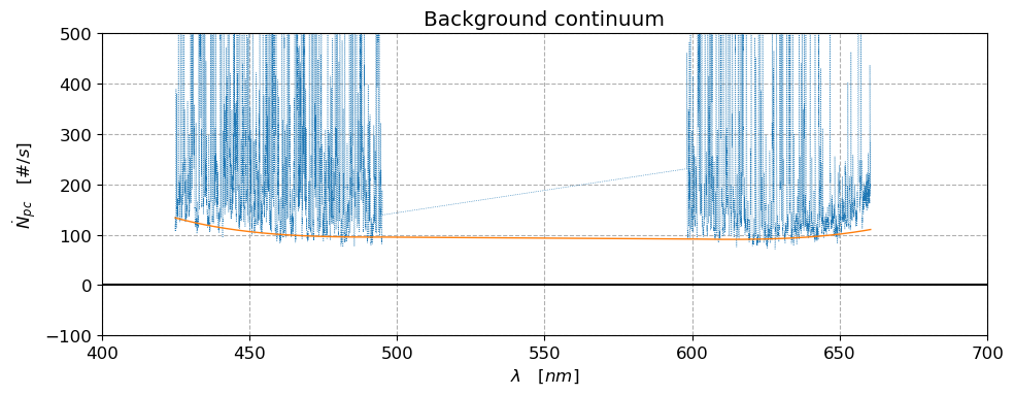}
	\caption[The instrumental broadening of a $H_{\alpha}$ line
	]{The instrumental broadening of a $H_{\alpha}$ line
	} \label{fig:background}	
\end{figure}

\begin{figure}[htbp]
	\centering
    \includegraphics[width=10cm]{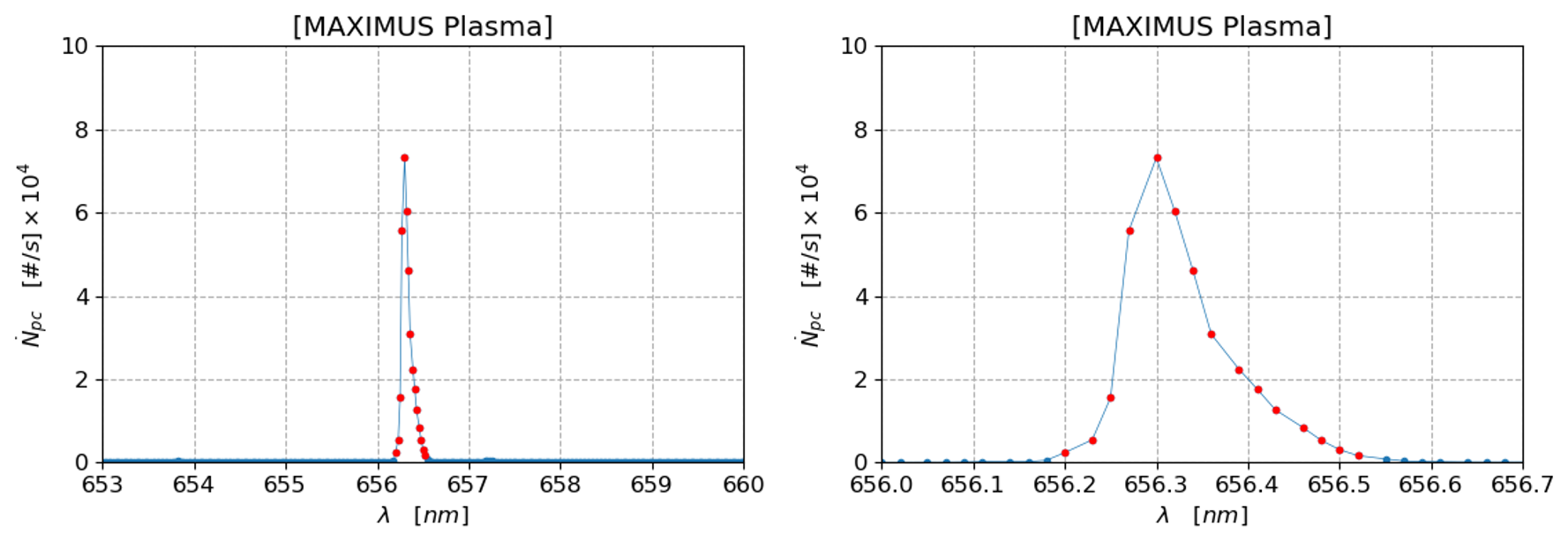}
	\caption[The instrumental broadening of a $H_{\alpha}$ line
	]{The instrumental broadening of a $H_{\alpha}$ line
	} \label{fig:The instrumental broadening}	
\end{figure}

The raw spectra of all cases are processed for the aforementioned important lines.
First, the calibration factor is calculated comparing the spectrum of the LED lamp(LSVN0407) measured by the experimental setting, and the intensity-accurate reference spectrum given (Figure \ref{fig:The reference spectrum(blue) and the measured spectrum(magenta) of the LSVN0407 lamp}).\footnote{The term `intensity' in this work is a quantity proportional to a number of photons, not energy.}
The reference spectrum(blue) is divided by the measured one(magenta) to calculate the calibration factor.
The reference spectrum is scaled so that its magnitude is comparable to the measured one.
Figure \ref{fig:The calibration factor calculated} shows the calibration factor calculated, and there are several jumps since the groove-dense grating has to rotate as many as four times for the EMCCD to measure the wavelength range of interest.
This calibration factors are multiplied to all the measured plasma spectra to calibrate their intensity.
Second, the small background continuum radiation (Figure \ref{fig:background}) is removed algorithmically.
The background continuum radiation is likely from the heated cathode.
Third, there are small instrumental broadenings for the peaks detected.
When peaks are zoomed in, broadenings smaller than $0.5 nm$ can be seen.
Figure \ref{fig:The instrumental broadening} shows the instrumental broadening of a $H_{\alpha}$ line.
Therefore, all lines of interest are integrated to find their accurate photon count rate.\footnote{All are divided by the exposure time to turn the photon count to the photon count rates. However, since only their relative quantities mattered, this process, or any other scaling processes would be irrelevant to the analysis.}

\begin{figure}[htbp]
	\centering
    \includegraphics[width=15cm]{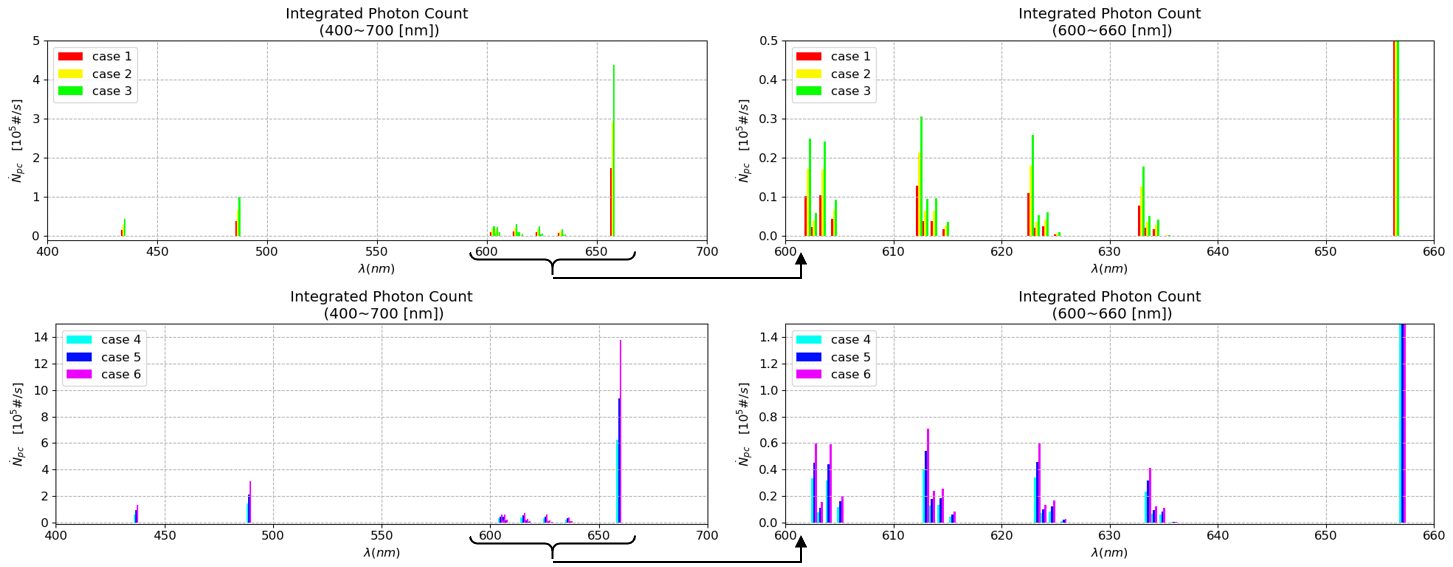}
	\caption[The processed photon count rates of all lines of interest
	]{The processed photon count rates of all lines of interest. Case 1, 2, \& 3 and Case 4, 5, \& 6 were grouped separately, and each group is plotted in the same plot to compare the magnitude. The location of the peaks were shifted slightly to avoid overlaps.
	} \label{fig:The processed photon count rates of all lines of interest}
\end{figure}

The result of these process gives photon count rates of interest.
Figure \ref{fig:The processed photon count rates of all lines of interest} are bar plots of the processed photon count rates of all lines of interest.
The plots on the right are those zoomed in at the Fulcher-$\alpha$ band.
Different cases are plotted together to compare their photon count rates.
As can be seen from the plots, the cases with higher gas pressure have stronger intensities.

\begin{figure}[htbp]
	\centering
    \includegraphics[width=8cm]{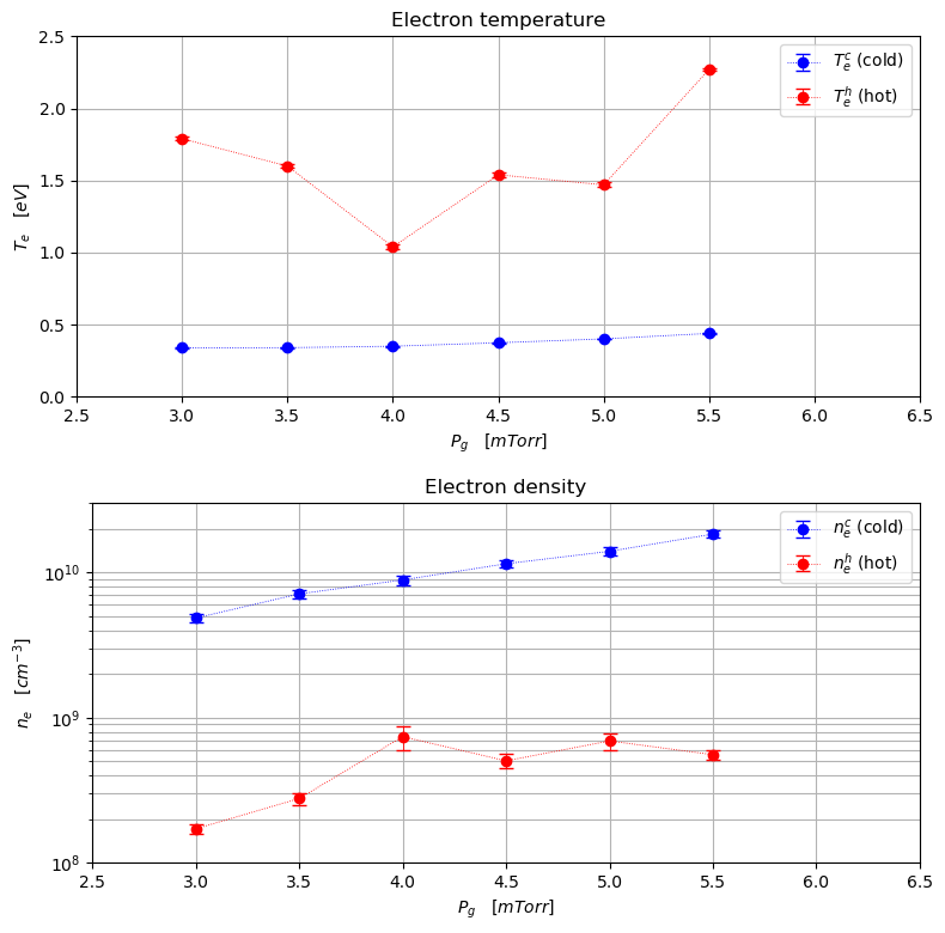}
	\caption[The four electron parameters obtained with LP
	]{The four electron parameters obtained with LP. The top and bottom plots are for electron temperature and electron density, respectively. The blue and red points are for cold and hot electrons respectively.
	}\label{fig:The four electron parameters obtained with LP}	
\end{figure}

\begin{figure}[htbp]
	\centering
    \includegraphics[width=10cm]{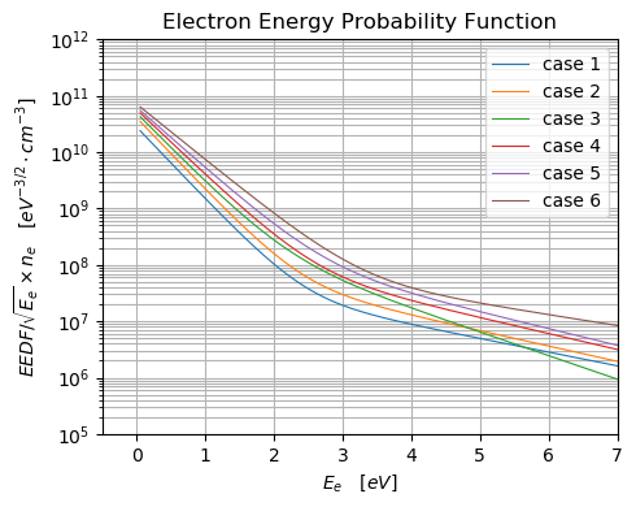}
	\caption[Electron energy probability functions for all cases
	]{Electron energy probability functions for all cases. They are drawn using the EEPF formula with measured parameters.
	}\label{fig:Electron energy probability functions for all cases}
\end{figure}

Electron parameters measured by the LP are presented in Figure \ref{fig:The four electron parameters obtained with LP}.
As explained earlier, there are four parameters, cold electron temperature and density, and hot electron temperature and density.
There are about one order more cold electrons than hot electrons, while the hot electrons are several times hotter than the cold ones.
Using the electron parameter measured, electron energy probability functions(EEPF) are drawn for all cases (Figure \ref{fig:Electron energy probability functions for all cases}).
In the log-scale plot, there are knees of curves visually present that signify the dominance transition from cold to hot electrons.
This means the tail shapes of the distributions are dominantly affected by hot electron parameters.
One speculation for the cause of this phenomenon is that inelastically scattered electrons that are originally primary electrons from the cathode, and ionized electrons from neutrals reach thermodynamic equilibrium separately, since electrons mostly collide with heavy particles, while the collision frequency among electrons is negligibly low.
Further investigations are needed to clarify this problem.

\begin{figure}[htbp]
	\centering
    \includegraphics[width=14cm]{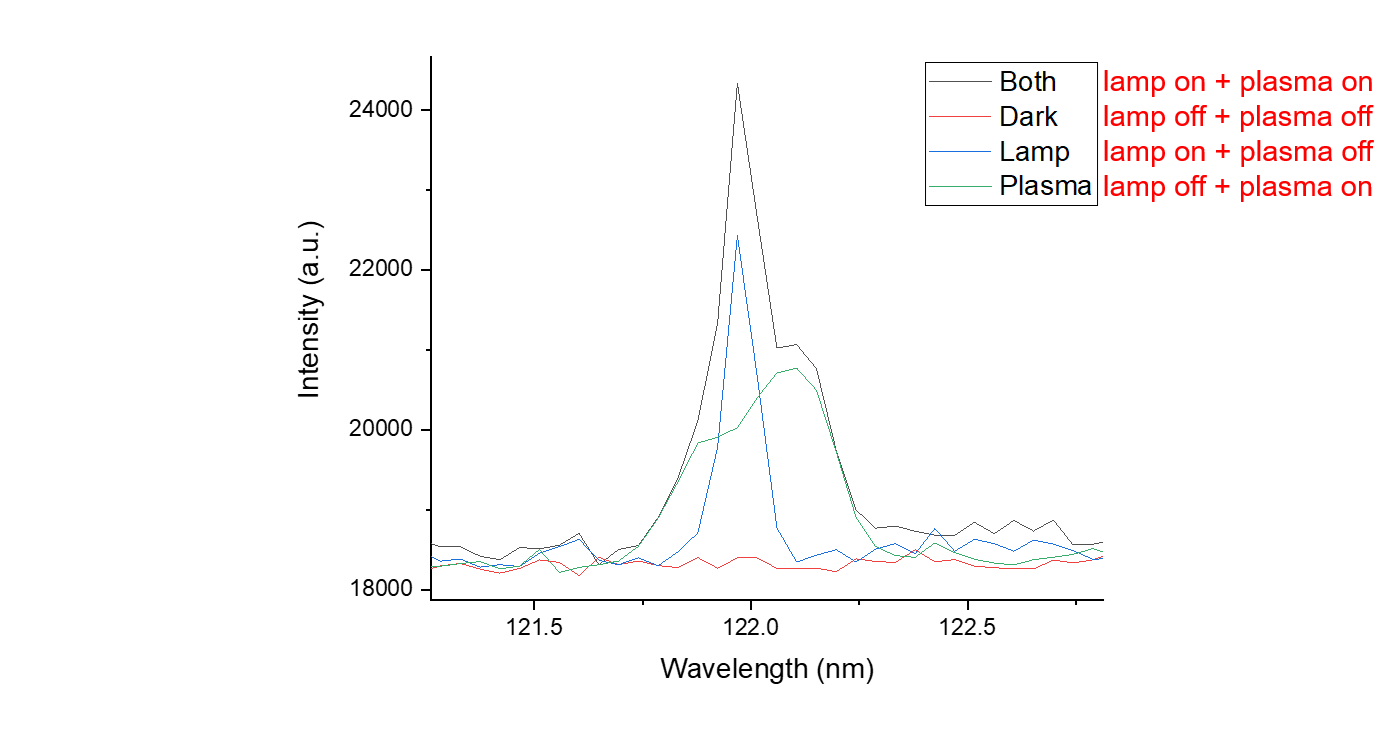}
	\caption[The spectra of various conditions measured by the VUV spectrometer
	]{The spectra of various conditions measured by the VUV spectrometer
	}\label{fig:The spectra of various conditions measured by the VUV spectrometer}	
\end{figure}

Figure \ref{fig:The spectra of various conditions measured by the VUV spectrometer} shows the measured spectra of the VUV spectrometer for the OAS.
For each case, spectra are measured for four different conditions as shown on the figure, in order to subtract the plasma light and background light in the analysis.
The incident and final Lyman photon intensities traveled from the $H$ lamp is $I_{i}=I_{Lamp}-I_{Dark}$ and $I_{f}=I_{Both}-I_{Plasma}-I_{Dark}$, respectively.
Then, the absorption is $\frac{I_{i}-I_{f}}{I_{i}}=\frac{I_{Lamp}+I_{Plasma}-I_{Both}}{I_{Lamp}-I_{Dark}}$, and this quantity is used later to calculate hydrogen atom density.

\chapter[Theoretical Background]{Theoretical Background}
The theoretical background is explained in detail in this chapter, prior to the explanation of the development of the analysis technique.
It is important to understand the physics of quantum states of neutrals to analyze them in plasmas.
The term, neutral, indicates any particle with neutral charge.
Atoms and molecules are neutrals.
Neutrals have certain number of bound electrons, and state of the bound electrons mostly determine the states of the neutrals.
Hydrogen atoms, $H$, and hydrogen molecules, $H_{2}$, have different kinds of electron orbitals, and so the nature of their states differ.
Energy of states and transition between states have to be considered for different neutrals separately.

Neutrals are not the only ones with bound electrons.
Ions such as $H^{-}$ and $H^{+}_{2}$ have bound electrons, and so they have quantum states as well.
Also, there are other neutrals such as triatomic hydrogen $H_{3}$.
They are all present in $H$ plasmas.
All these particles interact with one another and change from one to the other, and so all can be analyzed simultaneously.
However, only $H$ and $H_{2}$ are mainly dealt with in this work.
In this chapter, the nature of quantum states of these two species are described thoroughly.

\section{Hydrogen atom state}

\subsection{Energy level}

\begin{figure}[htbp]
	\centering
    \includegraphics[width=12cm]{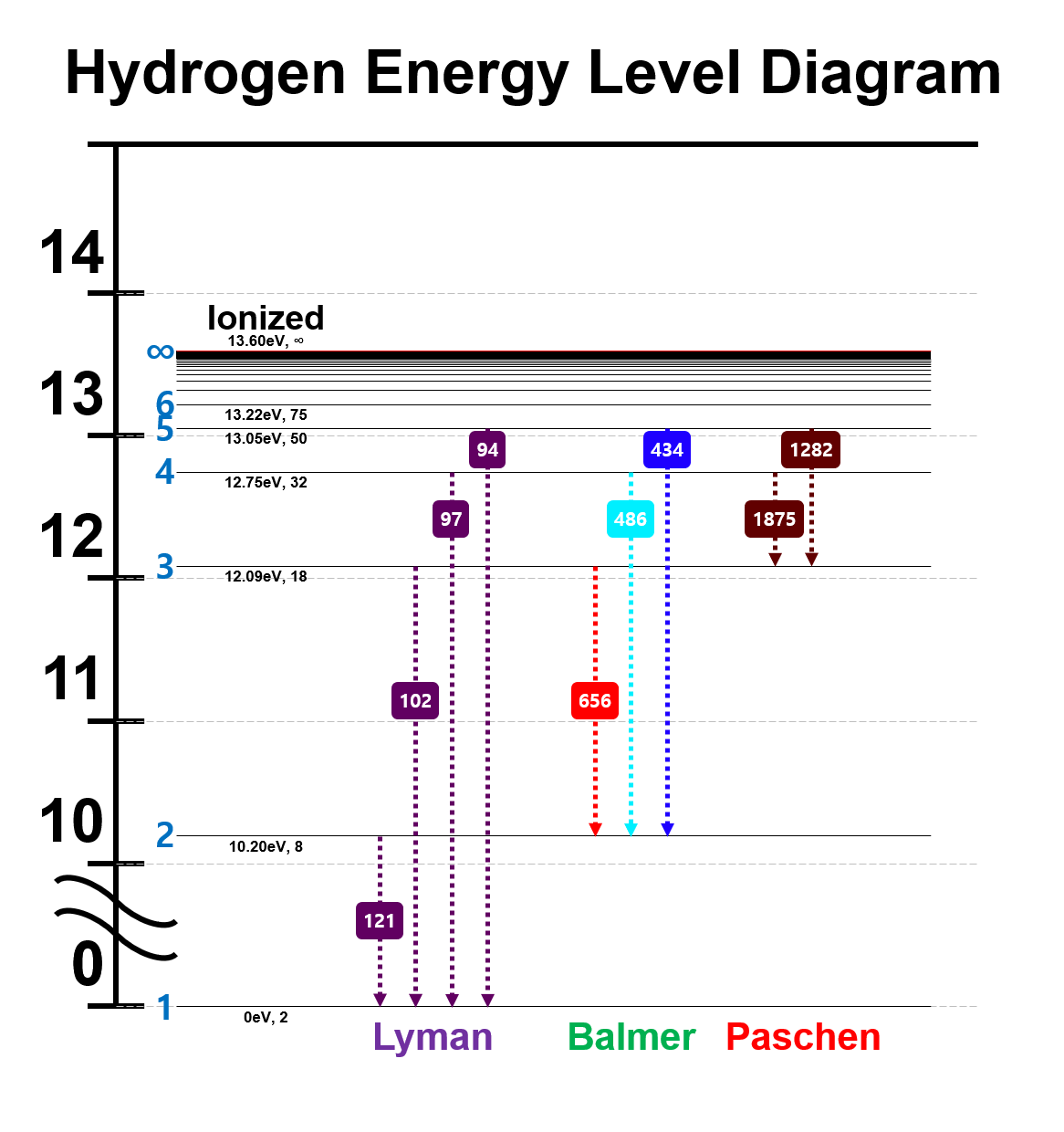}
	\caption[$H$ energy level diagram
	]{$H$ energy level diagram, and spontaneous emission of excited states
	}\label{fig:H energy level diagram}
\end{figure}

The energy of each electron orbit of $H$ is\cite{Tiesinga2018}
\begin{equation} \label{eq:rydberg}
E_{n}=-\frac{m_{e}e^{4}}{8h^{2}\epsilon^{2}_{0}}\frac{1}{n^{2}}=-\frac{13.605}{n^{2}}eV=-1Ry\times\frac{1}{n^{2}}
\end{equation}
where $m_{e}$ is the electron mass, $e$ is the electron charge, $h$ is the Planck constant, $\epsilon_{0}$ is the vacuum permittivity, and $n(=1, 2, ...)$ is the principal quantum number.
Bound electrons of $H$ can be in any orbit and have energies as described.
The value of the bound electron energy is the energy of the quantum state of $H$.
The ground state is when $n=1$, the ionized state is when $n=\infty$, and the states in-between($n=2, 3, ...$) are called excited states.
By letting the energy of the ground state be $0eV$, the energy of the ionized state become $1Ry$, and the energies of the excited states have positive values (Figure \ref{fig:H energy level diagram}).
The numbers next to the state energy in Figure \ref{fig:H energy level diagram} are `statistical weight' or `degeneracy'.
They indicate how many `rooms' are in each state.
Statistical weights of these states are calculated by $g_{n}=2n^{2}$.

Electron orbits are not only characterized by the principal quantum numbers, but more rigorously also by azimuthal quantum number $l(=0, 1, 2...<n)$, which is denoted by symbols $s, p, d, f,...$, respectively.
The electron configuration of the ground state is $1s^{1}$.
The electron configuration of the $n=2$ excited state can either be $1s^{0}2s^{1}p^{0}$ or $1s^{0}2s^{0}p^{1}$.
This means the $n=2$ state can have an electron in two different orbit, and they are different states that differently behave.

There is a notation scheme called the term symbol, which distinguish quantum states more rigorously.
The term symbol describes a $H$ state with $n^{M}L$, where $n$ is the principal quantum number, $M$ is the multiplicity(which is always $2$, for $H$), and $L(=S, P, D, F, ...)$ is the total orbital momentum quantum number.\footnote{For $H$, $L$ is the same as $l$ of the orbit the excited electron resides in.}
In this case, the term symbols for the above mentioned three electron configurations are $1^{2}S, 2^{2}S$ and $2^{2}P$, respectively.
Statistical weights for these angularly resolved states are $g=(2L+1)M$.($g=2, 2, 6$ for $1^{2}S, 2^{2}S, 2^{2}P$)\footnote{It is easy to make sense of these statistical weights by visualizing two opposite spin electrons occupying each shell of an orbital.}
It is however okay to group $2^{2}S$ and $2^{2}P$ states together as the $n=2$ state, because their energy gap is negligibly small.
Their energy difference is smaller than $0.1meV$, and so they are `strongly coupled' \cite{Wunderlich2009}. This means they reach a thermodynamic equilibrium very easily.
At a thermodynamic equilibrium, states follow the Boltzmann distribution:
\begin{equation} \label{eq:Boltzmann}
\frac{n_{H(u)}/g_{H(u)}}{n_{H(l)}/g_{H(l)}}=\exp{\left(-\frac{\Delta E_{H(l,u)}}{T}\right)}
\end{equation}
where $n_{H(u)}$ and $n_{H(l)}$ are number densities of upper and lower states, $\Delta E_{H(l,u)}$ is the energy gap, and the $T$ is an equilibrium temperature.
From this equation, it is easy to see that when the energy gap is very small, the number densities are only related by their statistical weights.
For example, if there are $80$ of $n=2$ states, it can be considered that $20$ of them are $2^{2}S$, and $60$ of them are $2^{2}P$.
The same is true for all excited states, and therefore, it is okay to only let the principal quantum numbers distinguish different states.
This rule may break when there is a strong magnetic field which can widen the energy gaps of this `fine structure.'

\subsection{Transition}

\subsubsection{Spontaneous emission}

Excited states are unstable and decay naturally into lower states, without any external interaction.
This is called `spontaneous emission' or `natural decay'.
Since there is a finite energy difference between initial and final states, a photon with the corresponding energy is emitted in the process.
Figure \ref{fig:H energy level diagram} shows some selected spontaneous emission processes and their corresponding photon wavelengths in $[nm]$.\footnote{These photons from distinct energy gap transitions are conventionally called `line radiations.'}
The Lyman series are the lines where the lower state of the transition is the ground state($n=1$).
Due to the high energy gap between the ground state and excited states, the emitted photons have VUV wavelengths.
The Balmer series are the lines for the $n=2$ final states.
These lines are in the VIS spectrum range, and therefore can be detected easily with the VIS spectrometer or human eyes.
There are more named series following this rule.

A spontaneous emission transition flux\footnote{Quantities: In this work, flux, rate, and rate coefficient have units of $[s^{-1}cm^{-3}]$, $[s^{-1}]$, and $[s^{-1}cm^{3}]$, respectively.} is
\begin{equation} \label{eq:spontaneousemission}
-\left(\frac{dn_{H(u)}}{dt}\right)_{s.e.}=\left(\frac{dn_{H(l)}}{dt}\right)_{s.e.}=\dot{N}_{PE,H(u \rightarrow l)}=A_{H(u \rightarrow l)}n_{H(u)}
\end{equation}
where $\dot{N}_{PE}$ is a photon emission flux, and $A$ is an Einstein A coefficient of this particular transition.
Equation \ref{eq:spontaneousemission} means the loss flux of the upper state, the gain flux of the lower state, and the photon emission flux are equal, and they equal to the product of the Einstein A coefficient and the upper state density.
Einstein A coefficients are fixed constants.
Some examples are given in Table \ref{tb:Example Einstein A coefficients}.

\begin{table}[htbp]
	\caption[Example Einstein A coefficients\cite{Wiese2009}
	]{Example Einstein A coefficients\cite{Wiese2009}
	}
	\label{tb:Example Einstein A coefficients}
	\begin{center}
		\begin{tabular} {ccccccccccc}
			\hline\hline
			& Upper state & 2 & $2^{2}S$ & $2^{2}P$ & 3 & 3 &\\
			& Lower state & 1 & $1^{2}S$ & $1^{2}S$ & 1 & 2 &\\
			\hline
			& $A_{ul} \quad 10^{6}[s^{-1}]$ & 470 & $\sim$0 & 626 & 56 & 44 &\\
			\hline\hline
		\end{tabular}
	\end{center}
\end{table}

It can be seen that $A_{2\rightarrow1}=A_{2^{2}S\rightarrow1^{2}S}/g_{2^{2}S}+A_{2^{2}P\rightarrow1^{2}S}/g_{2^{2}P}$, which means transition rates between states, separated by principal numbers, are the statistically weighted average values of corresponding angularly resolved transitions.
This relation is true for the same reason mentioned in Subsection 3.1.1, and is not limited to spontaneous emission and applies to other transitions.

There is a `natural lifetime' for each excited state, because of spontaneous emissions.
For example, without any external interaction, the state 3 decays to the state 1 or 2 with the distinct rates, so the lifetime of the state 3 is $\tau_{3}=(A_{3\rightarrow1}+A_{3\rightarrow1})^{-1} \cong 10ns$.

\subsubsection{Electron impact transition}

Free energetic electrons in plasmas collide with neutrals and transfer their kinetic energy partially to the bound electrons of the neutrals to excite them.
This is called `electron impact(collision) excitation.'
In order to obtain rates of electron impact excitations, `cross sections' must be considered first.
A total electron collision microscopic cross section of $H$, $\sigma^{total}_{e-H} [cm^{2}]$, is a cross sectional area an electron `sees' a $H$ atom in its trajectory, and so it is related to the probability of the collision.
The macroscopic cross section is $\Sigma^{total}_{e-H} [cm^{-1}]=\sigma^{total}_{e-H}n_{H}$.
Its inverse would be the `collision length', and the macroscopic cross section times the electron speed, $\Sigma^{total}_{e-H}v_{e}[s^{-1}]$ is the `collision frequency.'

\begin{figure}[htbp]
	\centering
    \includegraphics[width=12cm]{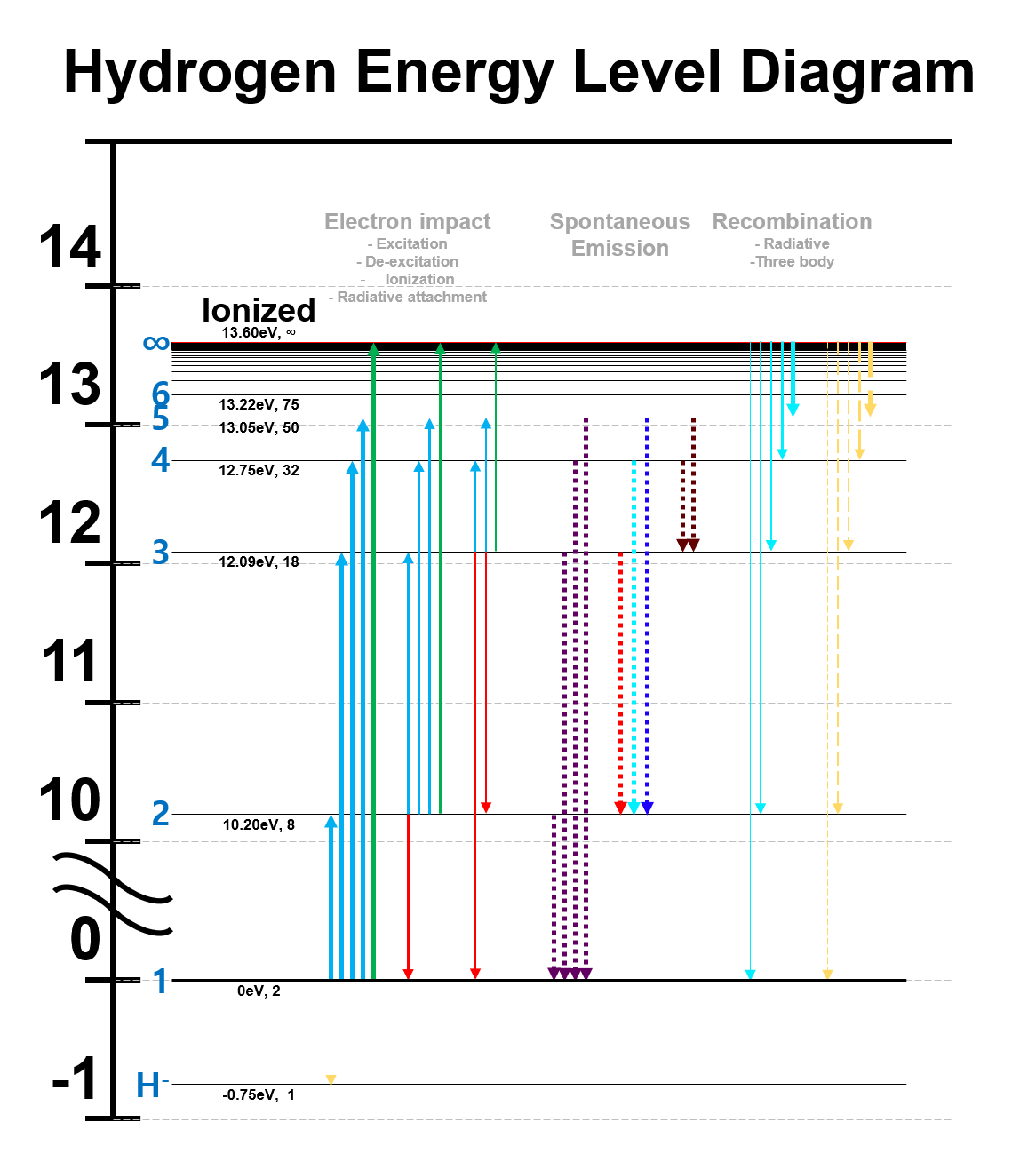}
	\caption[$H$ energy level diagram with important transitions
	]{$H$ energy level diagram with important transitions
	}\label{fig:H energy level diagram with important transitions}
\end{figure}

Out of all collisions taking place, some collisions are inelastic and may `excite' a $H$ of a certain state.
Excitations from any initial lower state to any final upper state are possible (Figure \ref{fig:H energy level diagram with important transitions}).
Therefore, parts of the cross section are impact excitation cross sections, $\sigma^{i.e.}$, and it is a function of electron energy, $E_{e}[eV]$.
Since impact excitations are taking place in plasma with an electron density, the expectation value of the collision frequency per volume is
\begin{equation} \label{eq:rcie}
\langle\sigma^{i.e.}_{H(l \rightarrow u)}(E_{e}) v_{e}n_{H(l)}\rangle_{f_{EEDF}}n_{e}=RC^{i.e.}_{H(l \rightarrow u)}(T_{e})n_{e}n_{H(l)}
=-\left(\frac{dn_{H(l)}}{dt}\right)_{i.e.}=\left(\frac{dn_{H(u)}}{dt}\right)_{i.e.}
\end{equation}
\begin{equation} \label{eq:rc}
RC(T_{e})=\langle \sigma(E_{e}) v_{e} \rangle_{f_{EEDF}}
\end{equation}
where Equation \ref{eq:rc} is the definition of the rate coefficient, which is a function of an electron temperature, $T_{e}[eV]$, since the $f_{EEDF}$ also is.
A more detailed formula for the Maxwellian rate coefficient is in Equation \ref{eq:rcdetail}.
Equation \ref{eq:rcie} describe the electron impact transition from a lower state to an upper state.
The flux of the transition is proportional to the densities of the lower state and electron, and the corresponding rate coefficient, and it is both the loss flux of the lower state and the gain flux of the upper state.

An electron impact may also `ionize' or `de-excite' a $H$ of a state.
An electron may be strong enough to ionize a bound electron of any state, turning $H(l)$ to $H^{+}$.
Similar to the above explanation, the electron impact ionization cross section, $\sigma^{i.i.}_{l}$, governs the rate of the transition.
On the other hand, for de-excitation, the `principle of detailed balance' \cite{Fujimoto2004} is typically used.
A de-excitation is the inverse process of an excitation.
The principle of detailed balance states that, in thermodynamic equilibrium, the fluxes of excitation and de-excitation of the same pair of lower and upper states should be equal, and it can be mathematically represented by
\begin{equation} \label{eq:detailedbalance}
\left[ C_{H(l \rightarrow u)}n_{H(l)}n_{e}=F_{H(u \rightarrow l)}n_{H(u)}n_{e} \right]_{T.E.}
\end{equation}
where  $C_{H(l \rightarrow u)}$ and $F_{H(u \rightarrow l)}$ are the excitation and the de-excitation rate coefficients (Equation \ref{eq:rc}).
Here, since the lower and the upper states are in thermodynamic equilibrium, the Boltzmann distribution (Equation \ref{eq:Boltzmann}) again holds.
So the following relation is used to obtain electron impact de-excitation rates.
\begin{equation} \label{eq:detailedbalance2}
\frac{C_{H(l \rightarrow u)}}{F_{H(u \rightarrow l)}}=\left[ \frac{n_{H(u)}}{n_{H(l)}} \right]_{T.E.}=\frac{g_{H(u)}}{g_{H(l)}}\exp{\left(-\frac{\Delta E_{H(l,u)}}{T_{e}}\right)}
\end{equation}

In addition, an electron impacting a $H^{+}$ may recombine with it to become a $H$.
A `three-body recombination' is where the excess energy of the recombination is transferred to the second (not recombining) electron.
This is the inverse process of the electron impact ionization.
A `radiative recombination' is where the excess energy is lost radiatively.
Also, an impacting electron may be attached to a $H$, turning it into a negative ion, $H^{-}$, and emitting excess energy radiatively.
This is called a `radiative attachment.'
However, these three transitions are negligible compared to excitation and ionization, unless the electron temperature is very low \cite{Wunderlich2016}, and these are therefore not considered much in this work.
All transitions mentioned above are shown with the energy levels in Figure \ref{fig:H energy level diagram with important transitions}.

\subsubsection{Rate coefficient calculation}

\begin{figure}[htbp]
	\centering
    \includegraphics[width=16cm]{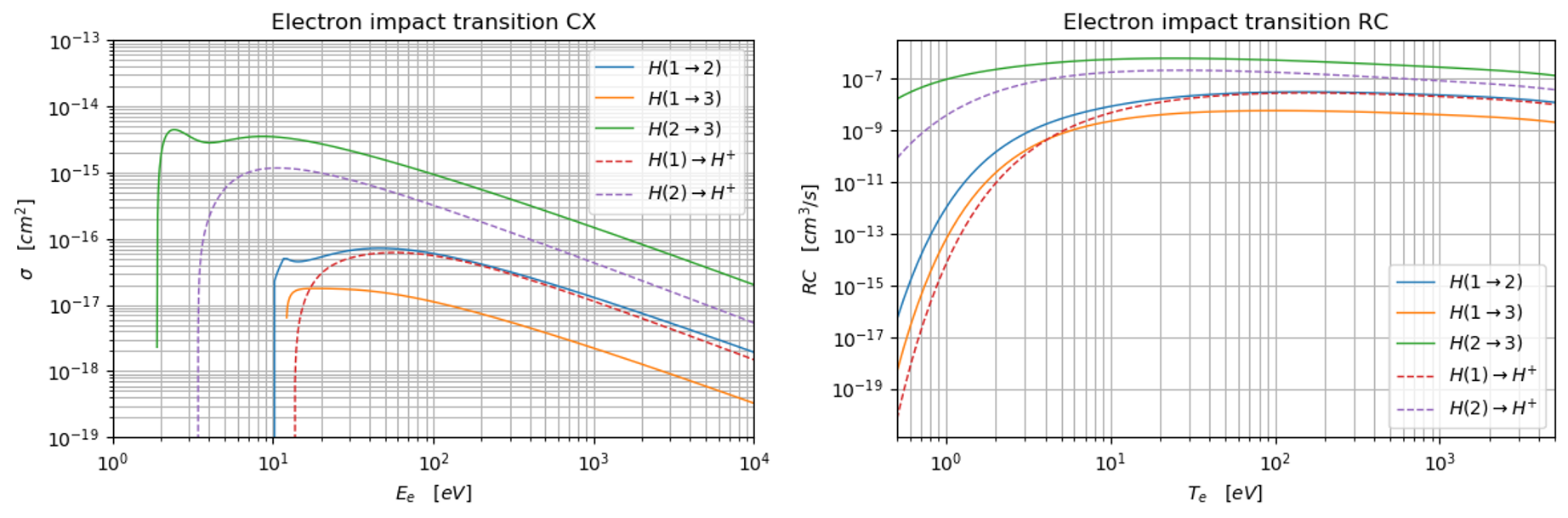}
	\caption[Example cross sections and corresponding rate coefficients\cite{Janev2003}
	]{Example cross sections and corresponding Maxwellian rate coefficients\cite{Janev2003}
	}\label{fig:Example cross sections and corresponding rate coefficients}	
\end{figure}

The Maxwellian distribution function has the forms of
\begin{equation} \label{eq:evdf}
f_{EVDF}(v_{e}, T_{e})dv_{e} = (\frac{m_{e}}{2\pi T_{e}})^{3/2}\exp(-\frac{\frac{1}{2}m_{e}v^{2}_{e}}{T_{e}})4\pi v^{2}_{e}dv_{e}
\end{equation}
\begin{equation} \label{eq:EEDF}
f_{EEDF}(E_{e}, T_{e})dE_{e} = \frac{2}{\sqrt{\pi}}\frac{1}{T^{3/2}_{e}}\sqrt{E_{e}}\exp(-\frac{E_{e}}{T_{e}})dE_{e}
\end{equation}
where $f_{EVDF}(v_{e})$ is the electron velocity distribution function(EVDF), and $f_{EEDF}(E_{e})$ is the electron energy distribution function(EEDF).
Typically, a $\sigma$ from a database is given as a function of $E_{e}$, and therefore, the expectation value is calculated with a EEDF.
The formula for the Maxwellian $RC$, is

\begin{equation} \label{eq:rcdetail}
	\begin{aligned}
	RC(T_{e})& = \langle \sigma (E_e)v_{e} \rangle _{f_{EEDF}}\\
	&=\int_{E_{thrs}}^{\infty}{\sigma (E_e)v_{e}f_{e}(E_e, T_{e})dE_e}\\
	&=\sqrt{\frac{8}{\pi m_{e}}}\frac{1}{T_{e}^{3/2}}\int_{E_{thrs}}^{\infty}{\sigma (E_e)E_{e}\exp (-\frac{E_e}{T_{e}})dE_e}
	\end{aligned}
\end{equation}
where $E_{thrs}$ is the threshold energy of the transition.
Figure \ref{fig:Example cross sections and corresponding rate coefficients} shows some example cross sections and Maxwellian $RC$s.

The population models that are explained in the later section involve numerous transition rates.
Since the calculation of $RC$s require integration processes, using the population models starting from cross section data every time for different EEDFs may become computationally intensive.
Therefore, in order to efficiently use the population models, databases for Maxwellian $RC$s for $T_{e}$ in the range of interest are usually constructed in advance.

\begin{figure}[htbp]
	\centering
    \includegraphics[width=10cm]{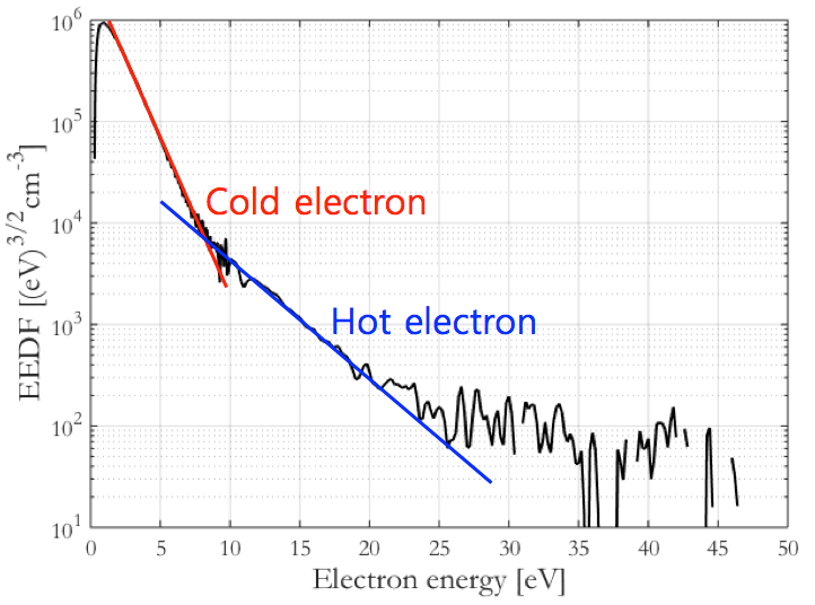}
	\caption[A bi-Maxwellian EEDF showing cold electron and hot electron dominant regions
	]{A bi-Maxwellian EEDF showing cold electron and hot electron dominant regions
	}\label{fig:A bi-Maxwellian EEDF showing cold electron and hot electron dominant regions}	
\end{figure}

EEDFs of interest are however not always Maxwellian and may deviate from it, as explained in Chapter 2.
The EEDFs are measured to bi-Maxwellian from the plasmas generated in this work.
One example bi-Maxwellian EEDF is shown in Figure \ref{fig:A bi-Maxwellian EEDF showing cold electron and hot electron dominant regions}.
The relative amounts of both cold electrons and hot electrons are important for $RC$s, since the transitions with bigger energy gaps are caused by more energetic electrons, while those with smaller energy gaps are caused by less energetic electrons.
Therefore, these detailed shape of EEDFs must be considered to obtain accurate $RC$s.

\begin{figure}[htbp]
	\centering
    \includegraphics[width=14cm]{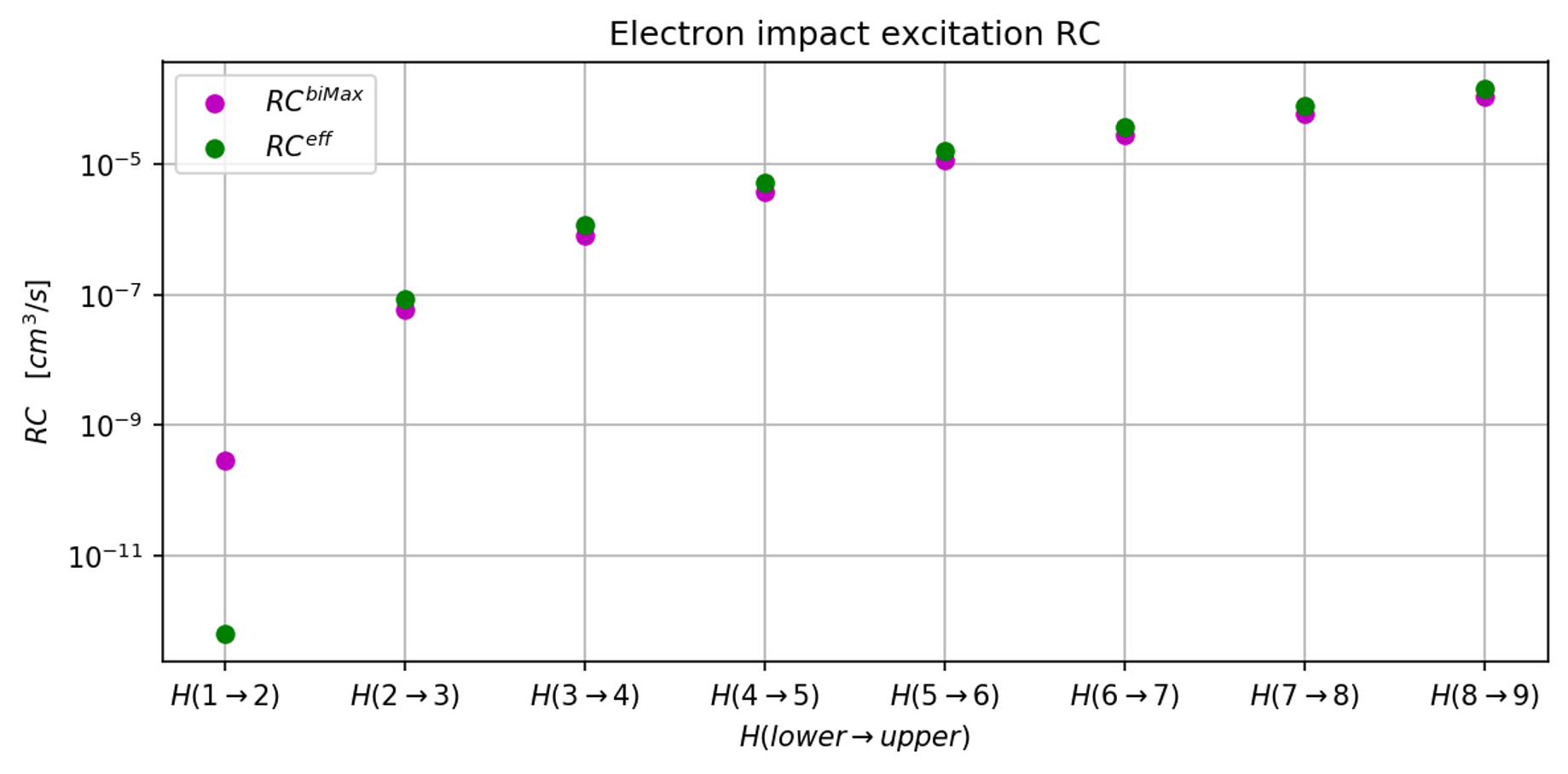}
	\caption[Compare bi-Maxwellian $RC$ and effective $RC$
	]{Compare bi-Maxwellian $RC$ and effective $RC$. $90\%$ cold electron with $T^{c}_{e}=0.5eV$, and $10\%$ hot electron with $T^{h}_{e}=5eV$
	}\label{fig:Compare RCs}
\end{figure}

In order to avoid inefficient calculation of $RC$s for every bi-Maxwellian EEDF for the population models, a simpler approach is developed in this work.
By utilizing the fact that a bi-Maxwellian EEDF is a weighted superposition of two mono-Maxwellian EEDF, the following formula is derived

\begin{equation} \label{eq:bimaxEEDF}
f_{biMax}(E_{e}, T_{e}) = \frac{n^{c}_{e}}{n_{e}}f_{Max}(E_{e}, T^{c}_{e})+\frac{n^{h}_{e}}{n_{e}}f_{Max}(E_{e}, T^{h}_{e})
\end{equation}
\begin{equation} \label{eq:bimaxrc}
RC^{biMax}(T_{e}) = \frac{n^{c}_{e}}{n_{e}}RC^{Max}(T^{c}_{e})+\frac{n^{h}_{e}}{n_{e}}RC^{Max}(T^{h}_{e})
\end{equation}
Here, $n_{e}=n^{c}_{e}+n^{h}_{e}$.
Equation \ref{eq:bimaxrc} allows a fast derivation of $RC^{biMax}(T_{e})$, with constructed $RC^{Max}(T_{e})$ database.
In the past, detailed shapes of EEDFs were not considered \cite{Lavrov2006}, rather an effective Maxwellian EEDF was used.
For an effective Maxwellian EEDF, an effective electron temperature, $T^{eff}_{e}=(n^{c}_{e}/n_{e})T^{c}_{e}+(n^{h}_{e}/n_{e})T^{h}_{e}$.
So, $RC^{eff}=RC^{Max}(T^{eff}_{e})$ is used for a simple approximation.
However, this may lead to significantly different $RC$ values for the reason mentioned above.
Figure \ref{fig:Compare RCs} compares the bi-Maxwellian $RC$ and the effective $RC$ for selected transitions, with $90\%$ cold electron with $T^{c}_{e}=0.5eV$, and $10\%$ hot electron with $T^{h}_{e}=5eV$.
Clearly, $RC^{eff}$ deviates from the correct values.
For $H(1 \rightarrow 2)$, $RC^{eff}$ underestimates the value since it underestimates the energetic electrons contributing to this big energy gap transition.
The opposite is true for other small energy gap transitions.
Thus, for this reason, Equation \ref{eq:bimaxrc} is adopted for rate coefficient calculations in this work.

\subsubsection{Radiation trapping effect}

There are other kinds of radiative transitions, along with spontaneous emission explained above.
The inverse transition of spontaneous emission is `photon absorption,' where a photon with an energy equivalent to a certain energy gap of states becomes absorbed by the lower state and produces the upper state.
Also, an emission may also occur with a photon with the same energy stimulating the upper state.
This is called `stimulated emission.'
Figure \ref{fig:The mechanisms of spontaneous emission, photon absorption, and stimulated emission} shows the mechanisms of the three radiative transitions.
Since some photons are not free to escape plasmas but re-absorbed by neutrals, this phenomenon is called `radiation trapping effect.'
Plasmas are said to be `optically thick,' if photon absorptions occur significantly in them.
Plasmas are `optically thin' for the opposite case.
The optical thickness depends on photons with different wavelengths.

\begin{figure}[htbp]
	\centering
    \includegraphics[width=12cm]{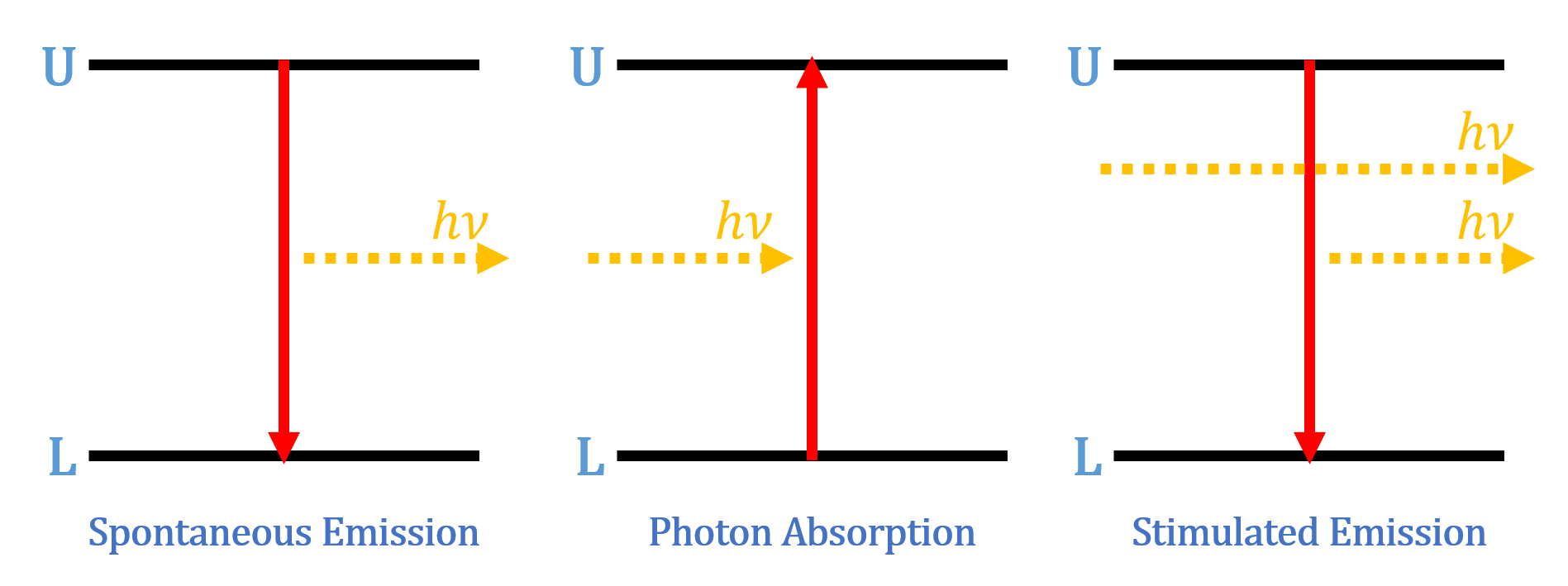}
	\caption[The mechanisms of spontaneous emission, photon absorption, and stimulated emission
	]{The mechanisms of spontaneous emission, photon absorption, and stimulated emission
	}\label{fig:The mechanisms of spontaneous emission, photon absorption, and stimulated emission}	
\end{figure}

The total flux of radiative transitions is then
\begin{equation} \label{eq:radiative}
-\left(\frac{dn_{H(u)}}{dt}\right)_{rad.}=\left(\frac{dn_{H(l)}}{dt}\right)_{rad.} = A_{H(u \rightarrow l)}n_{H(u)}
- B_{H(l \rightarrow u)}n_{H(l)}u_{\nu} + B_{H(u \rightarrow l)}n_{H(u)}u_{\nu}
\end{equation}
where $B_{H(l \rightarrow u)}[J^{-1}cm^{3}s^{-2}]$ and $B_{H(u \rightarrow l)}[J^{-1}cm^{3}s^{-2}]$ are Einstein B coefficients, and $u_{\nu}[Jcm^{-3}Hz^{-1}]$ is a spectral energy density. There is a different unit of Einstein B coefficient that is used, when spectral radiances are used instead of spectral energy densities. Einstein B coefficients are constants similar to Einstein A coefficients, and they are related by the following formula.
\begin{equation} \label{eq:Einsteincoefficientsrelation}
\frac{A_{ul}}{B_{ul}}=\frac{8\pi h \nu^{3}}{c^{3}}, \quad \frac{B_{lu}}{B_{ul}}=\frac{g_{u}}{g_{l}}
\end{equation}

Photon absorptions must be considered when optically thick.\footnote{Stimulated emissions are usually neglected since upper state densities are several orders smaller than lower state densities. Spontaneous emissions and photon absorptions are significant mostly.}
The calculation of the flux of radiative transitions is difficult, since it is nearly impossible to track all photons present in plasmas.
Therefore, an approximation technique, `optical escape factor(OEF)' method, is used to include photon absorption effect.
The idea is to reduce spontaneous emission rate by OEFs to compensate for the inverse absorption process \cite{Fujimoto2004, Holstein1947, Holstein1951}.

\begin{equation} \label{eq:effectiveA}
A^{eff}_{ul}=\Lambda_{ul} A_{ul}
\end{equation}
\begin{equation} \label{eq:OEF}
\Lambda_{ul}(k_{0, ul}R) \cong \frac{1.6}{(k_{0, ul}R)(\sqrt{\pi \log{(k_{0, ul}R)})}}
\end{equation}
\begin{equation} \label{eq:MOD}
k_{0, ul}R = \frac{1}{8 \pi^{3/2}}\frac{g_{u}}{g_{l}}\frac{A_{ul}\lambda^{3}_{ul}}{v_{n}}n_{l}R
\end{equation}
where $\Lambda_{ul}$ is the OEF($=0 \,-\, 1$)($\Lambda_{ul} \cong 1$ means the plasma is optical thin for this particular transition.), and $k_{0, ul}R$ is the mean optical depth(MOD), for Doppler-broadened line case.
$R$ is a plasma radius, $\lambda_{ul}$ is the wavelength of the photon with corresponding energy, and $v_{n}=\sqrt{\frac{2T_{n}}{m_{n}}}$ is the thermal velocity of the absorbing neutral.

\begin{figure}[htbp]
	\centering
    \includegraphics[width=12cm]{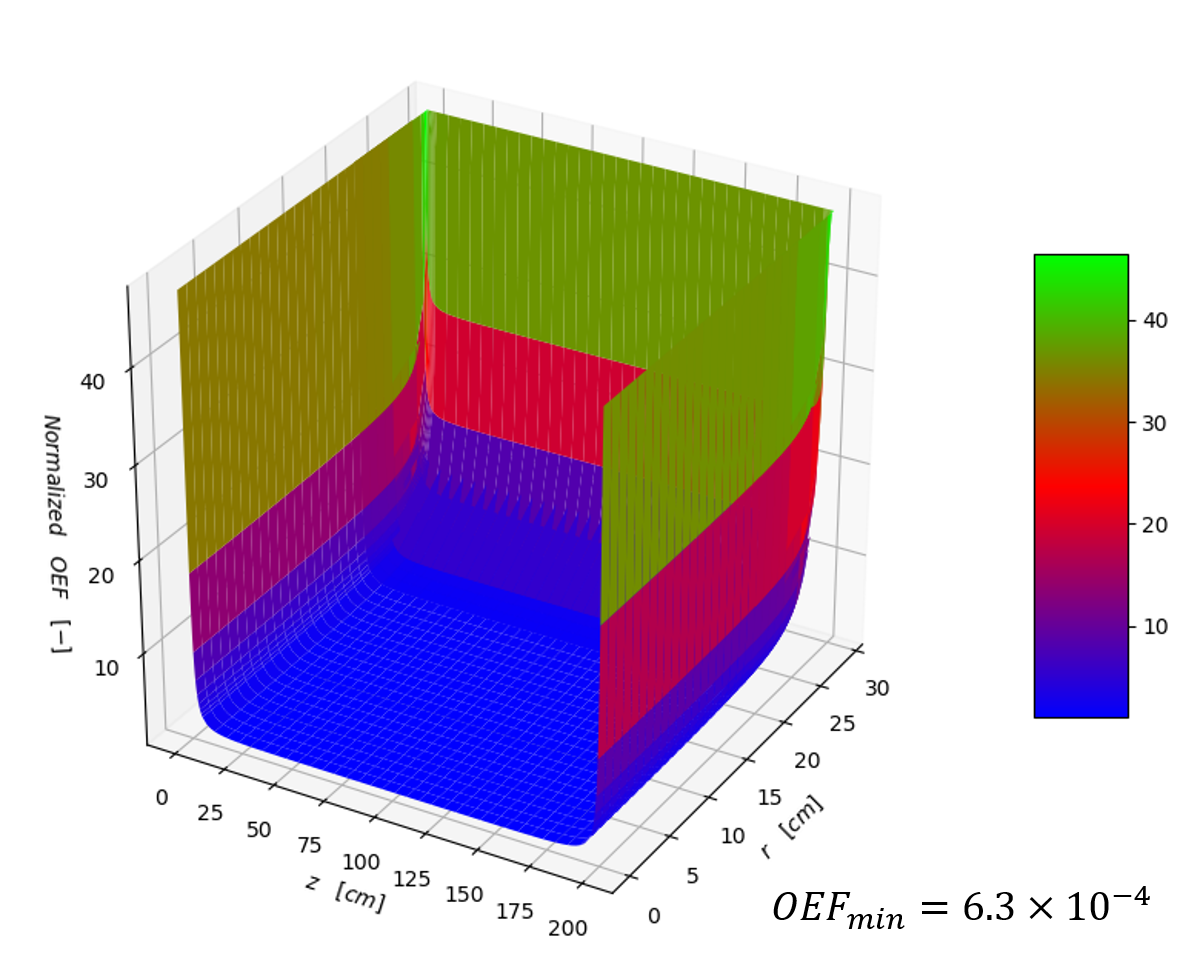}
	\caption[An arbitary normalized OEF using Equation \ref{eq:bahnOEF}
	]{An arbitary normalized OEF using Equation \ref{eq:bahnOEF}. The minimum of OEF is shown.
	}\label{fig:An arbitary normalized OEF using Equation}	
\end{figure}

Equation \ref{eq:OEF} is an approximate formula derived with the eigenmode analysis \cite{Fujimoto2004}, and it differs by references.
For a more rigorous and spatially-resolved formula, a detailed OEF is derived in this work, following similar steps of \cite{Golubovskii2009}.
A newly derived OEF formula for Doppler-broadened line of a finite-size cylindrical plasma is

\begin{equation} \label{eq:bahnOEF}
\Lambda(r,z)=\frac{1}{2\pi}\int_{0}^{\pi}{\left[K(\frac{L-z}{\xi},\frac{L-z}{q_{0}},\infty)+K(q_{0},\frac{-z}{q_{0}},\frac{L-z}{q_{0}})+K(\frac{z}{\xi},\frac{z}{q_{0}},\infty)\right]d\psi}
\end{equation}
\begin{equation} \label{eq:bahnK}
K(l, \xi_{1}, \xi_{2})=\int_{\xi_{1}}^{\xi_{2}}{T(l\sqrt{1+\xi^2})\frac{d\xi}{(1+\xi^2)^{3/2}}}
\end{equation}
\begin{equation} \label{eq:bahnq0}
q_{0}(r, \psi)=r\cos{\psi}+\sqrt{R^2-r^2\sin^2{\psi}}
\end{equation}
\begin{equation} \label{eq:bahnT}
T(\rho)=\int_{-\infty}^{\infty}{\frac{1}{\sqrt{\pi}}e^{-x^2}\exp[{k_{0}e^{-x^2}\rho}]dx}\\\approx \frac{1}{k_{0}\rho\sqrt{\pi\log{k_{0}\rho}}} 
\end{equation}
where $r, z$ are the position coordinate of interest, and $R, L$ are the radius and the length of the cylindrical plasma.
Since the position of the plasma to be analyzed in this work is at the radial center and reasonably away from the end of the chamber, Equation \ref{eq:effectiveA} and Equation \ref{eq:bahnOEF} do not differ much, but this formula is expected to be utilized for a future work involving a spatially-resolved plasma analysis.
An arbitrary OEF with respect to coordinates is plotted (Figure \ref{fig:An arbitary normalized OEF using Equation}).
It is normalized by dividing all OEFs by the minimum value($OEF_{min}=6.3\times 10^{-4}$).
It shows that at the edge of the cylindrical plasma, the OEF is reaching its peak value($=1$), and at the center of the cylinder, the OEF is many times smaller.
With a significantly small OEF, a spontaneous emission rate is reduced many orders of magnitude.
This approach of the radiation trapping effect is applied to all spontaneous emissions, of which OEFs are significantly small, which are mostly those involving the ground state as the lower state.

\section{Hydrogen molecule state}

\subsection{Energy level}

\begin{figure}[htbp]
	\centering
    \includegraphics[width=14cm]{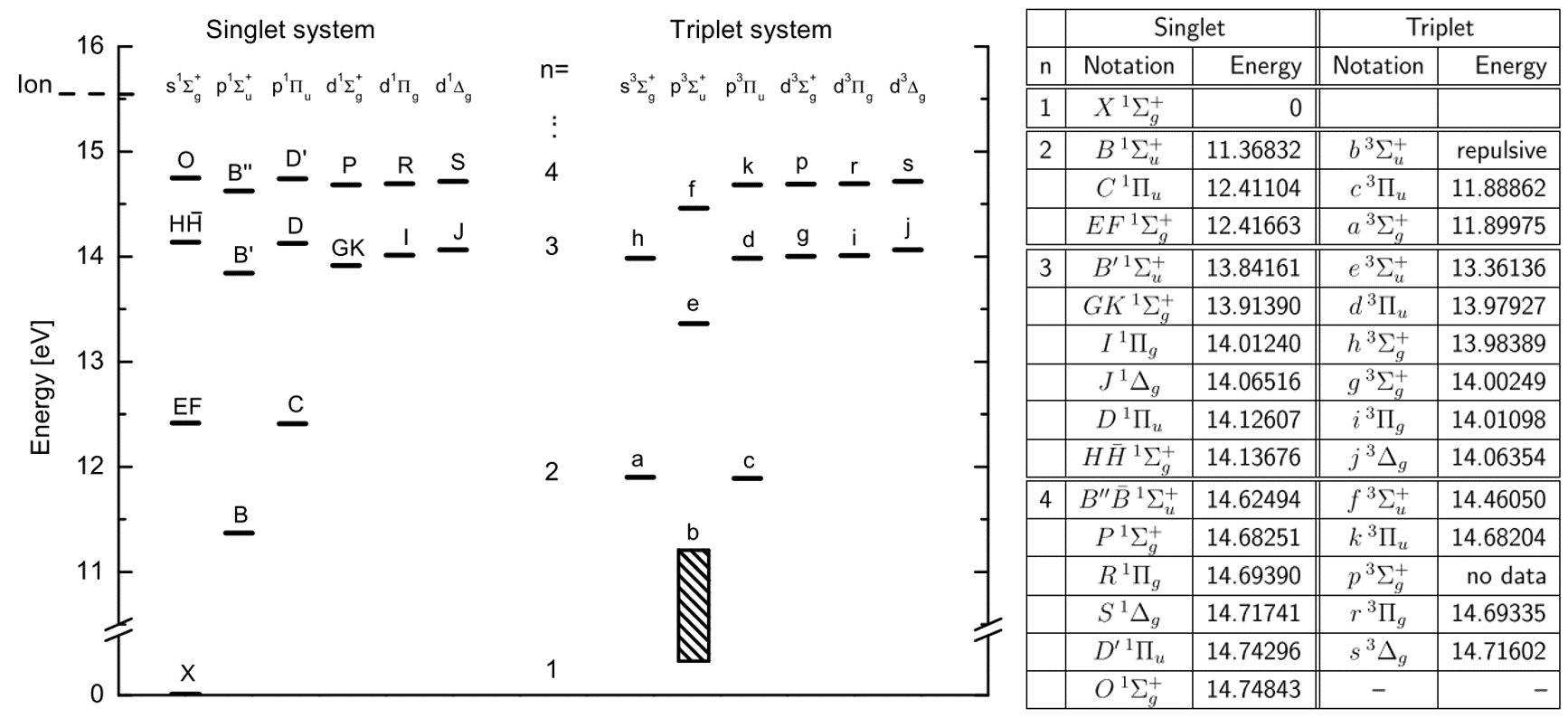}
	\caption[$H_{2}$ energy levels upto $n=4$
	]{$H_{2}$ energy levels upto $n=4$, and corresponding energies in $[eV]$\cite{Fantz2006}
	}\label{fig:H2 energy levels}	
\end{figure}

A hydrogen molecule, $H_{2}$, is a homonuclear diatomic molecule, and has two nuclei and two electrons bonding them together.
Because of its complex nature, $H_{2}$ states are not as simple as $H$ states.
The energy level diagram of $H_{2}$ is shown in Figure \ref{fig:H2 energy levels} \cite{Fantz2006}.

First, the principal quantum number divides the states.
For each principal quantum number, the azimuthal quantum number, $l(=0, 1, 2,...=s, p, d,...)$, divides them further.
The figure shows that the states are divided into two groups, `singlet' and `triplet' states.
This is due to the spins of two bound electrons in the system.

Two electrons occupying a shell have opposite spins by the Pauli exclusion principle.
When one electron is excited, it is in a different shell, and here it can either keep the opposite spin or have the same spin as the unexcited electron.
When the two electrons keep opposite spins(anti-parallel), only one state, $\{ (\uparrow \downarrow - \downarrow \uparrow)/\sqrt{2} \}$, is possible.
When the two electrons become the same spins(parallel), three states, $\{ \uparrow \uparrow, \downarrow \downarrow, (\uparrow \downarrow + \downarrow \uparrow)/\sqrt{2} \}$ , are possible.
Therefore, their multiplicity, $M$, are 1 and 3, respectively.
For $H$, there is only one electron in a shell initially, so two states, $\{ \uparrow, \downarrow \}$, are possible for the electron.
That is why all $H$ states are `doublet' and have multiplicity of 2, as explained in Section 3.1.

The molecular term symbol for states is much more complicated than the atomic one.
It has the form of $(state\_name)^{M}\Lambda^{+/-}_{g/u}$.
$\Lambda(=0, 1, 2...=\Sigma, \Pi, \Delta,...)$ is the total orbital angular momentum quantum number, similar to $L$ for the atomic one.
$(g/u)$ is the parity.
The orbital of a molecule is either symmetric($g$ for gerade) or anti-symmetric($u$ for ungerade), when inverted through the center.
$(+/-)$ is the reflection.
The orbital is either symmetric($+$) or anti-symmetric($-$) when reflected through a plane containg both nuclei.(For a state that has both $+$ and $-$, $(+/-)$ is omitted (Figure \ref{fig:H2 energy levels})).
Finally, each state is named with letters, in front of the term symbol for convenience.
Conventionally, the ground state is labeled the letter $X$, and singlets and triplets are labeled with upper and lower case letters, respectively.

\begin{figure}[htbp]
	\centering
    \includegraphics[width=16cm]{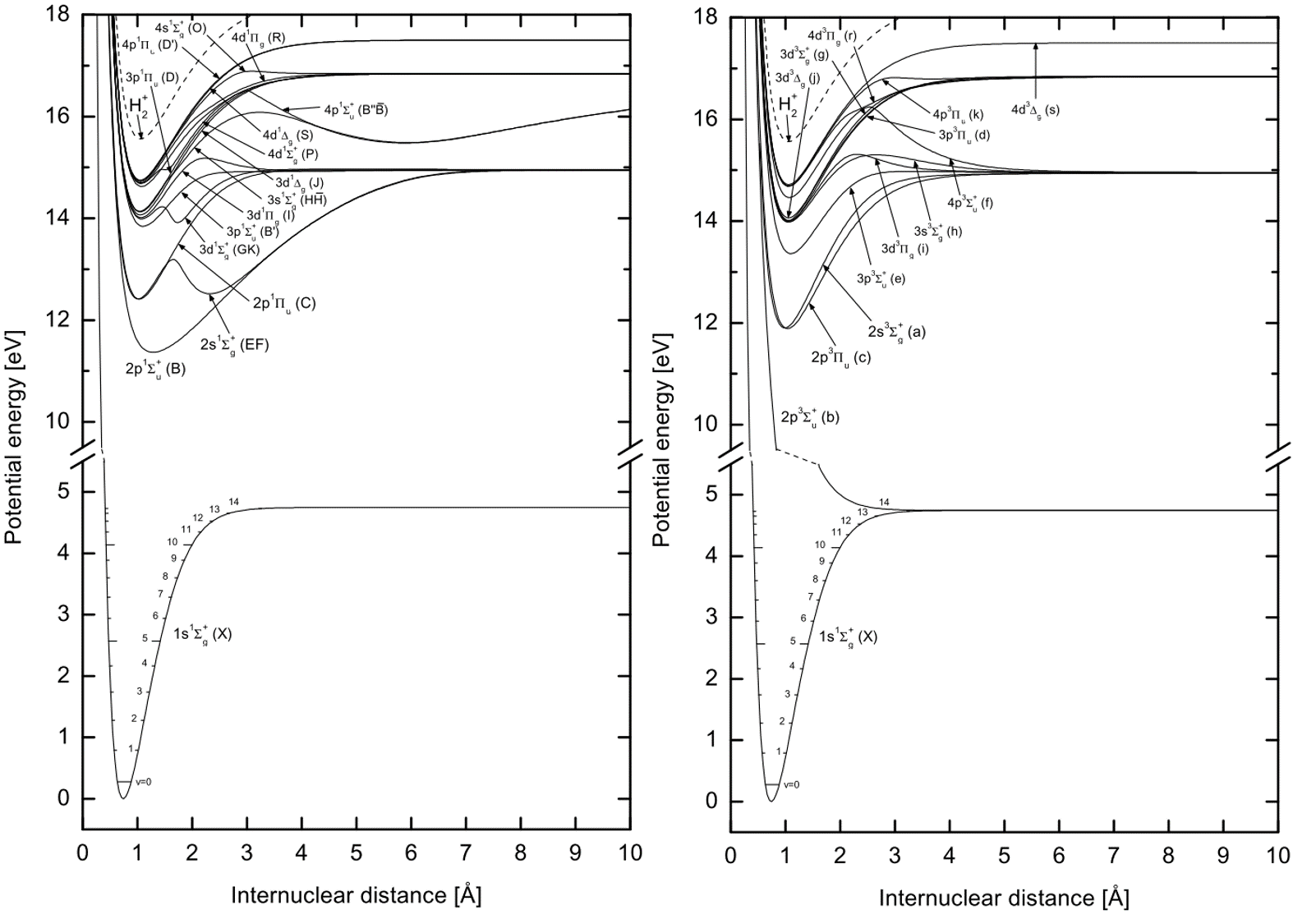}
	\caption[Potential curves of singlet(left) and triplet(right) states
	]{Potential curves of singlet(left) and triplet(right) states\cite{Fantz2006}
	}\label{fig:Potential curves of singlet(left) and triplet(right) states}
\end{figure}

\begin{figure}[htbp]
	\centering
    \includegraphics[width=16cm]{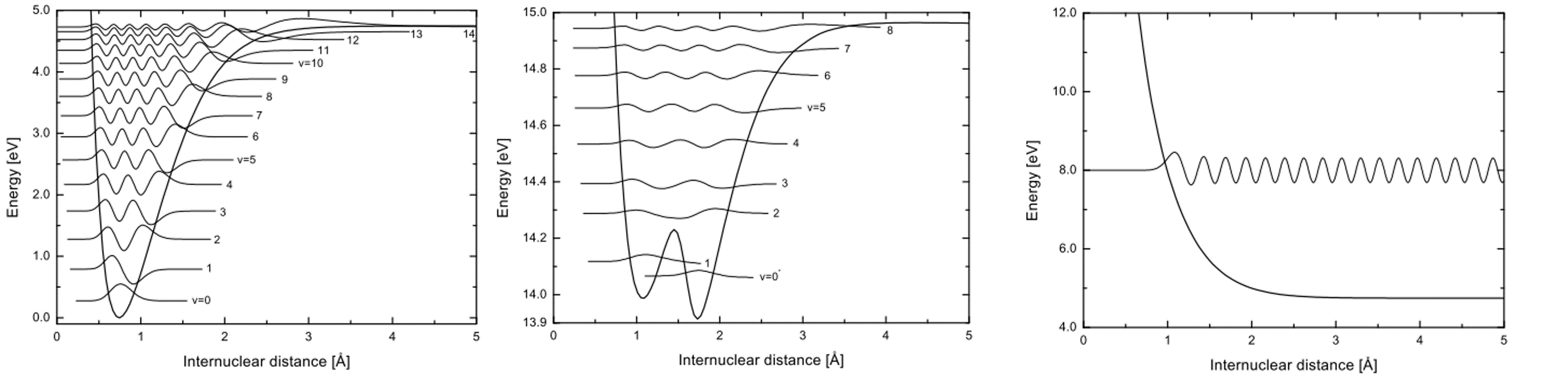}
	\caption[Potential curves and corresponding vibrational states, of $X$, $GK$, and $b$, respectively
	]{Potential curves and corresponding vibrational states, of $X$, $GK$, and $b$, respectively\cite{Fantz2006}
	}\label{fig:Potential curves and corresponding vibrational states}	
\end{figure}

Different kinds of $H_{2}$ states and their meanings are explained.
On the right of Figure \ref{fig:H2 energy levels} are the energies of states upto $n=4$.
There are states with two letters, and also, the $b$ state is `repulsive.'
These can be understood easily with Figure \ref{fig:Potential curves of singlet(left) and triplet(right) states} and \ref{fig:Potential curves and corresponding vibrational states}.

This figure shows the potential curves of $H_{2}$ states, with the horizontal axis as the distance between two bonded nuclei.
There are several important points here.

There is the minimum potential energy location for each state.
The ground, the $X$ state, has the minimum potential energy near $1 \AA$.
Classically, the internuclear distance of this should be kept at minimum if there is no vibrational motion.
However, the energy of a molecular vibrational motion is quantized and the energy is never zero.
There are finite discrete vibrational energies following the potential curve, and they are characterized by `vibrational quantum number', $v(=0, 1, 2,...v_{max})$.
`Excited' states described till now in this work are more rigorously `electronically excited' states, and vibrational quantum numbers describe additional `vibrational states.'
For the $X$ state, $v_{max}=14$, and so there are 15 discrete vibrational states (Figure \ref{fig:Potential curves and corresponding vibrational states}).
A vibrational state energy is calculated by
\begin{equation} \label{eq:venergy}
E_{vib}\frac{1}{hc} = \omega_{e} (v+\frac{1}{2}) - \omega_{e} \chi_{e} (v+\frac{1}{2})^{2}
\end{equation}
where $c$ is the speed of light, $\omega_{e}$ is the harmonic wavenumber, and $\chi_{e}$ is the anharmonicity constant.
Both $\omega_{e}$ and $\chi_{e}$ depend on the electronic state.
$E_{vib}(v_{max})$ would be the maximum vibrational energy.
However, a state may have higher vibrational energy and be at `vibrational continuum,' but this would immediately lead to the dissociation of the molecule.

Figure \ref{fig:Potential curves and corresponding vibrational states} shows some selected potential curves, and corresponding vibrational states.
The plot in the middle of Figure \ref{fig:Potential curves and corresponding vibrational states} is of the $GK$ state, and there are two minima present in the potential curves.
All states with two letter names have two minima.
The last plot of Figure \ref{fig:Potential curves and corresponding vibrational states} is of the $b$ state.
It is repulsive, and as its description suggests, there is no minimum for this state.
Therefore, there is no discrete vibrational state, but only vibrational continuum.
A $H_{2}$ that reach the $b$ state immediately dissociates into atoms.

\begin{figure}[htbp]
	\centering
    \includegraphics[width=10cm]{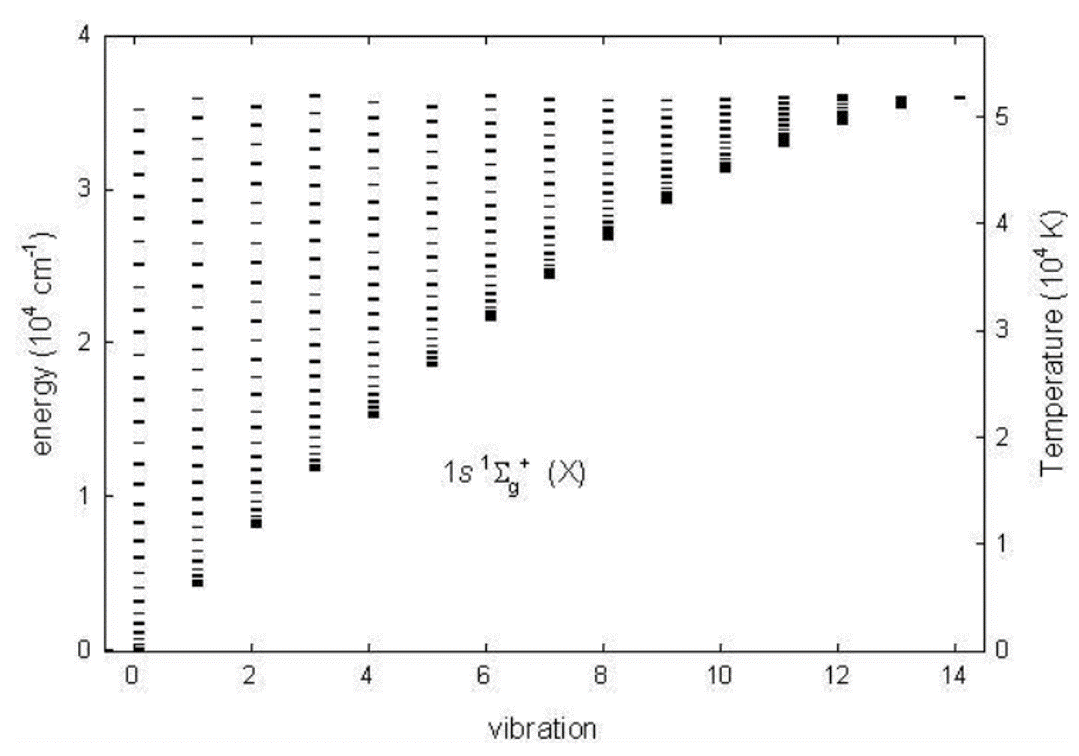}
	\caption[Rotational-vibrational levels of the $X$ state\cite{Shaw2005}
	]{Rotational-vibrational levels of the $X$ state\cite{Shaw2005}
	}\label{fig:Rotational-vibrational levels of the X state}	
\end{figure}

A molecule also `rotates' with a certain energy.
Rotational energies are also quantized, and rotational states are described by `rotational quantum numbers,' $J(=0, 1, 2,...)$.
A rotational energy gap is typically smaller than a vibrational energy gap.
Figure \ref{fig:Rotational-vibrational levels of the X state} shows rotational-vibrational energy levels of the electronic $X$ state.
A rotational state energy is calculated by
\begin{equation} \label{eq:renergy}
E_{rot}\frac{1}{hc} = B_{e} J(J+1) - D J^{2} (J+1)^{2}
\end{equation}
where $B_{e}$ is the rotational constant that depends on the electronic state, and $D$ is the centrifugal distortion constant.

Consequently, a state of a molecule ($H_{2}$) need electronic, vibrational, and rotational states to rigorously describe it.
There are conventional terms for describing coupled energy states.
When rotational and vibrational states are considered together, it is `ro-vibrational.'
For vibrational and electronic, it is `vibronic,' and for all three together, it is `ro-vibronic.'
The molecular term symbol can be modified into $(state\_name)^{M}\Lambda^{+/-}_{g/u}(v, J)$ or just $(state\_name)(v, J)$, to include vibrational and rotational numbers, for a ro-vibronic state.
The total energy of the state is then $E_{state}=E_{e}+E_{vib}+E_{rot}$.

\begin{table}[htbp]
	\caption[Statistical weights of molecule electronic states
	]{Statistical weights of molecule electronic states
	}
	\label{tb:Statistical weights of molecule electronic states}
	\begin{center}
		\begin{tabular} {ccccccccccc}
			\hline\hline
			& $ ^{M}\Lambda$ & $^{1}\Sigma$ & $^{2}\Sigma$ & $^{3}\Sigma$ & $^{1}\Pi$, $^{1}\Delta$ & $^{2}\Pi$, $^{2}\Delta$ & $^{3}\Pi$, $^{3}\Delta$ &\\
			\hline
			& $g_{e}$ & 1 & 2 & 3 & 2 & 4 & 6 &\\
			\hline\hline
		\end{tabular}
	\end{center}
\end{table}

Statistical weights for three kinds of states are calculated in different ways \cite{Ochkin2009}.
Here are the statistical weight formulas for specifically $H_{2}$.
Statistical weights of electronic states are shown in Table \ref{tb:Statistical weights of molecule electronic states}.
For a diatomic molecule, statistical weights of vibrational states are always $g_{v}=1$.
For a homonuclear diatomic molecules, statistical weights of rotational states are given by $g_{J}=g_{n}(2J+1)$, where $g_{n}$ is the statistical weight of the nuclear spin, and it is calculated in two forms (Equation \ref{eq:gns}).
The rule for choosing $g_{n}$ is given in Table \ref{tb:The rule for statistical weights of the nuclear spins of rotational states}.
Thus, the total statistical weight for a $H_{2}$ state is $g_{tot}=g_{e}g_{v}g_{J}$.

\begin{equation} \label{eq:gns}
g_{n}=
	\begin{cases}
	g_{n1}=I_{n}(2I_{n}+1) \\
	g_{n2}=(I_{n}+1)(2I_{n}+1)
	\end{cases}
\end{equation}

\begin{table}[htbp]
	\caption[The rule for statistical weights of the nuclear spins of rotational states \cite{Ochkin2009}
	]{The rule for statistical weights of the nuclear spins of rotational states \cite{Ochkin2009}
	}
	\label{tb:The rule for statistical weights of the nuclear spins of rotational states}
	\begin{center}
		\begin{tabular} {ccccccccccc}
			\hline\hline
			& $(+/-)$ & $+$ & $-$ & $-$ & $+$ &\\
			& $(g/u)$ & $g$ & $u$ & $g$ & $u$ &\\
			& $I_{n}$ & integer & half-integer & integer & half-integer &\\
			\hline
			& Even $J$ & $g_{n1}$ & $g_{n2}$ & $g_{n2}$ & $g_{n1}$ &\\
			& Odd $J$ & $g_{n2}$ & $g_{n1}$ & $g_{n1}$ & $g_{n2}$ &\\
			\hline\hline
		\end{tabular}
	\end{center}
\end{table}

\subsection{Transition}

\subsubsection{Spontaneous emission}

\begin{figure}[htbp]
	\centering
    \includegraphics[width=16cm]{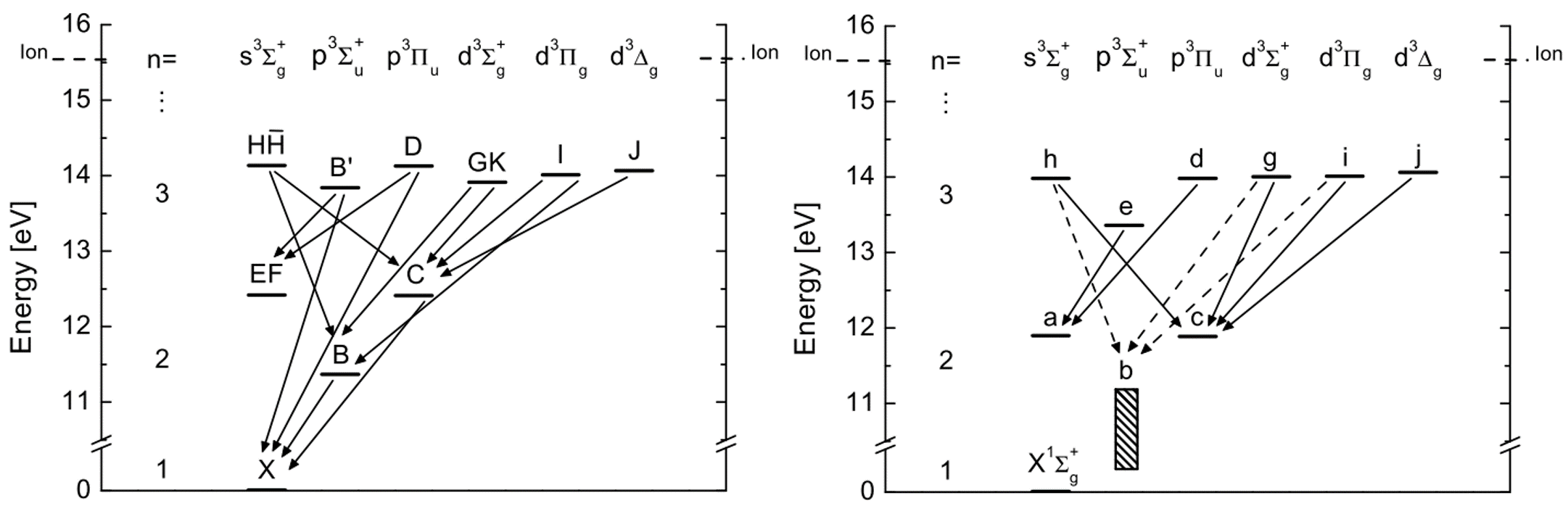}
	\caption[Optically allowed spontaneous emissions of singlet(left) and triplet(right) states\cite{Fantz2006}
	]{Optically allowed spontaneous emissions of singlet(left) and triplet(right) states\cite{Fantz2006}
	}\label{fig:Optically allowed spontaneous emissions of singlet(left) and triplet(right) states}	
\end{figure}

The basic idea of the spontaneous emission is explained in Subsection 3.1.2, and it is similar for $H_{2}$.
Figure \ref{fig:Optically allowed spontaneous emissions of singlet(left) and triplet(right) states} shows `optically allowed' spontaneous emissions of electronic states of $H_{2}$, with arrows.

The optically allowed transitions are the ones following the `selection rule' \cite{Herzberg1951}.
In this case, to be optically allowed, $l(=s, p, d,...)$ should only change by 1, and $M(=1, 3)$(multiplicity) should not change.
Other transitions not following this rule are said to be `forbidden.'
Forbidden spontaneous emissions are not impossible(Finite probabilities exist.) but many orders smaller than allowed ones, and thus negligible most of the time.
Therefore, the arrows in Figure \ref{fig:Optically allowed spontaneous emissions of singlet(left) and triplet(right) states} look diagonal.\footnote{The arrows are dashed for the lower $b$ state, to show they lead to dissociation.}
One quantitative example of comparing allowed and forbidden spontaneous emission rates is in Table \ref{tb:Example Einstein A coefficients}, where $A_{H(2^{2}S \rightarrow 1^{2}S)} \ll A_{H(2^{2}P \rightarrow 1^{2}S)}$.

\begin{figure}[htbp]
	\centering
    \includegraphics[width=8cm]{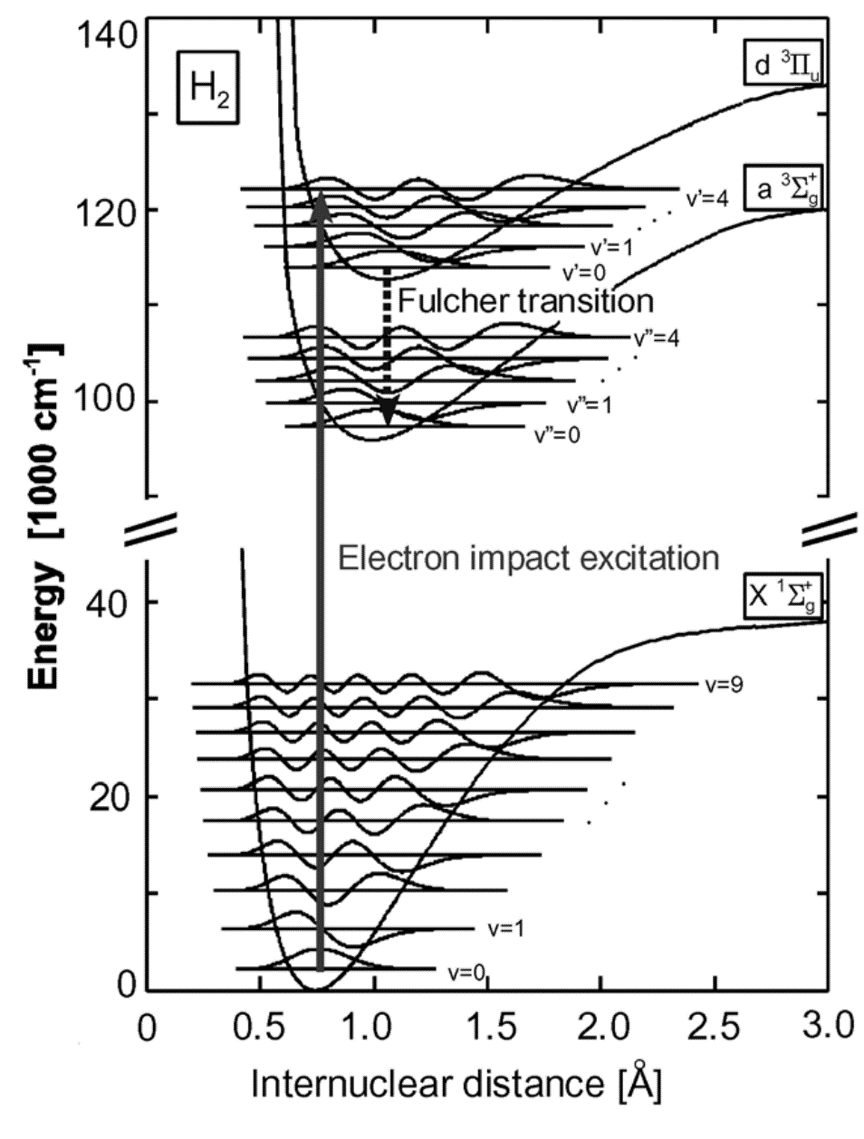}
	\caption[The Fulcher-$\alpha$ transition with potential curves and vibrational states\cite{Fantz2006b}
	]{The Fulcher-$\alpha$ transition with potential curves and vibrational states\cite{Fantz2006b}
	}\label{fig:The Fulcher transition with potential curves and vibrational states}	
\end{figure}

The influence of vibrational states on spontaneous emissions should be discussed as well.
For that, the `Franck-Condon principle' need to be explained first \cite{Herzberg1951}.
Since the $H_{2}$ spontaneous emission from the upper $d$ state to the lower $a$ state, the origin of the Fulcher-$\alpha$ spectrum, is of primary importance in this work, this transition/spectrum is used mostly as examples henceforth.
Figure \ref{fig:The Fulcher transition with potential curves and vibrational states} shows the Fulcher-$\alpha$ transition with potential curves.

The Franck-Condon principle states that, in a vibronic transition, a transition probability is proportional to the amount of overlaps of wave functions of corresponding vibrational states.
As explained earlier, a molecule in a vibrational state is not vibrating classically, but its energy is quantized and rather its internuclear distance due to the vibration is related to the corresponding wave function, quantum mechanically (Figure \ref{fig:Potential curves and corresponding vibrational states} and \ref{fig:The Fulcher transition with potential curves and vibrational states}).
And different electronic states have different potential curves and vibrational wave functions in them.
Since, the molecule tend to keep its internuclear distance when the electronic state change, the shape of wave functions of the initial and the final vibrational states and how much they overlap affects the transition probability.
If a pair of an initial and a final vibrational states overlap the most, the vibronic transition involving that two vibrational states is most likely to occur.
This rule applies to any vibronic transition, whether it is a spontaneous emission or an electron impact transition.

\begin{figure}[htbp]
	\centering
    \includegraphics[width=16cm]{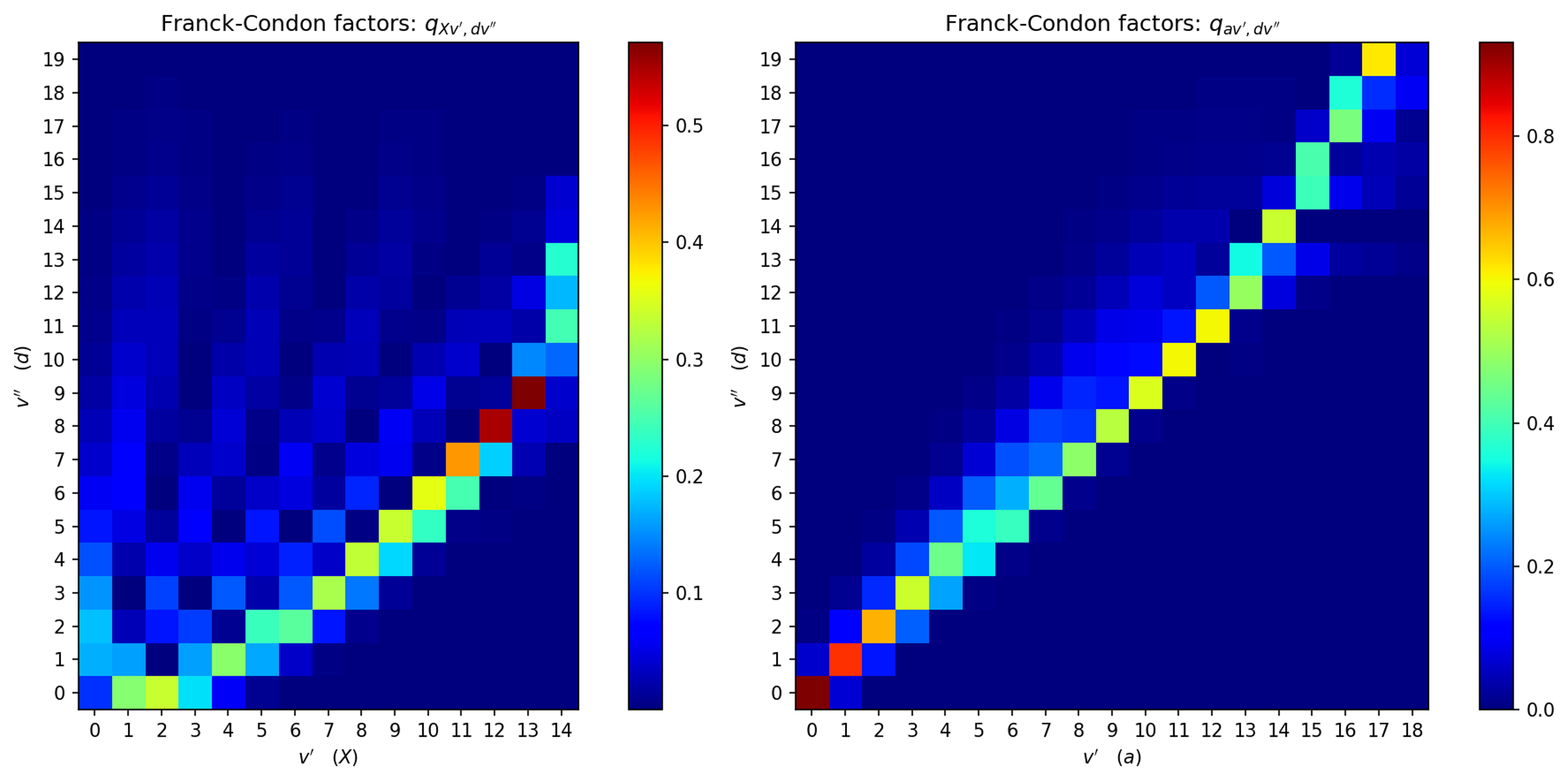}
	\caption[Franck-Condon factors of the $X, d$ states(left) and $a, d$ states(right)
	]{Franck-Condon factors of the $X, d$ states(left) and $a, d$ states(right)
	}\label{fig:Franck-Condon factors}	
\end{figure}

Spontaneous emission rates are therefore related to the Franck-Condon principle.
A vibrationally-resolved Einstein A coefficient of a molecule is proportional to \cite{Fantz2006}
\begin{equation} \label{eq:EinsteinAmolecule}
A_{E'(v') \rightarrow E''(v'')} \propto \frac{q_{E'(v'), E''(v'')}}{(\lambda_{E'(v') \rightarrow E''(v'')})^{3}}
\end{equation}
where $E', E''$ are the initial and final electronic state name(letter), $v', v''$ are the initial and final vibration number, and $\lambda_{E'(v') \rightarrow E''(v'')}$ is the corresponding wavelength of the emitted photon.
$q_{E'(v'), E''(v'')}$ is the `Franck-Condon factor(FCF)' that quantifies the Franck-Condon principle.
The directions of transitions do not matter for $q_{E'(v'), E''(v'')}$.
The FCFs of the $X$ state and the $d$ state, and the FCFs of the $a$ state and the $d$ state, for every possible pair of vibrational states are plotted in Figure \ref{fig:Franck-Condon factors}.

The $X, a, d$ states have $v$s upto 14, 18, 19, respectively.
Figure \ref{fig:Franck-Condon factors} shows which vibrational state pairs have the strongest wave function overlaps.
Due to the shift of potential curves between the $X$ state and the $d$ state(The molecular bond lengths change much.) (Figure \ref{fig:The Fulcher transition with potential curves and vibrational states}), the transitions from the $X(v)$ states with higher $v$s to the $d(v)$ states are more probable, as Figure \ref{fig:Franck-Condon factors}(left) shows the apparent diagonal line, indicating the largest FCFs, shited toward higher $v$s of the $X$ state.
On the other hand, since the $a$ state and the $d$ state have potential curves of a similar shape and size, $v$s are unlikely to change much in the transitions, as can be seen in Figure \ref{fig:Franck-Condon factors}(right).

\begin{figure}[htbp]
	\centering
    \includegraphics[width=15cm]{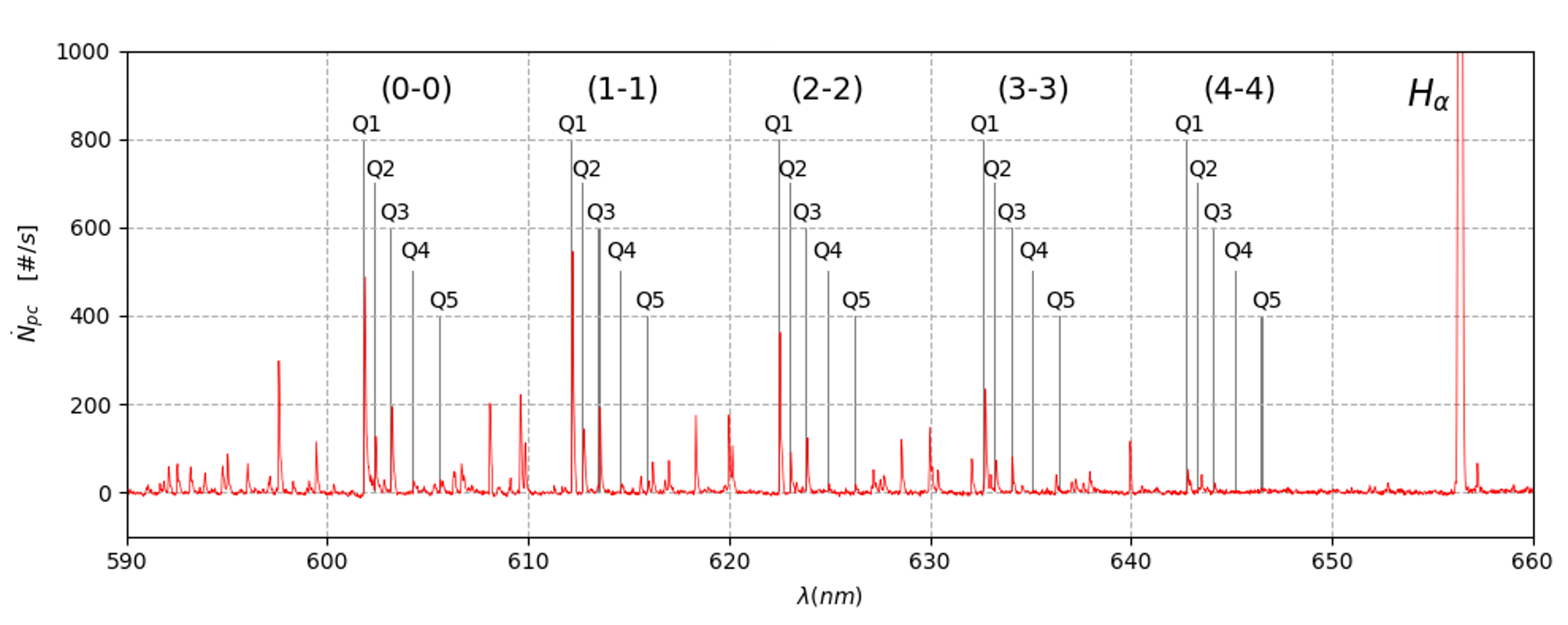}
	\caption[The measured raw spectrum zoomed in at the Fulcher-$\alpha$ band
	]{The measured raw spectrum zoomed in at the Fulcher-$\alpha$ band
	}\label{fig:The measured raw spectrum zoomed in at the fula band}	
\end{figure}

It can be therefore inferred from Equation \ref{eq:EinsteinAmolecule} and Figure \ref{fig:Franck-Condon factors}(right) that the strongest $d \rightarrow a$ spontaneous emissions are those with the same initial and final $v$s, for $v<5$.
This phenomenon can be seen in the measured spectra.
Figure \ref{fig:The measured raw spectrum zoomed in at the fula band} is the measured raw spectrum (Figure \ref{fig:measured_raw_spectrum}) zoomed in at the Fulcher-$\alpha$ band.
$(0-0)$, $(1-1)$, ... in Figure \ref{fig:The measured raw spectrum zoomed in at the fula band} are the representation of initial and final $v$s as $(v'-v'')$.
The strong lines in the Fulcher-$\alpha$ band are shown to be of $\Delta v = 0$, as explained.

Each $(v'-v')$ transition is also divided into several lines, denoted by $Q$.
This is due to the rotational structure.
The letter $Q$ means $\Delta J = 0$ for a decay.
For example, the $(2-2)Q3$ line is of $d(v=2, J=3) \rightarrow a(v=2, J=3)$.
Since the energy gaps differ depending on the rotational numbers, $J$s, the photon wavelengths for $Q$ lines differ as well.\footnote{In addition, there are $P$(poor) and $R$(rich) lines, corresponding to $\Delta J = +1$ and $\Delta J = -1$ for decays, respectively. Also, more rigorously, $Q$ lines are from $d^{3} \Pi^{-}_{u}$ while $P$ and $R$ lines are from $d^{3} \Pi^{+}_{u}$. And $d^{3} \Pi^{+}_{u}$ is said to be perturbed by other states, and so strong and clear $Q$ lines from $d^{3} \Pi^{-}_{u}$ are usually used to analyze the Fulcher-$\alpha$ band \cite{Qing1996}.}

\subsubsection{Other transition}

The electron impact excitation from the $X$ state to the $d$ state will be considered mostly in this work.
The basic idea of electron impact transition of $H_{2}$ and its rate calculation method are the same as explained in Subsection 3.1.2.
However, cross sections of molecular electronic transitions are usually given in vibrationally-resolved ways \cite{Janev2003}.
For example, there is an electron impact vibronic excitation cross section from the $X(v')$ state of a certain $v'$ to the $d(v'')$ state with all $v''$s, $\sigma^{i.e.}_{X(v') \rightarrow d}$.
Then, the following relations hold.
\begin{equation} \label{eq:CXwithFCFs}
\sigma^{i.e.}_{X(v') \rightarrow d(v'')} = {q_{X'(v'), d''(v'')}} \times \sigma^{i.e.}_{X(v') \rightarrow d}
\end{equation}
\begin{equation} \label{eq:CXwithFCDs}
\sigma^{i.e.}_{X(v') \rightarrow d(vcon)} = {q_{X'(v'), d''(vcon)}} \times \sigma^{i.e.}_{X(v') \rightarrow d}
\end{equation}
\begin{equation} \label{eq:FCFs1}
\sum_{v''}^{vmax''}{{q_{X'(v'), d''(v'')}}}+{q_{X'(v'), d''(vcon)}}=1
\end{equation}
where $vcon$ is the vibrational continuum.
These equations mean, out of the total excitation, the rates for each final $v''$ is determined by the FCFs.
The sum of all FCFs for a fixed $v'$ is smaller than 1.
The remainder is the Franck-Condon continuum factor.\footnote{This is named so in this work. It is related to the Franck-Condon density \cite{Janev2003}.}
So, the sum of all FCFs plus this continuum factor is 1, and all these factors are dividing the total cross section(or the rate) to vibrationally-resolved ones.
Therefore, with the knowledge of the total cross section and the corresponding FCFs, the upper vibrational distribution for the vibronic excitation can be known.

Equation \ref{eq:CXwithFCDs} is the cross section of the excitation to the continuum, which leads to dissociation as explained in the Subsection 3.2.1.
This kind of transitions is called `dissociative excitation,' and it is particularly important in that it affect the production of excited $H$.
Dissociations of $H_{2}$ at different electronic states produce $H$ at different states, with one of the two produced $H$ always being the ground, $1^{2}S$ (Table \ref{tb:Electronic state pairs singlet} and \ref{tb:Electronic state pairs triplet}) \cite{Janev2003}.

\begin{table}[htbp]
	\caption[Electronic state pairs of $H_{2}$ and $H$ for dissociations (Singlet $H_{2}$)]{
	Electronic state pairs of $H_{2}$ and $H$ for dissociations (Singlet $H_{2}$)}
	\label{tb:Electronic state pairs singlet}
	\centering
	\begin{tabular}{ccccccccccc}
		\hline\hline
		& $n$ & \multicolumn{6}{c}{$H_{2}(E)(\text{singlet}) \rightarrow H(n^{2}L)$} & & & \\ 
		\hline
		& $2$ & $EF \rightarrow 2^{2}S$ & $B \rightarrow 2^{2}P$ & $C \rightarrow 2^{2}P$ & & & & & & \\ 
		& $3$ & $H \bar{H} \rightarrow 3^{2}D$ & $B' \rightarrow 2^{2}S$ & $D \rightarrow 3^{2}D$ & $GK \rightarrow 2^{2}P$ & $I \rightarrow 2^{2}P$ & $J \rightarrow 3^{2}D$ & & & \\
		& $4$ & $O \rightarrow 3^{2}P$ & $B'' \rightarrow 3^{2}S$ & $D' \rightarrow 4^{2}P$ & $P \rightarrow 3^{2}D$ & $R \rightarrow 3^{2}P$ & $S \rightarrow 4^{2}F$ & & & \\
		\hline\hline
	\end{tabular}
\end{table}

\vspace{0.5cm} % small vertical space between tables if needed

\begin{table}[htbp]
	\caption[Electronic state pairs of $H_{2}$ and $H$ for dissociations (Triplet $H_{2}$)]{
	Electronic state pairs of $H_{2}$ and $H$ for dissociations (Triplet $H_{2}$)}
	\label{tb:Electronic state pairs triplet}
	\centering
	\begin{tabular}{ccccccccccc}
		\hline\hline
		& $n$ & \multicolumn{6}{c}{$H_{2}(E)(\text{triplet}) \rightarrow H(n^{2}L)$} & & & \\ 
		\hline
		& $2$ & $a \rightarrow 2^{2}S$ & $b \rightarrow 1^{2}S$ & $c \rightarrow 2^{2}P$ & & & & & & \\ 
		& $3$ & $h \rightarrow 2^{2}P$ & $e \rightarrow 2^{2}S$ & $d \rightarrow 3^{2}P$ & $g \rightarrow 3^{2}S$ & $i \rightarrow 2^{2}P$ & $j \rightarrow 3^{2}D$ & & & \\
		& $4$ & $o \rightarrow ?$ & $f \rightarrow 2^{2}P$ & $k \rightarrow 3^{2}D$ & $p \rightarrow 3^{2}D$ & $r \rightarrow 3^{2}D$ & $s \rightarrow 4^{2}D$ & & & \\
		\hline\hline
	\end{tabular}
\end{table}

\begin{figure}[htbp]
	\centering
    \includegraphics[width=15cm]{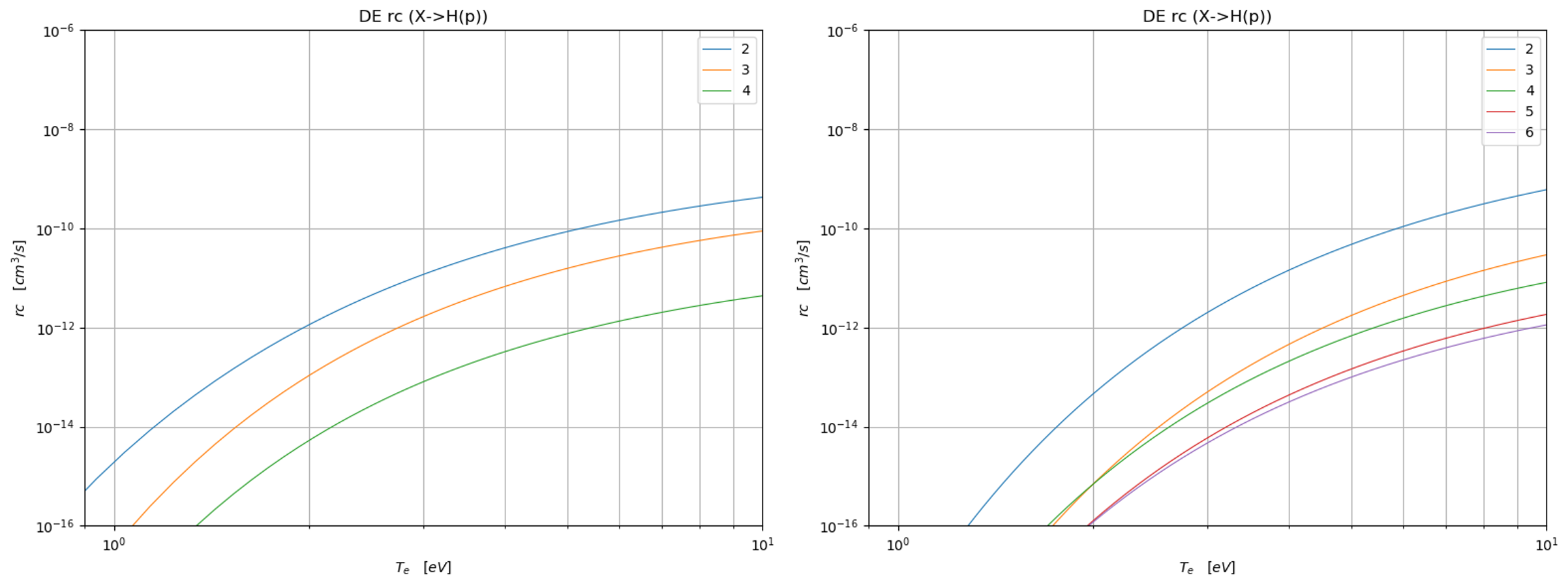}
	\caption[Rate coefficients of dissociative excitations of $H_{2}(X) \rightarrow H(p)$, from \cite{Janev2003}(left) and \cite{Miles1972}(right)
	]{Rate coefficients of dissociative excitations of $H_{2}(X) \rightarrow H(p)$, from \cite{Janev2003}(left) and \cite{Miles1972}(right)
	}\label{fig:derc}	
\end{figure}

Dissociations of $H_{2}$ into $H$ are mostly from $H_{2}$ reaching the $b$ state \cite{Sawada1995}, and the both produced atoms are at the ground state \ref{tb:Electronic state pairs triplet}.
However, dissociative excitations from the $X$ state produce excited $H$, and affect the state distribution of $H$.
Figure \ref{fig:derc} shows the rate coefficients of dissociative excitations to excited $H$.
The plot on the left is of $RC$s obtained by using the data given in \cite{Janev2003} with Table \ref{tb:Electronic state pairs singlet} and \ref{tb:Electronic state pairs triplet} and FCF data.
The plot on the right is of $RC$s from \cite{Miles1972}.

\section{Population model}

Population distributions of excited states for neutrals can be calculated using population models.
In the previous sections, the energy levels of states and various kinds of transitions among them, for $H$ and $H_{2}$, are introduced.
Using this information, appropriate population models can be constructed for the neutral state analysis.
Population model kinds vary depending on the thermodynamic equilibria of the systems \cite{Wunderlich2009}.
For $H$ plasmas, if the electron density is very high($n_{e}>10^{18}cm^{-3}$), the system is at the local thermodynamic equilibrium(LTE) and the states follow the Boltzmann distribution (Equation \ref{eq:Boltzmann}).
If the electron density is very low($n_{e}<10^{11}cm^{-3}$), the system is at the corona equilibrium, where electron impact excitations from the ground state and radiative decays from excited states are the only significant production and depopulation sources for excited states.
If the electron density is between these two limits, the collisional-radiative model becomes necessary.
This section provides a thorough explanation of the corona model and the collisional-radiative model.

\subsection{Corona model}

The corona equilibrium is where an electron density of a plasma is so low that step-wise transitions are negligible.
For an excited state of $H$, spontaneous emissions, electron de-excitations and recombinations from higher states, and excitations from lower excited states are negligible for its production, compared to electron impact excitation from the ground state.
This is true because, with very low electron densities, most of $H$ are in the ground state, and not many excited states present.
Therefore, excitations from the ground dominantly affect the production.
Also, electrons are too few to contribute to the loss of the state.
As explained in section 3.1.2, the state has a natural lifetime due to spontaneous emissions.
So the natural decay of the states is the dominant loss mechanism for excited states in the corona equilibrium.
With these assumptions, the corona model(CoronaM) is formulated.
\begin{equation} \label{eq:coronamodel1}
C_{H(1 \rightarrow p)}n_{e}n_{H(1)} = \sum_{q<p}{A_{H(p \rightarrow q)}n_{H(p)}}
\end{equation}
\begin{equation} \label{eq:coronamodel2}
n_{H(p)} = \frac{C_{H(1 \rightarrow p)}n_{e}n_{H(1)}}{\sum_{q<p}{A_{H(p \rightarrow q)}}} = \tau_{H(p)} C_{H(1 \rightarrow p)}(T_{e})n_{e}n_{H(1)}
\end{equation}
where $C$ is the $RC$ of electron impact excitation, and $\tau$ is the lifetime of an excited states.
Equation \ref{eq:coronamodel1} means the flux of electron impact excitation from the ground to the $p$ state is balanced by the flux of radiative decays of the state.
This is rearranged to Equation \ref{eq:coronamodel2}, and the relationship between the densities of the ground and an excited state is formulated.
With the knowledge of EEDF($T_{e}, n_{e}$), the distribution of states can be calculated.

The same idea is applied to $H_{2}$.
The density of the $d$ state is obtained by balancing the electron impact excitation from the $X$ state and the radiative decays of the $d$ state.
Then, the CoronaM is simply $n_{H_{2}(d)} = \tau_{H_{2}(d)} C_{H_{2}(X \rightarrow d)}(T_{e})n_{e}n_{H_{2}(X)}$.
However, vibrational substates can be considered to construct a more detailed CoronaM for $H_{2}$.
\begin{equation} \label{eq:vresolvedCoronaM1}
\begin{bmatrix} { n }_{ { d }(0) } \\ { n }_{ { d }(1) } \\ { n }_{ { d }(2) } \\ { n }_{ { d }(3) } \\ { n }_{ { d }(4) } \\ \vdots \end{bmatrix}=\begin{bmatrix} { \tau  }_{ { d }(0) } \\ { \tau  }_{ { d }(1) } \\ \tau _{ { d }(2) } \\ \tau _{ { d }(3) } \\ \tau _{ { d }(4) } \\ \vdots \end{bmatrix}\bigodot \begin{bmatrix} { q }_{ 0,0 } & \cdots  & { q }_{ 14,0 } \\ { q }_{ 0,1 } & \cdots  & { q }_{ 14,1 } \\ { q }_{ 0,2 } & \cdots  & { q }_{ 14,2 } \\ { q }_{ 0,3 } & \cdots  & { q }_{ 14,3 } \\ { q }_{ 0,4 } & \cdots  & { q }_{ 14,4 } \\ \vdots & \vdots & \vdots  \end{bmatrix}\times \left( { n }_{ e }\begin{bmatrix}  \langle { v }_{ e }{ \sigma  }^{ tot }_{ { { X }(0)\rightarrow d } } \rangle  \\  \langle { v }_{ e }{ \sigma  }^{ tot }_{ { { X }(1)\rightarrow d } } \rangle  \\  \langle { v }_{ e }{ \sigma  }^{ tot }_{ { { X }(2)\rightarrow d } } \rangle  \\ \vdots  \\  \langle { v }_{ e }{ \sigma  }^{ tot }_{ { { X }(12)\rightarrow d } } \rangle  \\  \langle { v }_{ e }{ \sigma  }^{ tot }_{ { { X }(13)\rightarrow d } } \rangle  \\  \langle { v }_{ e }{ \sigma  }^{ tot }_{ { { X }(14)\rightarrow d } } \rangle  \end{bmatrix}\bigodot \begin{bmatrix} { n }_{ { X }(0) } \\ { n }_{ { X }(1) } \\ { n }_{ { X }(2) } \\ \vdots  \\ { n }_{ { X }(12) } \\ { n }_{ { X }(13) } \\ { n }_{ { X }(14) } \end{bmatrix} \right)
\end{equation}

Equation \ref{eq:vresolvedCoronaM1} is the `vibrationally-resolved corona model(v-resolved CoronaM)', newly constructed in this work.
In Equation \ref{eq:vresolvedCoronaM1}, some detailed subscript marks are omitted, and `$\bigodot$' is the element-wise multiplication operator.
It takes account the vibronic excitations with the Franck-Condon principle, by including the FCF matrix in the CoronaM.
Therefore, with the knowledge of EEDF, this model provides the relationship between the absolute vibrational distributions of the $d$ state and the $X$ state.

Equation \ref{eq:vresolvedCoronaM1} included all vibrational states for the both electronic states, but $v>3$ of the $d$ state is not of interest in this work, since those informations are barely measurable in the experiment, and the lifetime values are not as accurate.\footnote{$(3-3)Q$ lines deviate from the Boltzmann distribution, likely due to predissociation \cite{Tsankov2012}.}
So, those states are excluded from the formula.
Also, there is no available data for $\langle { v }_{ e }{ \sigma  }^{ tot }_{ { { X }(v)\rightarrow d } } \rangle$ with $v>0$ in the literatures, and only  $\langle { v }_{ e }{ \sigma  }^{ tot }_{ { { X }(0)\rightarrow d } } \rangle$ is available \cite{Janev2003, Miles1972}.
Thus, others are replaced with this value.
Since high vibrational states have smaller energy gaps to excite, the rates would be underestimated in this way.
The resultant v-resolved CoronaM is the following.

\begin{equation} \label{eq:vresolvedCoronaM2}
\begin{bmatrix} { n }_{ { d }(0) } \\ { n }_{ { d }(1) } \\ { n }_{ { d }(2) } \\ { n }_{ { d }(3) } \end{bmatrix}=\begin{bmatrix} { \tau  }_{ { d }(0) } \\ { \tau  }_{ { d }(1) } \\ \tau _{ { d }(2) } \\ \tau _{ { d }(3) } \end{bmatrix}\bigodot \begin{bmatrix} { q }_{ 0,0 } & \cdots  & { q }_{ 14,0 } \\ { q }_{ 0,1 } & \cdots  & { q }_{ 14,1 } \\ { q }_{ 0,2 } & \cdots  & { q }_{ 14,2 } \\ { q }_{ 0,3 } & \cdots  & { q }_{ 14,3 } \end{bmatrix}\times \left( { n }_{ e } \langle { v }_{ e }{ \sigma  }^{ tot }_{ { { X }(0)\rightarrow d } } \rangle \times \begin{bmatrix} { n }_{ { X }(0) } \\ { n }_{ { X }(1) } \\ { n }_{ { X }(2) } \\ \vdots  \\ { n }_{ { X }(12) } \\ { n }_{ { X }(13) } \\ { n }_{ { X }(14) } \end{bmatrix} \right)
\end{equation}

\subsection{Collisional-Radiative model}

A collsional-radiative model(CRM) is a more rigorous population model that takes account most of the significant transitions.
Accordingly, the CRM provides a more complete description of population dynamics than the CoronaM.

A state populates and depopulates with various transitions, most of which are introduced in Subsection 3.1.2.
All states are changing back and forth continuously with finite rates, mostly depending on the rates of significant transitions.
Therefore, a set of differential equations are built to quantitatively analyze the phenomena.
First, the total flux of a state density is equated to all the significant transition rates, and form a complete `rate equation' \cite{Fujimoto2004}.

\begin{equation} \label{eq:rateequation}
	\begin{aligned}
	\frac { d }{ dt } n_{ p } = & [ n_{ e }\sum_{ q<p } C_{ qp }+n_{ e }\sum _{ q>p}F_{qp}+\sum _{ q>p}A_{qp}]n_{q}\\
	& - [ n_{ e }\sum _{ q>p } C_{ pq }+n_{ e }\sum _{ q<p } F_{ pq }+\sum _{ q<p } A_{ pq }+n_{ e }S_{ p } ] n_{ p }\\
	& + [n_{e}\alpha_{p}+\beta_{p}]n_{e}n_{+}\\
	= & \sum_{ q<p } U_{ qp }n_{ q }+\sum_{ q>p } D_{ qp }n_{ q }+O_{ p }n_{ p }+R_{p}n_{+}
	\end{aligned}
\end{equation}
where $p$ is the state of interest, $C, F, S$ are $RC$s of electron impact excitation, de-excitation, and ionization, respectively, and $\alpha, \beta$ are $RC$s of three-body and radiative recombination, respectively.
The rate of change of the $p$ state density due to all significant transitions is formulated by the rate equation.
The rate equation is simplified further for easier rearangement later, with newly defined grouped rates,
\begin{equation} \label{eq:newrates}
\begin{cases} { U }_{ qp }={ n }_{ e }{ C }_{ qp } \\ { O }_{ p }=-\left[ n_{ e }\sum _{ q>p } C_{ pq }+n_{ e }\sum _{ q<p } F_{ pq }+\sum _{ q<p } A_{ pq }+n_{ e }S_{ p } \right] \\ { D }_{ qp }=n_{ e }{ F }_{ qp }+{ A }_{ qp }\\R_{p}=n_{e}[n_{e}\alpha_{p}+\beta_{p}] \end{cases}
\end{equation}
where they are names `up' rate, `out' rate, `down' rate, and `recomb' rate, respectively, in this work.

The `quasi-steady state approximation(QSSA)' can be applied for simpler analyses of the rate equations.
First, here is the explanation of the QSSA \cite{Fujimoto2004}.
Suppose a group of neutrals of some initial state is submerged in a `sea' of electrons with some EEDF.
All neutral state densities will reach the final state with some characteristic time, ($n^{ini}_{p} \xrightarrow{\tau} n^{fin}_{p}$).\footnote{In this explanation, the term `state' is used to refer to both the quantum state and the temporal state of a density.} To be more precise, Equation \ref{eq:rateequation} is further simplified to be expressed as
\begin{equation} \label{eq:rateequationbucket}
\frac{dn}{dt} = \chi+On
\end{equation}
where $p$ for the state label is omitted, and $\chi$ is the sum of rest of terms for all populating the state density.
Then,
\begin{equation} \label{eq:finalstate}
n^{fin} = -O/\chi
\end{equation}
\begin{equation} \label{eq:nperturb}
n(t) = n^{fin}+\Delta n(t)
\end{equation}
where $n^{fin}$ is the final state, and $\Delta n(t)$ is the deviation from the final state.
with Equation \ref{eq:rateequationbucket} and \ref{eq:nperturb},
\begin{equation} \label{eq:nperturb2}
	\begin{aligned}
		& \frac{\Delta n(t)}{dt} = O\Delta n(t) &\\
		& \Delta n(t) = \Delta n(0) \exp(Ot) &\\
		& n^{ini} = n^{fin} + \Delta n(0) &\\
		& \rightarrow n(t) = n^{fin} + \Delta n(0) \exp(Ot) &
	\end{aligned}
\end{equation}

Since $O$ is a negative value for the depopulation flux of the state, the characteristic time of reaching the final state, as known as the `relaxation time,' is defined as
\begin{equation} \label{eq:relaxationtime}
\tau^{relax} = -1/O
\end{equation}

The relaxation time tells us how fast the state density reach the final steady state.
From Equation \ref{eq:newrates} and \ref{eq:relaxationtime}, the relaxation time is largely determined by the Einstein A coefficient, unless the electron density is very high.
As explained earlier in the chapter, Einstein A coefficients, or spontaneous emission rates, are dominantly high for excited states, and thus, their relaxation times are very short.
Therefore, it is correct to assume that the excited states reach the final steady state instantaneously in a `sea' of electrons.
On the other hand, the ground state or the ionized state do not have very short relaxation time(no natural decay), and the same assumption cannot be applied to these state.\footnote{The quantitative analysis of relaxation times are not explained here, but in \cite{Fujimoto2004}.}

The QSSA therefore states
\begin{equation} \label{eq:qssa}
\frac { dn_{p} }{ dt } = 0 \quad for \quad p = (excited)
\end{equation}
This means the spatial inhomogeneity and the transport effect of the plasmas can be ignored, since the neutrals that reach the specific local plasma of interest become instantly equilibrated and the information of other local plasma in the neutrals become negligible.
Consequently, this makes neutral state population analysis much simpler.

The series of the rate equations for the all states is a set of coupled differential equations.
However, with the QSSA, it becomes a set of coupled linear equations.
\begin{equation} \label{eq:qssatorateequations}
	\begin{aligned}
	& { \begin{bmatrix} O_{ 2 } & { D }_{ 3,2 } & \cdots  & \cdots  & { D }_{ p,2 } & \cdots  & \cdots  & { D }_{ 19,2 } & { D }_{ 20,2 } \\ U_{ 2,3 } & O_{ 3 } & \cdots  & \cdots  & { D }_{ p,3 } & \cdots  & \cdots  & { D }_{ 19,3 } & { D }_{ 20,3 } \\ \vdots  & \vdots  &  & \vdots  & \vdots  & \vdots  &  & \vdots  & \vdots  \\ \vdots  & \vdots  & \cdots  & O_{ p-1 } & { D }_{ p,p-1 } & { D }_{ p+1,p-1 } & \cdots  & \vdots  & \vdots  \\ U_{ 2,p } & U_{ 3,p } & \cdots  & U_{ p-1,p } & O_{ p } & { D }_{ p+1,p } & \cdots  & { D }_{ 19,p } & { D }_{ 20,p } \\ \vdots  & \vdots  & \cdots  & U_{ p-1,p+1 } & U_{ p,p+1 } & O_{ p+1 } & \cdots  & \vdots  & \vdots  \\ \vdots  & \vdots  &  & \vdots  & \vdots  & \vdots  &  & \vdots  & \vdots  \\U_{ 2,19 } & U_{ 3,19 } & \cdots  & \cdots  & U_{ p,19 } & \cdots  & \cdots  & O_{ 19 } & { D }_{ 20,19 } \\ U_{ 2,20 } & U_{ 3,20 } & \cdots  & \cdots  & U_{ p,20 } & \cdots  & \cdots  & U_{ 19,20 } & O_{ 20 } \end{bmatrix} }\begin{bmatrix} n_{ 2 } \\ n_{ 3 } \\ \vdots  \\ n_{ p-1 } \\ n_{ p } \\ n_{ p+1 } \\ \vdots  \\ n_{ 19 } \\ n_{ 20 } \end{bmatrix} &\\
	& +\begin{bmatrix} U_{ 1,2 } & R_{2}\\ U_{ 1,3 }& R_{3} \\ \vdots & \vdots \\ U_{ 1,p-1 }& R_{p-1} \\ U_{ 1,p }& R_{p} \\ U_{ 1,p+1 }& R_{p+1} \\ \vdots & \vdots \\ U_{ 1,19 }& R_{19} \\ U_{ 1,20 }& R_{20} \end{bmatrix} \begin{bmatrix} n_{ 1 } \\ n_{+} \end{bmatrix} = \frac{d}{dt}\begin{bmatrix} n_{ 2 } \\ n_{ 3 } \\ \vdots  \\ n_{ p-1 } \\ n_{ p } \\ n_{ p+1 } \\ \vdots  \\ n_{ 19 } \\ n_{ 20 } \end{bmatrix} = 0 &
	\end{aligned}
\end{equation}
where the total of 20 neutral states plus ionized state are considered.\footnote{The justification for considering 20 states is in the appendix Chapter A.}
Finally, rearranging the coupled linear equations yield
\begin{equation} \label{eq:crm}
n_{H(p)} = R_{H(1 \rightarrow p)}(T_{e}, n_{e})n_{e}n_{H(1)}+R_{H^{+} \rightarrow H(p)}(T_{e}, n_{e})n_{e}n_{H^{+}}
\end{equation}
where $R_{H(1 \rightarrow p)}$ and $R_{H^{+} \rightarrow H(p)}$ are named the ionizing and recombining population coefficients, respectively.
Because of the electron impact excitations included in the rate equations, the population coefficients depend of electron parameters.
The formula for the densities of excited states for a plasma is now set up, and the state distribution can now be calculated, with known electron parameters, and densities of the ground and the ionized.
Equation \ref{eq:crm} is the collisional-radiative model(CRM) of $H$.
Various analyses of the $H$ CRM are in the appendix Chapter A.

\begin{figure}[htbp]
	\centering
    \includegraphics[width=16cm]{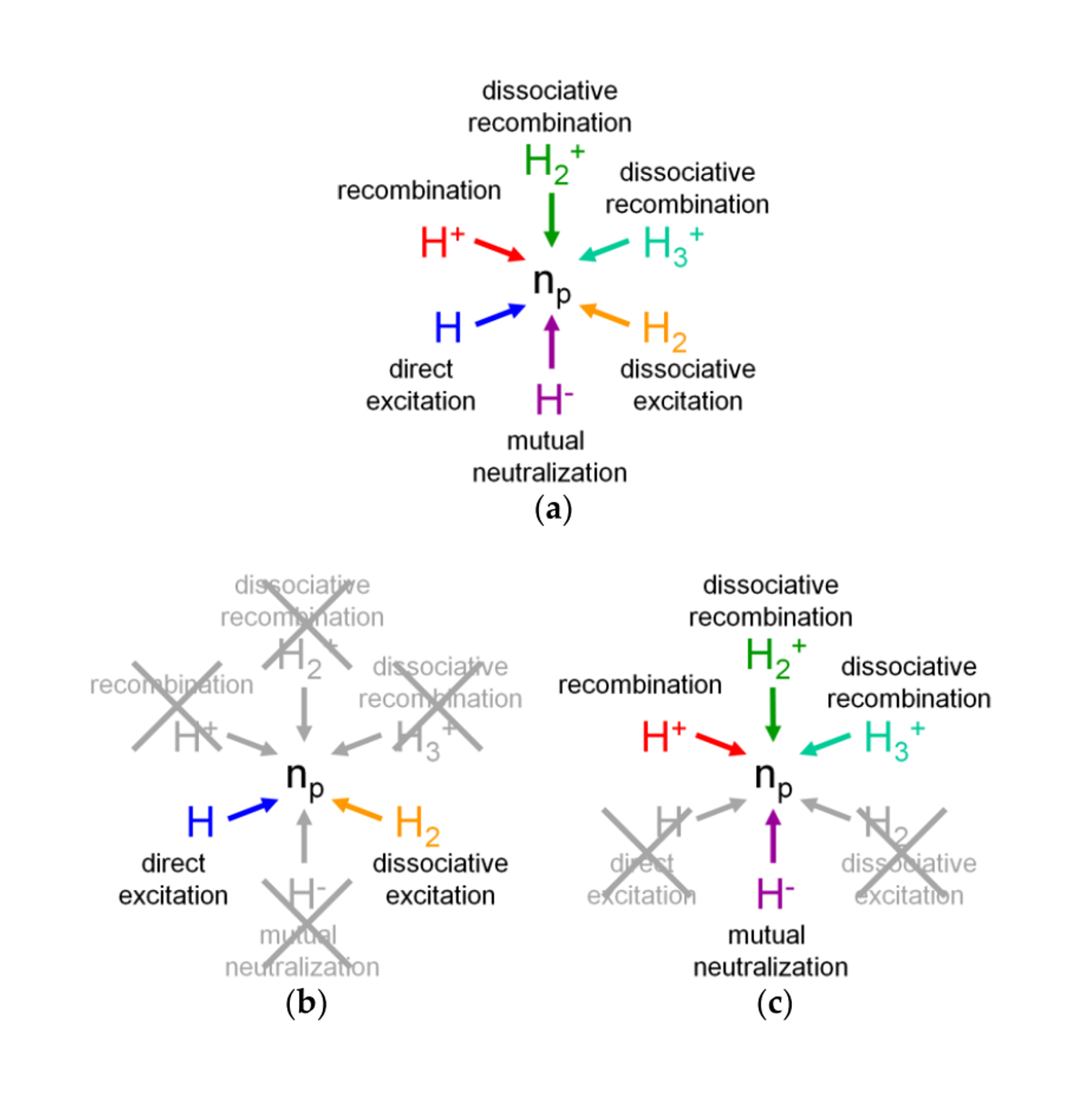}
	\caption[List of contributing channels for $H$, and ionizing(left) and recombining(right) case \cite{Wunderlich2016}
	]{List of contributing channels for $H$, and ionizing(left) and recombining(right) case \cite{Wunderlich2016}
	}\label{fig:List of contributing channels}
\end{figure}

The derivation of CRM tells us that each term of Equation \ref{eq:crm} correspond to those, which the QSSA cannot be applied.
These different species are referred to as the contributing channels, or the excitation channels.
According to \cite{Wunderlich2016}, there are six significant channels for $H$ (Figure \ref{fig:List of contributing channels}).
Besides the direct excitation and the atomic ion recombination by $H$ and $H^{+}$, there are the dissociative excitation from $H_{2}$, as explained in the previous section, and more various channels.
The plasmas of interest in this work are weakly collisional with low electron density, and so, they are `ionizing' plasmas, where the direct excitation and dissociative excitation channels are assumed to be dominant.
Therefore, only these two channels are considered in the analysis technique developed in this work.
Also, the consequence of ignoring other contributing channels is discussed in the next two chapters.

The CRM for $H_{2}$ can also be constructed, but, due to lack of vibrationally-resolved molecular process data in literatures, it is difficult to improve the accuracy of the population model.
So, in this work, the $H_{2}$ CRM is not constructed, and the analysis are performed with the $H_{2}$ v-resolved CoronaM.
Using the CoronaM is also justifiable since the plasmas of interest are weakly collisional.
Finally, the $H_{2}$ v-resolved CoronaM and the $H$ CRM, with the techniques for the bi-Maxwellian rate calcualtion and the radiation trapping effect consideration, are ready to be used in the analysis.

\chapter[Analysis Technique \& Result]{Analysis Technique \& Result}

The analysis technique for the inference of DOD is presented in this chapter.
There are number of steps for the analyses of the measured plasmas in this technique.
The flowchart of the technique is shown first for the overview.

\begin{figure}[htbp]
	\centering
    \includegraphics[width=16cm]{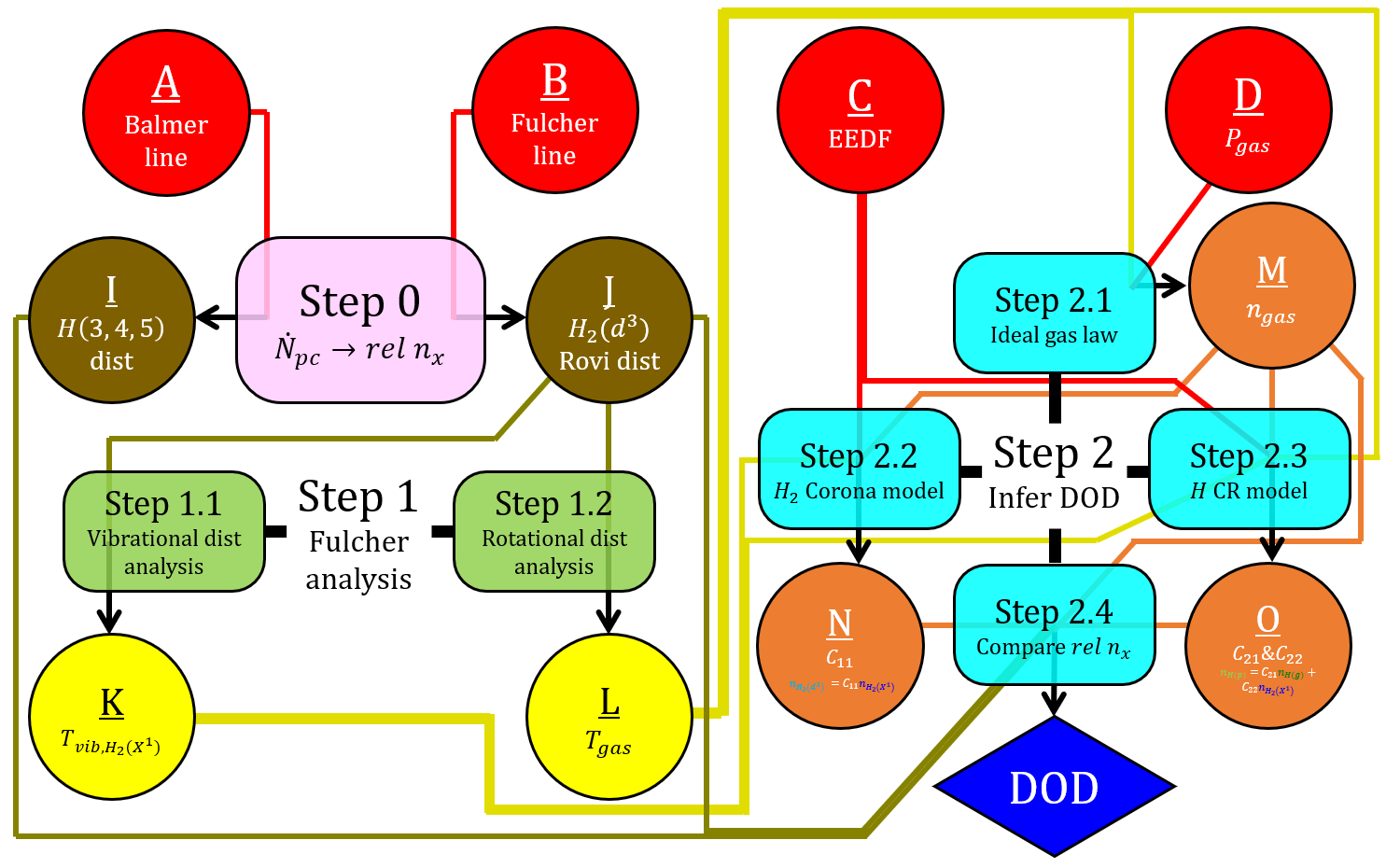}
	\caption[The flowchart of the analysis technique
	]{The flowchart of the analysis technique
	}\label{fig:The flowchart of the analysis technique}	
\end{figure}

There are initially four kinds of information available from the experiment.
$\underbar{A}$ and $\underbar{B}$ are the photon count rate of the Balmer lines and the Fulcher lines from the processed spectra, obtained from the OES.
$\underbar{C}$ is the EEDF information, which is a set of four electron parameters, obtained with the LP.
$\underbar{D}$ is the gas pressure measured by the baratron gauge in the MAXIMUS.
These go through series of steps to produce new information.
Step 0  is the `photon count rate to relative density.'
$\underbar{A}$ and $\underbar{B}$ pass Step 0 to produce $\underbar{I}$, $H$ excited state distribution, and $\underbar{J}$, $H_{2}(d)$ ro-vibrational distribution.
$\underbar{J}$ then pass Step 1, the Fulcher-$\alpha$ analysis to produce $\underbar{K}$, the ground vibrational temperature, and $\underbar{L}$, the gas temperature.
Step 2 is the inference of DOD.
With the available information, using the ideal gas law, the $H_{2}$ v-resolved CoronaM, and the $H$ CRM, the relationships between the ground and the excited states of $H_{2}$ and $H$ are calculated.
Since the measured excited state distribution is already available from $\underbar{I}$ and $\underbar{J}$, the measured and the calculated distributions can be compared to find the most reasonable DOD for the plasma.
This is the overview of the analysis technique.
The detailed steps are presented in the following sections.

\section{Step 0: State density distribution from spectrum}
Step 0 is the acquisition of relative state densities from photon count rates.
As explained in subsection 3.1.2, the photon emission flux is the product of the Einstein A coefficient and the upper state density (Equation \ref{eq:spontaneousemission}).
The photon count rate by the VIS spectrometer is
\begin{equation} \label{eq:photoncountrate}
\dot{N}_{PC, ul} = A_{ul}n_{u}G, \quad G = \frac{\Omega}{4\pi}VO(\lambda_{ul})
\end{equation}
where $G$ is the geometrical factor, $\Omega$ is the solid angle subtended by the detector, $V$ is the volume of the plasma measured, and $O$ is the optical factors of the system.
This means the spectrometer actually measures only a portion of photons emitted from the plasma due to various factors.
Since the measured spectra are intensity-calibrated, $G$ is independent of wavelengths.
Therefore, the photon count rates are divided by the corresponding Einstein A coefficients, with canceling out $G$, to obtain relative densities of the upper states.
\begin{equation} \label{eq:Npctorn}
\frac{\dot{N}_{PC, u_{1}l_{1}}}{A_{u_{1}l_{1}}} : \frac{\dot{N}_{PC, u_{2}l_{2}}}{A_{u_{2}l_{2}}} : \frac{\dot{N}_{PC, u_{3}l_{3}}}{A_{u_{3}l_{3}}} : ... = n_{u_{1}} : n_{u_{2}} : n_{u_{3}} : ...
\end{equation}

Using this method, the density distribution of $H_{2}$ and $H$ excited states is obtained.

\begin{figure}[htbp]
	\centering
    \includegraphics[width=12cm]{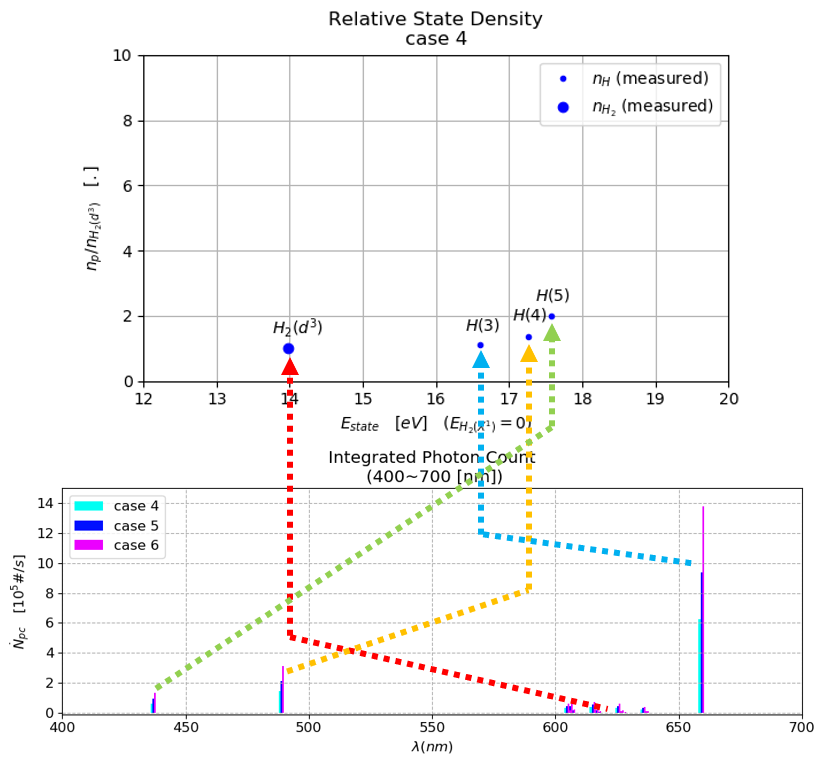}
		\caption[The relative densities of $H(3), H(4), H(5)$ and $H_{2}(d)$ obtained from the measured spectra (Case 4)
		]{The relative densities of $H(3), H(4), H(5)$ and $H_{2}(d)$ obtained from the measured spectra (Case 4)
		} \label{fig:to_relative_density}
	\end{figure}

Figure \ref{fig:to_relative_density} shows that the relative densities of $H(3), H(4), H(5)$ and $H_{2}(d)$ are obtained from the measured spectra of Case 4 for example, using Equation \ref{eq:Npctorn}.
Letting the $H_{2}(d)$ state as the reference($=1$), the rest state densities are plotted together.
Apparently, the higher states have higher densities, and their densities only differ by about two times at most.

\section{Step 1: Fulcher-$\alpha$ band analysis}
Step 1 is the vibrational and rotational analyses of the Fulcher-$\alpha$ band.
To perform those, the ro-vibrational distribution of the $d$ state is needed.
In the previous section for Step 0, the $d$ state is treated as a single state when obtaining the density.
To be more rigorous, every ro-vibrational states of the $d$ state goes through Step 0 to obtain the ro-vibrational state distribution.

\begin{figure}[htbp]
	\centering
    \includegraphics[width=12cm]{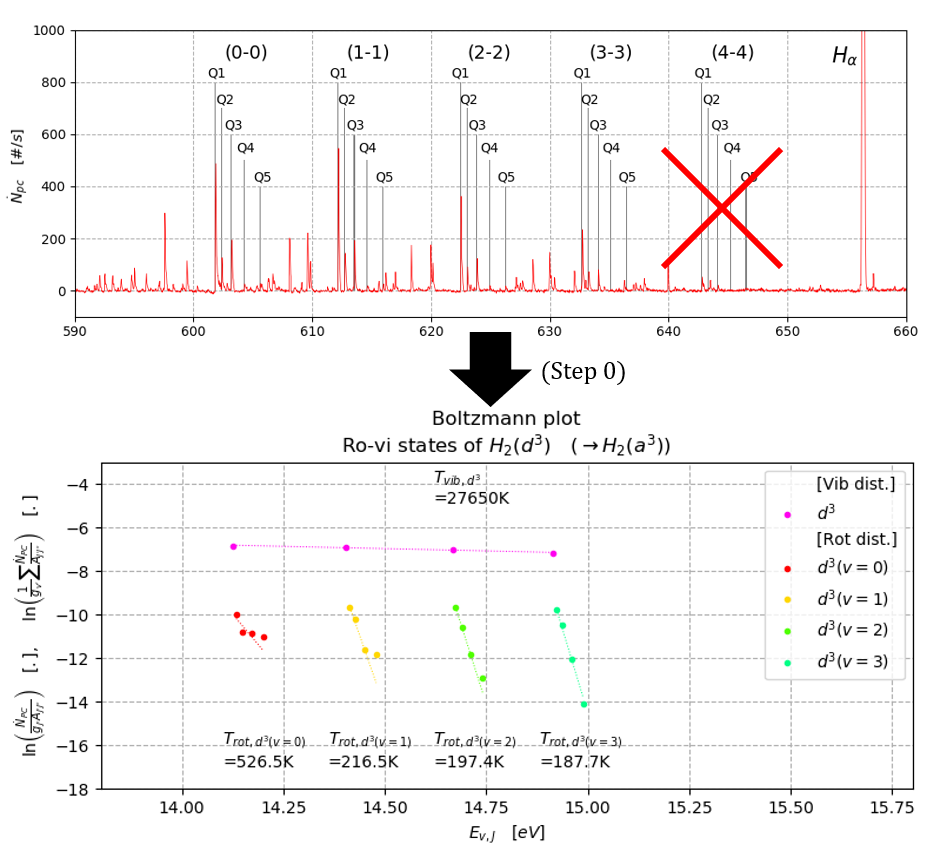}
	\caption[The Boltzmann plot of ro-vibrational states of $H_{2}(d)$ (Case 4)
	]{The Boltzmann plot of ro-vibrational states of $H_{2}(d)$ (Case 4)
	} \label{fig:Boltzmann_rovi}
\end{figure}

Figure \ref{fig:Boltzmann_rovi} shows the Boltzmann plot of ro-vibrational relative densities of the $d$ state obtained from the Fulcher-$\alpha$ band.
$v=0, 1, 2, 3$ and $J=1, 2, 3, 4$ are considered since other lines are not as clearly detected.
Consequently, 16 ro-vibrational states are plotted.
States are plotted by their energies in the Boltzmann plot.
The rotational states of the same vibrational number, $v$, are grouped to form the rotational distribution for each $v$.
And their densities can be summed together to become the total density of a vibrational state, which are plotted with pink points in Figure \ref{fig:Boltzmann_rovi}.
They are the vibrational distribution for the $d$ state.
Finally, the densities of all $v$ are summed to become the total density of the electronic $d$ state (Figure \ref{fig:to_relative_density}).

\subsubsection{Vibrational distribution analysis}

The vibrational distribution (pink points in Figure \ref{fig:to_relative_density}) is analyzed.
The inference of the ground vibrational temperature is the goal of this analysis.
As can be seen in Figure \ref{fig:to_relative_density}, a line is fitted to the vibrational states.
Since this is the Boltzmann plot, where the vertical axis is proportional to $\ln{(n_{v}/g_{v})}$, fitting a line would give a slope that corresponds to $-1/T$, where $T$ is an equilibrium temperature, if the distribution is assumed to be at an equilibrium (Equation \ref{eq:Boltzmann}).\footnote{The unit $[eV]$ of a temperature is converted to Kelvin $[K]$ with $1eV/k_{B} = 11605K/eV$, where $k_{B}$ is the Boltzmann constant.}
This temperature is a `vibrational temperature,' $T_{vib}$, of a specific electronic state.
$T_{vib, d} = 27650K$ is obtained in Figure \ref{fig:to_relative_density}.
However, the vibrational temperature of the $d$ state is not very important, but the origin of the $d$ vibrational distribution is. The $d$ vibrational distribution is actually decided by the ground vibrational temperature.

From Equation \ref{eq:vresolvedCoronaM2}, it is already known that the $d$ vibrational distribution is the result of electron impact vibronic excitation from the ground.
Therefore, the ground $X$ vibrational distribution can be inferred from the known $d$ vibrational distribution.
If the $X$ vibrational distribution is assumed to at a thermodynamic equilibrium, then the density distribution of 15 different vibrational states of the $X$ state can be decided by one parameter, the vibrational temperature of the $X$ state, $T_{vib, X}$.
Here, since only the relative vibrational distribution of the $d$ state is known, the absolute relation between the $d$ and $X$ vibrational density ditribution is not of interest.
Only their proportional relation matters.
So, Equation \ref{eq:vresolvedCoronaM2} is modified further to give the proportional relation with the ground vibrational temperature.
\begin{equation} \label{eq:vresolvedCoronaM3}
\begin{bmatrix} { n }_{ { d }(0) } \\ { n }_{ { d }(1) } \\ { n }_{ { d }(2) } \\ { n }_{ { d }(3) } \end{bmatrix} \propto \begin{bmatrix} { \tau  }_{ { d }(0) } \\ { \tau  }_{ { d }(1) } \\ \tau _{ { d }(2) } \\ \tau _{ { d }(3) } \end{bmatrix}\bigodot \left( \begin{bmatrix} { q }_{ 0,0 } & \cdots  & { q }_{ 14,0 } \\ { q }_{ 0,1 } & \cdots  & { q }_{ 14,1 } \\ { q }_{ 0,2 } & \cdots  & { q }_{ 14,2 } \\ { q }_{ 0,3 } & \cdots  & { q }_{ 14,3 } \end{bmatrix}  \times \begin{bmatrix} \exp{(-E_{v=0}/T_{vib, X})} \\ \exp{(-E_{v=1}/T_{vib, X})} \\ \exp{(-E_{v=2}/T_{vib, X})} \\ \vdots \\ \exp{(-E_{v=12}/T_{vib, X})} \\ \exp{(-E_{v=13}/T_{vib, X})} \\ \exp{(-E_{v=14}/T_{vib, X})} \end{bmatrix} \frac{n_{X}}{(NORM)} \right)
\end{equation}
where $(NORM)$ is the normalizing factor that make the sum of all exponentials in the last matrix become 1, and $n_{X}$ is the total density of the $X$ state.
So, the last matrix is distributing $n_{X}$ into each vibrational state with a certain $T_{vib, X}$.

The ground vibrational temperature, $T_{vib, X}$ can be inferred, using this proportion relation.
A $T_{vib, X}$ that fits the calculated $d$ vibrational distribution to the measured one best is the most reasonable $T_{vib, X}$.

\begin{figure}[htbp]
	\centering
    \includegraphics[width=16cm]{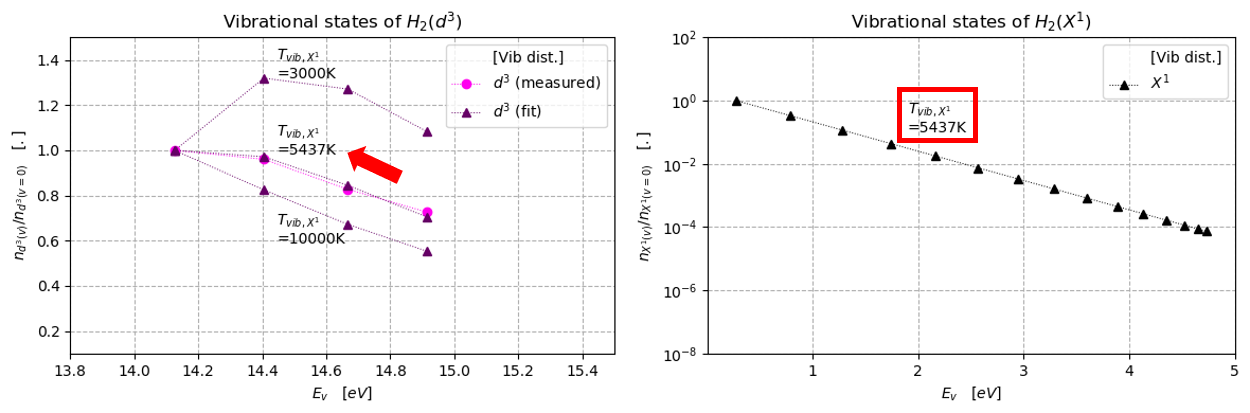}
	\caption[The measured and fitted vibrational distribution of the $d$ state(left) and the corresponding ground vibrational distribution(right)
	]{The measured and fitted vibrational distribution of the $d$ state(left) and the corresponding ground vibrational distribution(right)
	}\label{fig:Tvib_fit}
\end{figure}

Figure \ref{fig:Tvib_fit}(left) shows the measured $d$ vibrational distribution and the calculated ones.
When $T_{vib, X}=5437K$ (Figure \ref{fig:Tvib_fit}(right)), the corresponding excited $d$ vibrational distribution fits the measured distribution best, so $T_{vib, X}=5437K$ is the ground vibrational temperature of the measured plasma.
In Figure \ref{fig:Tvib_fit}(left), there are other cases shown together, which have quite different shapes, but the inferred distribution fit the measured one nicely.
This shows that the assumptions involved in constructing Equation \ref{eq:vresolvedCoronaM3} is acceptable.

The ground vibrational temperature inferred is an important parameter.
As explained, it determines the ground vibrational density distribution, and this affect the calculation of the total density of the $d$ state with the v-resolved CoronaM (Equation \ref{eq:vresolvedCoronaM2}) and the calculation of dissociative excitations from the ground, which affect excited densities of $H$.

\begin{figure}[htbp]
	\centering
    \includegraphics[width=14cm]{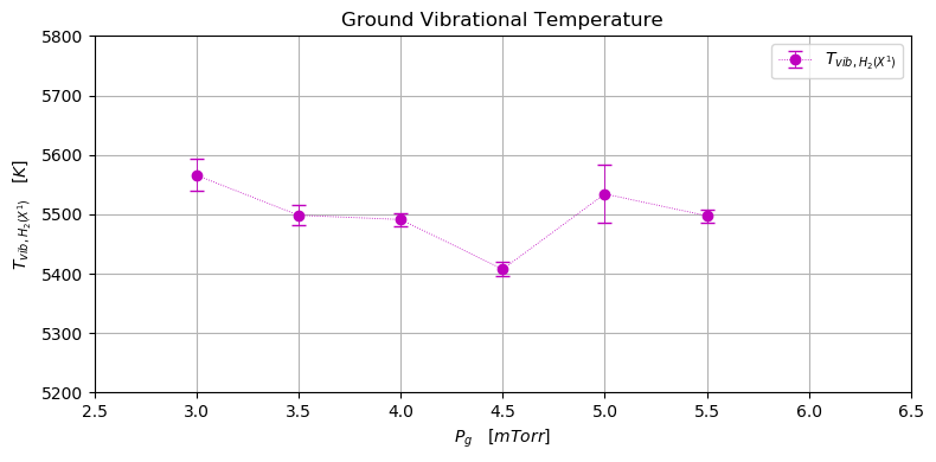}
	\caption[Ground vibrational temperatures inferred for all cases
	]{Ground vibrational temperatures inferred for all cases
	}\label{fig:TvibXs}
\end{figure}

The ground vibrational temperatures of all cases of the experiment are inferred in the same way (Figure \ref{fig:TvibXs}).
The inferred values are around $5500K$.

\subsubsection{Rotational distribution analysis}

The rotational distributions (not pink points in Figure \ref{fig:to_relative_density}) are analyzed.
The calculation of the gas temperature is the goal of this analysis.
As can be seen in Figure \ref{fig:to_relative_density}, the rotational distributions follow the Boltzmann distributions and are at thermodynamic equilibria, which means temperatures can be assigned to them.
The assigned temperature is `rotational temperature,' $T_{rot}$, of a $v$ of the $d$ state.

More precisely, the fitting function is (using Equation \ref{eq:Boltzmann})
\begin{equation} \label{eq:rbolzmannfit}
\ln{(\frac{\dot{N}_{PC}}{g_{J'}A_{J'J''}})} \propto -\frac{1}{T_{rot, d}}E_{dv'J'}
\end{equation}

Similar to the previous analysis, the origin of these rotational distributions is from the ground state.
The assumption is that, rotational distributions in the excited electronic state are images from those of the ground electronic state, since the radiative lifetimes of electronically excited states are often much shorter than the characteristic time of the rotational relaxations \cite{Qing1996}.
Therefore, the rotational temperature of the ground can be inferred from that of the excited state.
Though the rotational distribution of the excited state is imaged from the ground, their rotational temperatures differ, since the energy levels of rotational states of the excited and the ground differ.
In order to compensate for that difference, the energy term of the fitting function (Equation \ref{eq:rbolzmannfit}) is changed.
\begin{equation} \label{eq:dtoXrenergy}
E_{dv'J'} \cong E_{e, d}+E_{vib, dv'}+B^{d}_{e}J(J+1) \quad \rightarrow \quad E_{Xv'J'} \cong E_{e, X}+E_{vib, Xv'}+B^{X}_{e}J(J+1)
\end{equation}
where Equation \ref{eq:renergy} is used but the distortion term is ignored.

Equation \ref{eq:rbolzmannfit} is a proportional relation that is used with a fixed vibronic state, and only used to obtain the slope of the fit.
So, vibronic energies ($E_{e, d}+E_{vib, dv'}$ and $E_{e, X}+E_{vib, Xv'}$) can be ignored, and $B^{d}_{e} \rightarrow B^{X}_{e}$ is the only change needed in Equation \ref{eq:rbolzmannfit} to find the ground rotational temperature.
Using these facts, the simple relation between the ground and the excited rotational temperatures can be found.
\begin{equation} \label{eq:2.0027}
\frac{T_{rot, X}}{T_{rot, d}} = \frac{B^{X}_{e}}{B^{d}_{e}} = \frac{60.809cm^{-1}}{30.364cm^{-1}} \cong 2.0027
\end{equation}
With this relation, the ground rotational temperature can be inferred straight from the $d$ rotational temperature, without having to perform the fitting procedure again.

Another assumption is that, as $H_{2}$ particles in motion collide with one another, their ground rotational distributions reach the thermodynamic equilibrium with the translation motion of the particles, since the rotational energy gaps are very small.
This means the gas temperature is equilibrated with the ground rotational temperature.\footnote{If the characteristic time of the rotational relaxations for the $d$ state is much shorter than its radiative lifetime, the gas temperature is equilibrated with the $d$ rotational temperature, $T_{rot, d} \cong T_{gas}$ \cite{Cho2018}. This would be true if the collision frequency of $H_{2}$ particles is very high.}
Therefore, the following relation holds \cite{Iordanova2011, Tsankov2012, Majstorovic2007, Iordanova2008}.\footnote{There is the inequality of different temperatures, $T_{gas(trans)} \leq T_{rot} \leq T_{vib} \leq T_{ex} \leq T_{e}$, where $T_{ex}$ is the excitation temperature of electronic states \cite{Bruggeman2014}.}
\begin{equation} \label{eq:inferTgas}
2.0027 \times T_{rot, d} \cong T_{rot, X} \cong T_{gas}
\end{equation}

\begin{figure}[htbp]
	\centering
    \includegraphics[width=14cm]{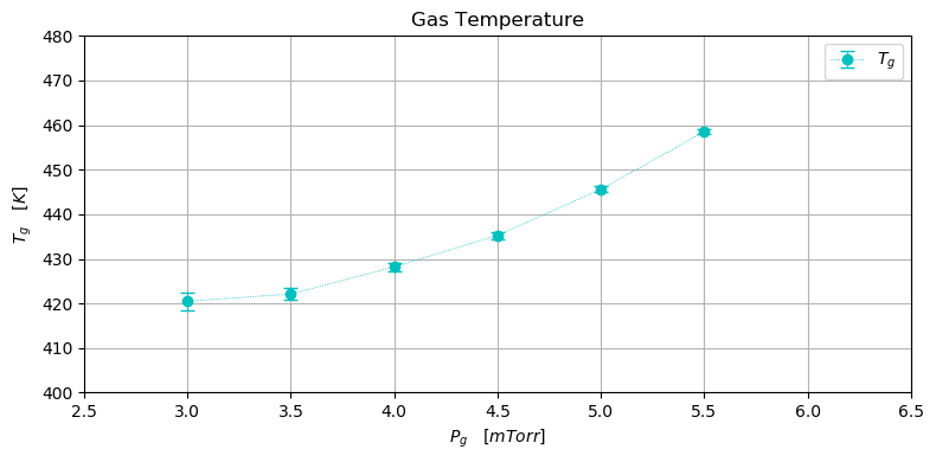}
	\caption[Gas temperatures inferred for all cases
	]{Gas temperatures inferred for all cases
	}\label{fig:Tgass}
\end{figure}

However, there are a number of different $T_{rot, d}$ depending on the $v$.
The choice of the $v$ for $T_{rot, d}$ differs by literatures \cite{Iordanova2011, Tsankov2012, Majstorovic2007, Iordanova2008}.
In this work, the rotational distribution of $v=1$ is chosen to infer the gas temperature.
The gas temperatures inferred for all cases are shown in Figure \ref{fig:Tgass}.

The gas temperature inferred increases with the gas pressure (Figure \ref{fig:Tgass}).
Since the discharge current increases with the gas pressure during the experiment (Table \ref{tb:pressure_Idischarge}), this result is expected.

\section{Step 2: Inference of DOD}

Step 2 is the inference of the degree of dissociations(DOD) for the measured plasmas.
In this step, all kinds of measured and produced information are used with the constructed population models.

Here is the procedure with substeps for the inference of DOD.

\bigskip
\bigskip
(1) Guess DOD

Suppose that the DOD of the plasma of interest is known at first.\footnote{$2 \times$ in the formula compensates for the fact that a $H_{2}$ consists of two $H$s. This way $DOD=50\%$ when $50\%$ of $H_{2}$ are broken into $H$, for example.}
\begin{equation} \label{eq:DOD}
DOD = \frac{1}{2n_{H_{2}(X)}/n_{H(1)}+1}
\end{equation}

It is assumed that $n_{H_{2}} \cong n_{H_{2}(X)}$ and $n_{H} \cong n_{H(1)}$, since the ground electronic states are typically dominant in weakly collisional plasmas.

\bigskip
\bigskip
(2) The ideal gas law \& ground state densities

\begin{equation} \label{eq:idealgaslaw}
n_{g} = \frac{P_{g} [mTorr]}{T_{g} [K]} \frac{1}{k_{B}}, \quad k_{B}=1.04 \times 10^{-16} \frac{mTorr}{K}cm^{3}
\end{equation}

The gas pressure is already known by the Baratron gauge, and the gas temperature is inferred in Section 4.2.
Therefore, the gas number density, $n_{g}$, is readily calculated with the ideal gas law (Equation \ref{eq:idealgaslaw}).
And also by the above assumption, $n_{g} \cong n_{H_{2}(X)}+n_{H(1)}$.

Since the DOD is already known from Substep (1), the absolute densities of the ground states of both $H_{2}$ and $H$, $n_{H_{2}(X)} \& n_{H(1)}$ are calculated.

\bigskip
\bigskip
(3) The population models to produce calculated state distribution

This is the substep that make uses of the population models constructed in this work. Equation \ref{eq:vresolvedCoronaM2} and \ref{eq:vresolvedCoronaM3} provides the following relation.
\begin{equation} \label{eq:H2staterelation}
n_{H_{2}(d)} = C_{11}(T^c_{e}, n^c_{e}, T^h_{e}, n^h_{e}, T_{vib, X})n_{H_{2}(X)}
\end{equation}
where $C_{11}$ is the suppressed coefficient.

The $H$ CRM (Equation \ref{eq:crm}) is modified so that it exclude the $H^{+}$ atomic ion recombination channel and include the significant $H_{2}$ dissociative excitation channel, since the plasma is assumed to be ionizing, as explained in the previous chapter. Therefore, the following relation holds.
\begin{equation} \label{eq:Hstaterelation}
n_{H(p)} = C_{21}(T^c_{e}, n^c_{e}, T^h_{e}, n^h_{e}, T_{g}, P_{g})n_{H(1)} + C_{22}(T^c_{e}, n^c_{e}, T^h_{e}, n^h_{e}, T_{vib, X})n_{H_{2}(X)}
\end{equation}
where $C_{21} \& C_{22}$ are again the suppressed coefficients.
The suppressed coefficient, $C$, is just a simplified term that tells us what kinds of parameter the relation depends on.

$T_{vib, X}$ is in both $C_{11}$ and $C_{22}$, since the ground vibrational distribution affect both the electronic excitation to the $H_{2}(d)$ state and the dissociative excitation to $H(p)$.

$T_{g}, P_{g}$ are in $C_{21}$ for the radiation trapping effect of $H$.
The photon absorptions are not as significant for $H_{2}$, since the emitted photons from $H_{2}$ have much wider range of wavelengths due to the ro-vibrational structure, as explained.
Therefore, the radiation trapping effect is not considered for $H_{2}$.

And finally, as can be seen in both equations, the bi-Maxwellian rate coefficients are utilized in both models, for accurate calculations of electron impact transition rates.

All parameters in Equation \ref{eq:H2staterelation} and \ref{eq:Hstaterelation} are already available with $n_{H_{2}(X)} \& n_{H(1)}$ from Substep (2).
Therefore, the excited state densities, or the excited state distribution of interest can be calculated.
\begin{equation} \label{eq:calculated_distribution}
\left[ n_{H(3)}:n_{H(4)}:n_{H(5)}:n_{H_{2}(d)} \right]^{calculated}
\end{equation}

\bigskip
\bigskip
(4) Compare the calculated and measured distributions

From the measured spectra (Figure \ref{fig:The processed photon count rates of all lines of interest}), the excited state distribution is already known (Figure \ref{fig:to_relative_density}).
\begin{equation} \label{eq:measured_distribution}
\left[ n_{H(3)}:n_{H(4)}:n_{H(5)}:n_{H_{2}(d)} \right]^{measured}
\end{equation}

Now, the calculated one and the measured one can be compared, and calculate the error value.
\begin{equation} \label{eq:errorfunc}
\epsilon = \sum_{p=3}^{5} \left( \left[ \frac{n_{H(p)}}{n_{H_{2}(d)}} \right]^{cal} / \left[\frac{n_{H(p)}}{n_{H_{2}(d)}}\right]^{mes} -1 \right)^{2}
\end{equation}

\bigskip
\bigskip
(5) Minimize the error to find the DOD

Repeat Substep (1-4) and find the DOD that minimize the error value in Substep (4).
Then, this DOD is the most reasonable DOD inferred for the plasma of interest.

\bigskip
\bigskip

\begin{figure}[htbp]
	\centering
    \includegraphics[width=14cm]{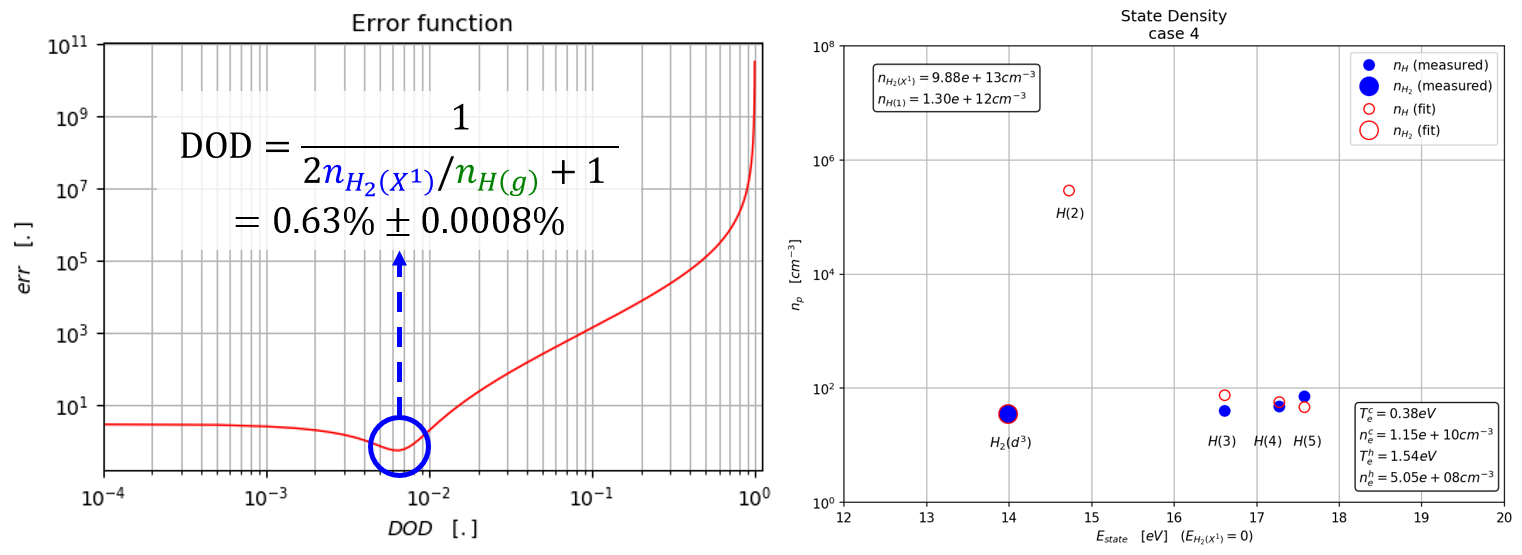}
	\caption[The plot for the error function with inferred DOD(left), and the calculated and measured state density plot(right) (Case 4)
	]{The plot for the error function with inferred DOD(left), and the calculated and measured state density (Case 4) plot(right)
	}\label{fig:infer_dod_example}
\end{figure}

This is the inference procedure for finding the DOD of the plasma.
An example is presented in Figure \ref{fig:infer_dod_example}.

Figure \ref{fig:infer_dod_example}(left) shows the error function for $DOD=0 \,-\, 100\%$.
The minimum of the error function is at around $DOD=0.1\%$, and the exact DOD is written on the plot.
Figure \ref{fig:infer_dod_example}(right) shows how the calculated values are fitted to the measured values.
Since the $d$ state density is the reference for comparing distribution (Equation \ref{eq:errorfunc}), the points for the $d$ state coincide.
The densities for $H(3, 4, 5)$ are fitted, but the trends are slightly different.
(The density for $H(2)$ cannot be obtained from the VIS spectrum measurement.)

\begin{figure}[htbp]
	\centering
    \includegraphics[width=14cm]{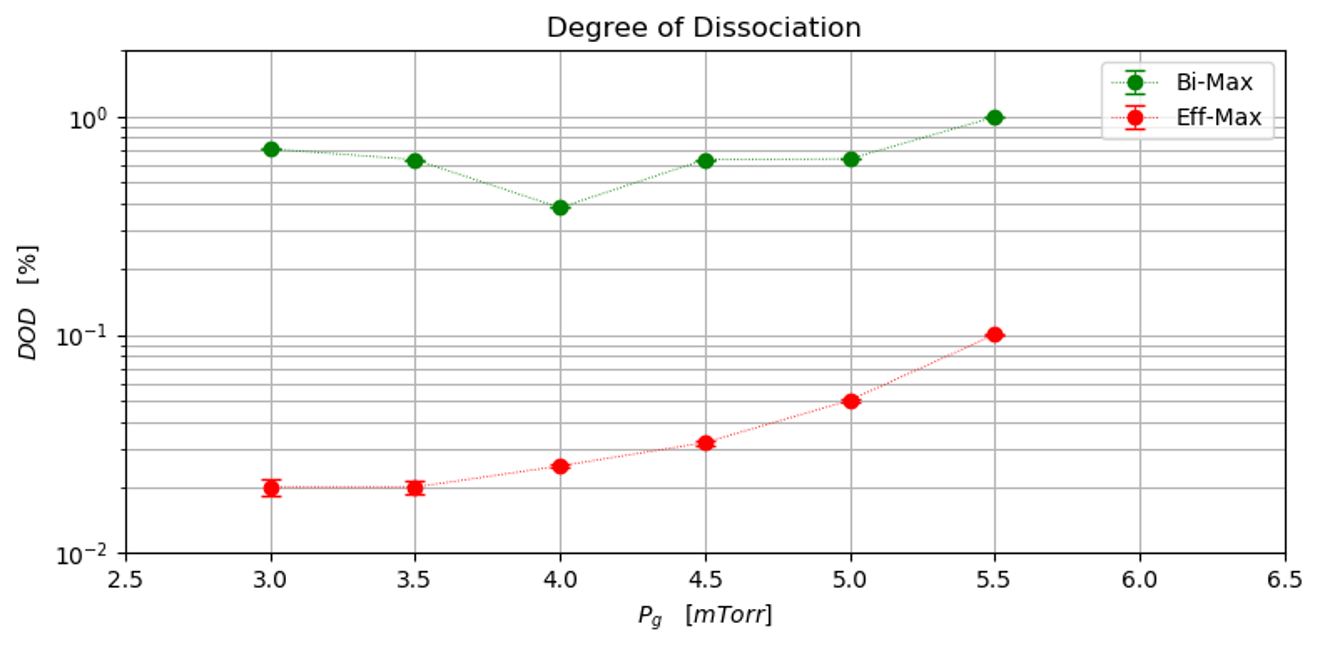}
	\caption[Degree of dissociation inferred for all cases
	]{Degree of dissociation inferred for all cases
	}\label{fig:dod_all}
\end{figure}

Finally, the DODs for the measured plasmas of all cases are inferred and plotted in Figure \ref{fig:dod_all}.
For comparison, the DODs are also inferred with another method involving the effective Maxwellian EEDF parameters calculated to be used as inputs in the inference procedure.
This comparison is similar to the one done in Figure \ref{fig:Compare RCs}.

\begin{figure}[htbp]
	\centering
    \includegraphics[width=14cm]{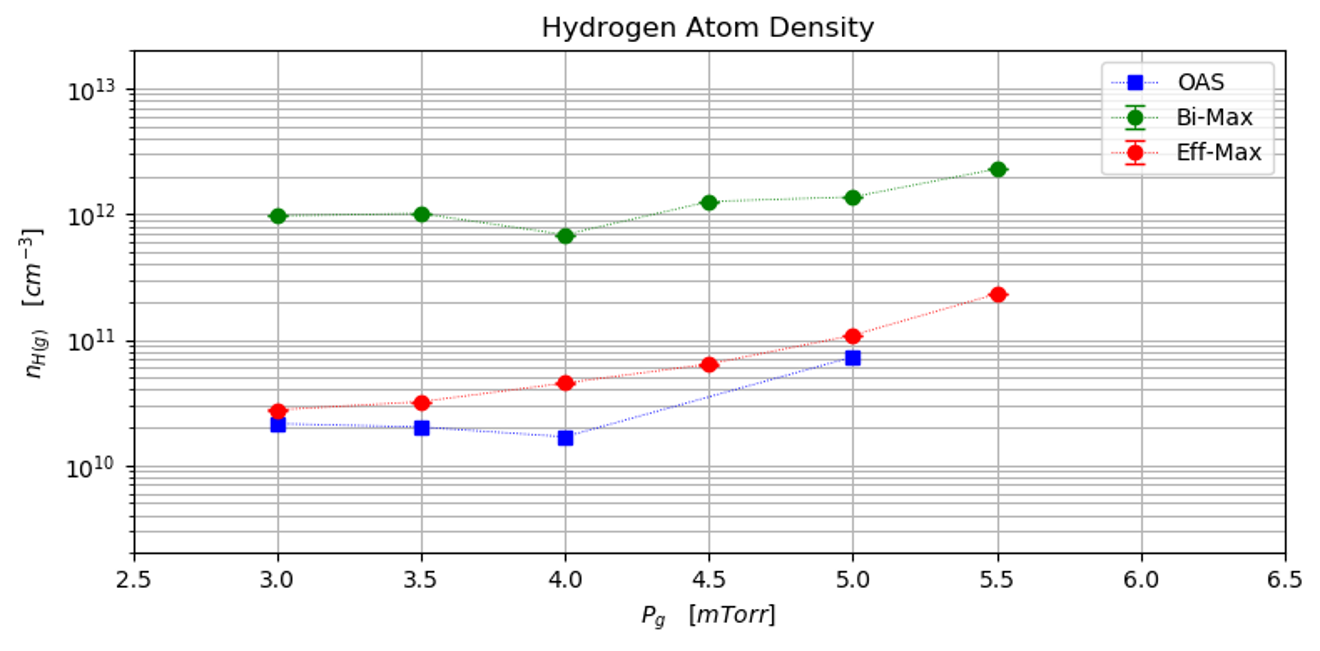}
	\caption[$H$ density for all cases
	]{$H$ density for all cases
	}\label{fig:nH_all}
\end{figure}

The $H$ densities for the inferred plasmas are also plotted with both bi-Maxwellian and Maxwellian method in Figure \ref{fig:nH_all}.
The result of the OAS analysis for the $H(1)$ density is also plotted, but there is no data for Case 4 and 6, due to unexpected data corruption.
The results from the analysis technique in these figures and possible causes of discrepancy in the results will be discussed in the next chapter.

\chapter[Discussion \& Conclusion]{Discussion \& Conclusion}

The results and the problems of the analysis technique are discussed, and the conclusion is given in this chapter.
As can be seen in Figure \ref{fig:dod_all} and \ref{fig:nH_all}, the DODs inferred with the bi-Maxwellian method and the effective Maxwellian method are different by some orders of magnitude.
And the OAS result shows that its trend of being minimum at $4mTorr$ coincides with that of the DODs by the bi-Maxwellian method, but its magnitude is comparable to that of the effective Maxwellian method.

Many kinds of information are involved in the analysis technique, and some are not as accurate as others due to various inevitable factors.
Lack of accurate atomic and molecular database may cause discrepancies in the rates calculated for both population models.
The low electron temperature of the plasmas may affect the precise calculation of excitation transition rates and dominance of contributing channels for $H(p)$.
There can be spatial inconsistencies in the data measured by the OES, the OAS, and the LP, due to the line-of-sight nature of the optical measurement.

\begin{figure}[htbp]
	\centering
    \includegraphics[width=12cm]{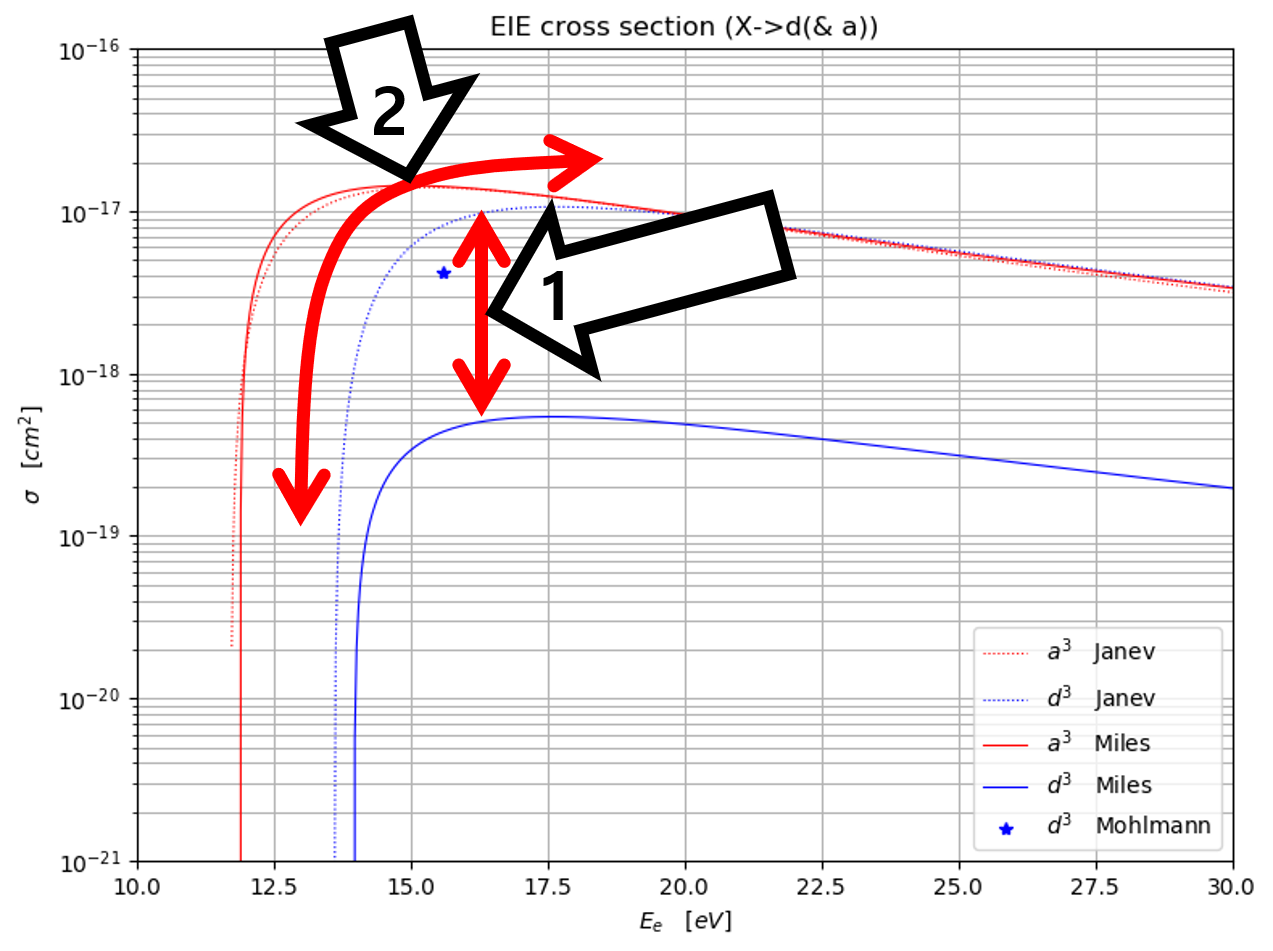}
	\caption[Electron impact vibronic excitations from the $X$ state to the $a$ and $d$ states \cite{Janev2003, Miles1972, Mohlmann1976}
	]{Electron impact vibronic excitations from the $X$ state to the $a$ and $d$ states \cite{Janev2003, Miles1972, Mohlmann1976}
	}\label{fig:discuss_eiecx}
\end{figure}

The following figure is an example of cross section data inaccuracy.
In Figure \ref{fig:discuss_eiecx}, there are total electron impact vibronic excitation cross sections, from the ground $X$ electronic state to the $a$ and $d$ states, $\sigma^{tot}_{X(0) \rightarrow a}$, $\sigma^{tot}_{X(0) \rightarrow d}$.
The cross section datas are obtained from several literatures \cite{Janev2003, Miles1972, Mohlmann1976}.

The data by Janev \cite{Janev2003} is plotted with the dashed lines.
This database contain vast amount of hydrogen atomic and molecular collisional cross sections and rate coefficients, collected from various sources.
The majority of the atomic and molecular data required in population models are therefore gathered from this source.
Figure \ref{fig:discuss_eiecx} shows that, for $T_{e}>20eV$, $\sigma^{tot}_{X(0) \rightarrow a}$ and $\sigma^{tot}_{X(0) \rightarrow d}$ are almost equal.
This similarity is not expected, since the energy difference between the $a$ and $d$ states is as large as $\sim 2eV$.

The cross sections are also gathered from the literature by Miles \cite{Miles1972}, which are plotted with solid lines in Figure \ref{fig:discuss_eiecx}.
The shape and magnitude for $\sigma^{tot}_{X(0) \rightarrow a}$ from Miles is similar to that from Janev.
However, there is more than one order difference for $\sigma^{tot}_{X(0) \rightarrow d}$, as the arrow number 1 in Figure \ref{fig:discuss_eiecx} indicates.

Also, another data is by Mohlmann \cite{Mohlmann1976}, which is experimentally obtained cross section.
In the literature by Mohlmann, the cross section data is provided with points with respect to electron energy, which is about $10eV$ apart.
This makes the integration of interpolated data to calculate the rate coefficients imprecise.
In Figure \ref{fig:discuss_eiecx}, only the maximum value is plotted with a star, and this value by Mohlmann is also different from cross sections by the other two sources.

This discrepancy in the cross sections can make a huge difference to the outcomes of the population models.
From Equation \ref{eq:vresolvedCoronaM2}, it is apparent that the $d$ state density calculated with the v-resolved CoronaM is directly proportional to the excitation cross section used.
The data by Miles is used for this specific cross section in this work since the ones by Janev and Mohlmann seem incorrect for the above reasons, but it is difficult to decide which source actually provides atomic and molecular data with better accuracy, and thus, the calculation results of the population models can be unreliable.

Another similar example is shown in Figure \ref{fig:derc}.
As explained, these are rate coefficients of dissociative excitations of $H_{2}(X(0)) \rightarrow H(p)$.
They are also derived from the data provided by Janev and Miles.
The left plot of the figure shows the rate coefficients calculated with the vibronic excitation cross section data and dissociation state pair information (Table \ref{tb:Electronic state pairs singlet} and \ref{tb:Electronic state pairs triplet}) from the literature by Janev, and the FCF database \cite{Fantz2006}.
The right plot of the figure show the rate coefficients directly provided in Miles.
Similarly, there are discrepancies for all dissociative excitation cross sections, and also, the cross sections for dissociations to $H$ excited state higher than 4 and 6 are unavailable from the sources by Janev and Miles, respectively.
Again, it is hard to choose sources for more accurate data, and there are many missing data needed for precise calculations.

The plasmas generated in this work have unexpectedly low electron temperatures, $T^{c}_{e} < 0.5eV$ and $T^{h}_{e} < 2eV$.
This may have amplified the inaccuracy of calculation even more.
Figure \ref{fig:derc} also shows that the discrepancies of cross sections are larger for low $T_{e}$.
Usually, inaccuracy of rate coefficients for low $T_{e}$ is especially high for the following reason.

As the arrow number 2 indicates in Figure \ref{fig:discuss_eiecx}, the slopes of the cross section values are much stiffer for low electron energy.
This is because the cross sections become zero at the threshold.
Plus, the cross section curve shapes are more stochastic near the threshold region because of the resonance effect, which is not apparent in the figure.
So, the inaccuracy of cross sections is magnified even more in the low energy region.
Since the calculation of rates involves multiplying the cross sections by EEDFs and integrating the products, the threshold region cross section values dominantly affect the calculated rate, and thus, affect the results of the population models.

Besides the fact that low $T_{e}$ worsens the reliability of the cross section data, low $T_{e}$ may also affect the calculation with the population models because of other possibly significant contributing channels.

It is explained in the previous chapters, two channels, $H_{2}$ and $H$, are included in the $H$ CRM.
This approach is used since the plasmas analyzed are assumed to be ionizing plasmas, where $H_{2}$ and $H$ are dominant.
However, since the plasmas generated actually have low $T_{e}$, and so, other channels may be contributing significantly (Figure \ref{fig:List of contributing channels}).
Recombinations from atomic and molecular ions may significantly affect the production of $H$ excited states.
Figure \ref{fig:crm_anal_1_2} in the appendix shows that the recombining population coefficients are dominant over the ionizing population coefficients when $T_{e}<1eV$.
To see how this affect the inference of DOD, the inference is also performed with additional atomic ion recombination channel, but its effect is found to be negligible (the result not shown), likely due to the presence of hot electron with $T_{e}>2eV$.
However, the contributions from molecular ion recombination are not checked in this work, and so, they may have affected the accuracy of the analysis result.
Recombinations primarily increase the population of higher states, since their energy is closer to ions.
Therefore, the increasing trend of measured state densities present in Figure \ref{fig:infer_dod_example} may be caused by significant rates of molecular ion recombinations.

\begin{figure}[htbp]
	\centering
    \includegraphics[width=10cm]{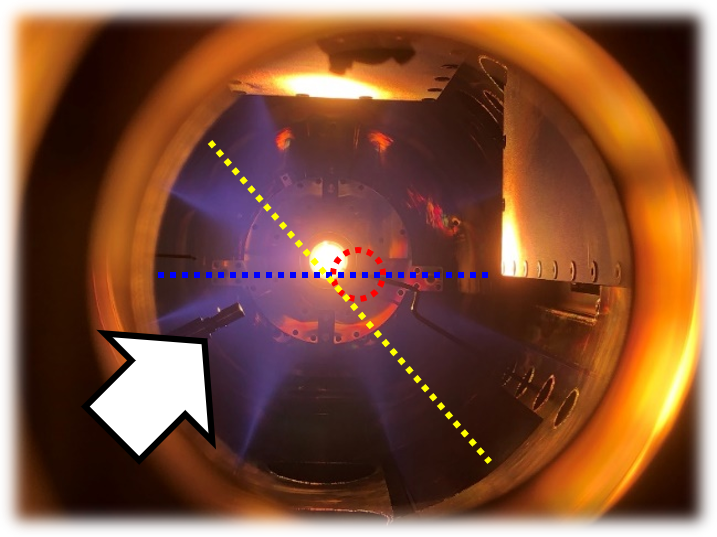}
	\caption[Line-of-sight of optical measurements and other probes
	]{Line-of-sight of optical measurements and other probes
	}\label{fig:discuss_los}
\end{figure}

Lastly, the spatial effect in the optical measurements may be a problem as well.
Figure \ref{fig:discuss_los} is the same figure shown in Chapter 2, with the line-of-sight of the optical measurement drawn on it.
The yellow dotted line is the line-of-sight of the VIS spectrometer, and the blue dotted line is the path of the Lyman photons shot from the H lamp to the VUV spectrometer.

The measured emission spectra are thoroughly analyzed in this work for the inference of DOD, but the spectra are line-integrated data that contain information of spatially dependent plasmas.
It is assumed that the measured spectra contain the information of the plasmas at the radial center, since the radiations from the center are the strongest.
However, rigorously, this assumption should be checked with an image reconstruction technique.\footnote{Alternatively, a local emission measurement is attempted in this work as well, with a local optical probe shown with the arrow in Figure \ref{fig:discuss_los}. This probe only measures spectra close to the probe by blocking other spectra. However, the signal-to-noise ratio is too low using this probe, and the probe is also damaged by the heat from the cathode.}
Due to the azimuthally asymmetric and large geometry of plasmas in the MAXIMUS as can be seen in Figure \ref{fig:discuss_los} (star-shape plasmas), a sophiticated reconstruction scheme need to be designed to solve this problem, which is one of the future works.

Similarly, the OAS analysis is affected by the geometry of the plasmas.
When the Lyman photons pass through the blue line in Figure \ref{fig:discuss_los}, the photons are mostly absorbed in the center, where a DOD is high and many $H$s are present.
Therefore, the gradient information of the DOD along the blue line should be considered for the OAS analysis.
However, since the shape of plasmas is not precisely known, the chamber radius is assumed to be the effective radius of the plasmas, when $H$ densities are calculated.
For that reason, the $H$ densities of the OAS in Figure \ref{fig:nH_all} are likely underestimated values.

Another problem of the OAS is that the emission spectra of the measured plasmas are too strong.
This is also caused by the large geometry of the generated plasmas.
The Lyman photon detection signal is overwhelmed by the strong plasma emission spectrum, and so, the absorption data is compromised due to the high noise.
Therefore, the absorption data is not clear enough for the OAS analysis, and it is impossible to analyze the data of Case 4 and 6 at all to obtain $H$ densities in this experiment.

In conclusion, the main results of this research are summarized as follows.
The analysis technique for the inference of DOD of $H$ plasmas was developed and tested.
The technique was algorithmically superior to the methods in the past, since the technique made use of newly constructed population models for $H_{2}$ and $H$, and the Fulcher-$\alpha$ analysis, with various improved approaches.
For the $H_{2}$ state analysis, the corona model was modified to consider vibrational states of each electronic state, and was used to obtain the density of electronically excited states and the ground vibrational temperatures.
Also, through the rotational distribution analysis of the Fulcher-$\alpha$ band, gas temperatures were calculated.
For the $H$ state analysis, the collisional-radiative model was constructed, considering 20 bound states and the significant transitions that affect their populations, including dissociative excitations from $H_{2}$.
The simple approach for handling bi-Maxwellian EEDF was adopted to precisely and efficiently calculate the rates for electron impact transition rates, in both the vibrationally-resolved corona model and the collisional-radiative model.
The new formula for spatially-resolved optical escape factor was derived to consider the radiation trapping effect present in the plasmas.
Finally, the population models and information from these analyses were used to infer DOD and $H$ density of the $H$ plasmas generated in the experiment.
The verification of the analysis technique was hindered by three main factors: the absence of an accurate atomic and molecular database, the low electron temperature leading to non-negligible contributions from molecular ions and other channels, and optical measurement challenges.
Nevertheless, the underlying issues were identified, and corresponding improvements have been suggested.
Future work to further enhance and verify the inference technique includes: identifying accurate atomic and molecular databases for use in population models, generating $H$ plasmas with higher electron temperatures $T_{e}$, and developing precise methods for local optical measurements.
The developed inference technique is expected to enable comprehensive investigations of neutrals in $H$ plasmas with these future improvements.

\chapter*{Acknowledgment}
\addcontentsline{toc}{chapter}{Acknowledgment}

I would like to express my sincere gratitude to my advisor, Prof. Young-chul Ghim, for his guidance and support throughout this research.
I am also grateful to Prof. Wonho Choe and Prof. Seungryong Cho for their thorough evaluation of this work.
Special thanks are extended to Dr. Yegeon Lim, Yong Sung You, and the experimental team led by Prof. Se Youn Moon for their collaboration and assistance with the experimental work.
Finally, I sincerely thank my friends for the support and encouragement.

\appendix
\chapter[Population model analysis]{Population model analysis}

\section{Hydrogen atom collisional-radiative model}

The $H$ CRM constructed in Section 3.3 is evaluated in this section of the appendix.

The contributions from the ionizing and recombining components of the CRM are compared for different electron temperature and density.
The CRM and the CoronaM for $H$ are compared for their calculation precision.
Number of states included in the CRM is varied to see its significance.
And line photon count ratios are plotted to see their sensitivities.

All following analyses involve $T_{e}, n_{e}$ of the Maxwellian EEDF.

\subsection{Ionizing vs. recombining}

In Equation \ref{eq:crm}, there are ionizing and recombining components, which are channels from the $H$ ground and $H^{+}$, respectively.
There are conditions when one of them become dominant over the other.
To analyze the behavior, the following plots are built.

\begin{figure}[htbp]
	\centering
    \includegraphics[width=14cm]{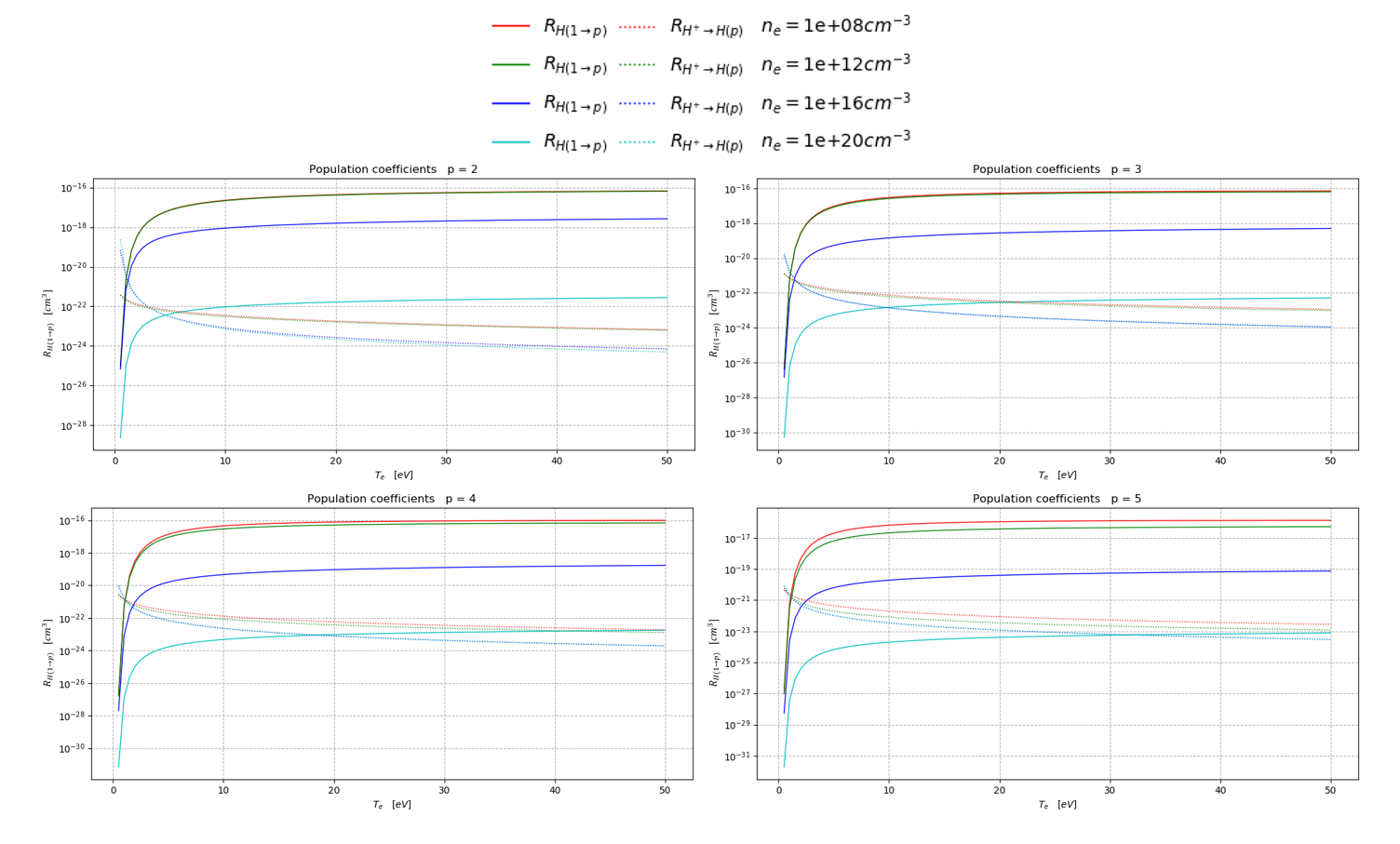}
	\caption[Ionizing vs. recombining coefficients ($T_{e}=0.5 \,-\, 50eV$)
	]{Ionizing vs. recombining coefficients ($T_{e}=0.5 \,-\, 50eV$)
	}\label{fig:crm_anal_1_1}
\end{figure}

In Figure \ref{fig:crm_anal_1_1}, the population coefficients for ionizing and recombining components are plotted together with respect to electron temperature for different electron densities.
The population coefficients that correspond to the excited state $H(2, 3, 4, 5)$ are plotted separately.
As can be seen in the figure, the ionizing population coefficients are dominant for $T_{e}>1eV$ mostly.
This means, even when $n_{H^{+}}$ is comparable to $n_{H(1)}$, the contribution from $H^{+}$ is negligible if  $T_{e}$ is high.

\begin{figure}[htbp]
	\centering
    \includegraphics[width=14cm]{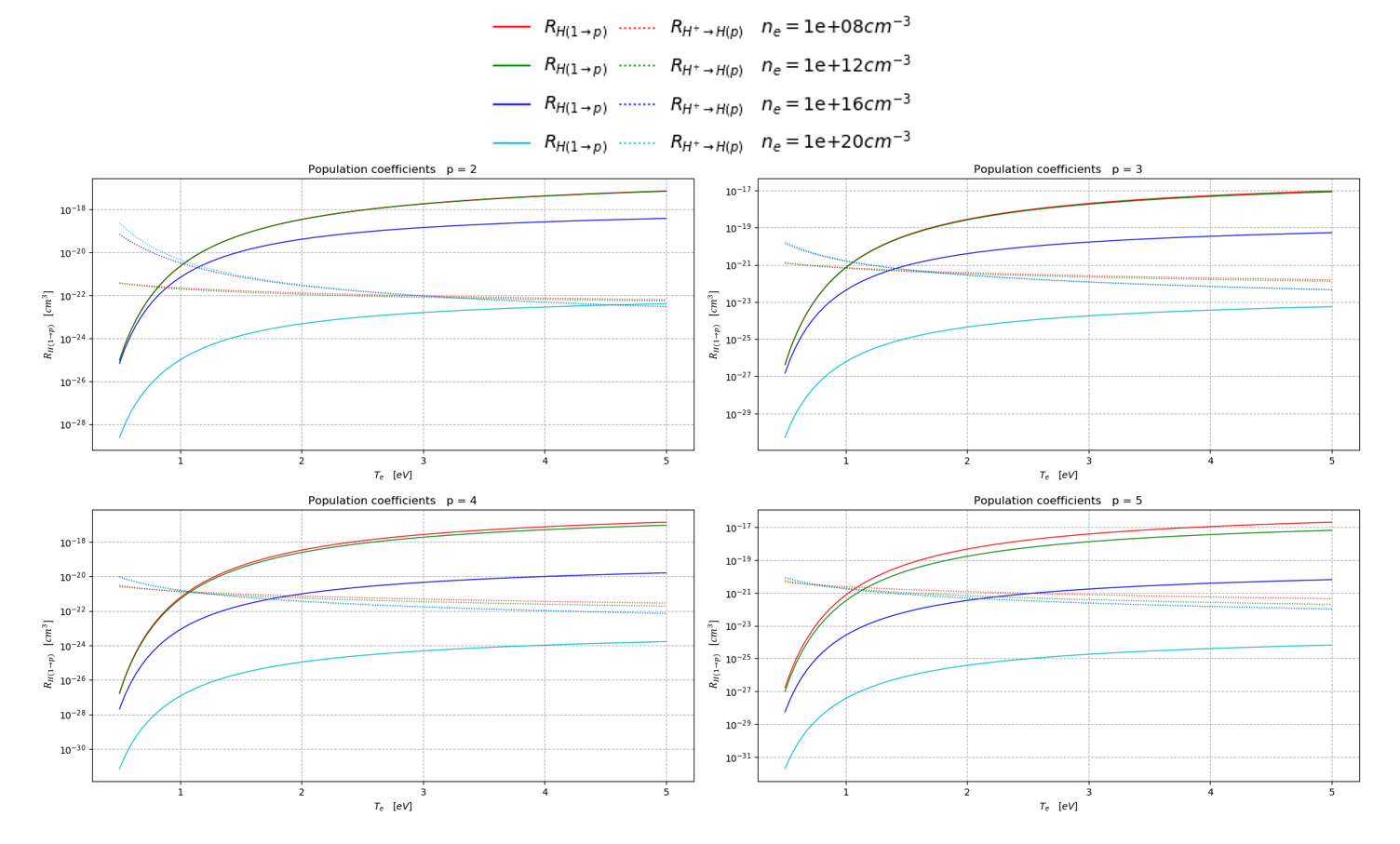}
	\caption[Ionizing vs. recombining coefficients ($T_{e}=0.5 \,-\, 5eV$)
	]{Ionizing vs. recombining coefficients ($T_{e}=0.5 \,-\, 5eV$)
	}\label{fig:crm_anal_1_2}
\end{figure}

Figure \ref{fig:crm_anal_1_2} is the zoomed-in plot for low $T_{e}$.
The recombining population coefficients are dominant over the ionizing ones when $T_{e}<1eV$.
And it is also true, when $n_{e}$ is very high.
This is due to the fact that both three-body and radiative recombination rates are high with low $T_{e}$ and high $n_{e}$, since many less energetic electrons are likely to recombine with ions.

\subsection{CRM vs. CoronaM}

\begin{figure}[htbp]
	\centering
    \includegraphics[width=14cm]{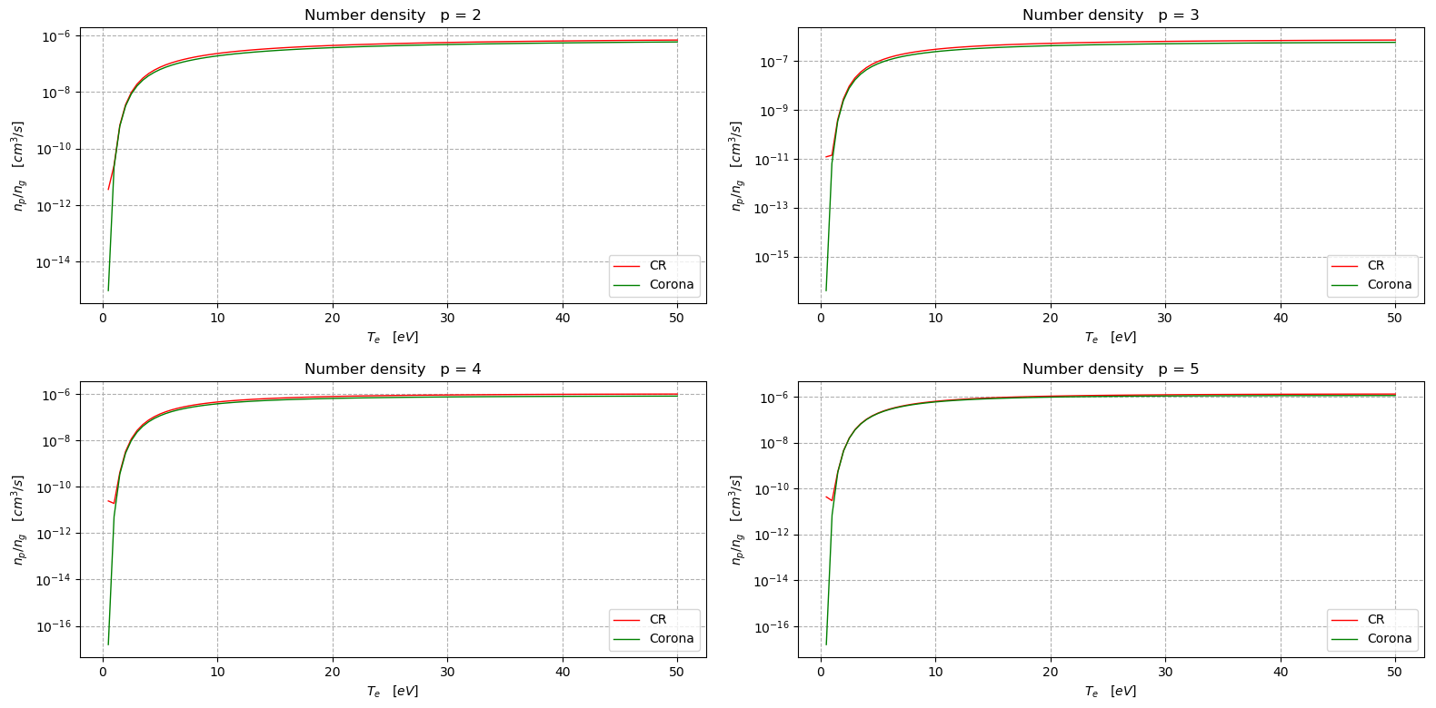}
	\caption[Excited state densities calculated by the CRM and the CoronaM ($n_{e}=10^{10}cm^{-3}$)
	]{Excited state densities calculated by the CRM and the CoronaM ($n_{e}=10^{10}cm^{-3}$)
	}\label{fig:crm_anal_2_1}
\end{figure}

\begin{figure}[htbp]
	\centering
    \includegraphics[width=14cm]{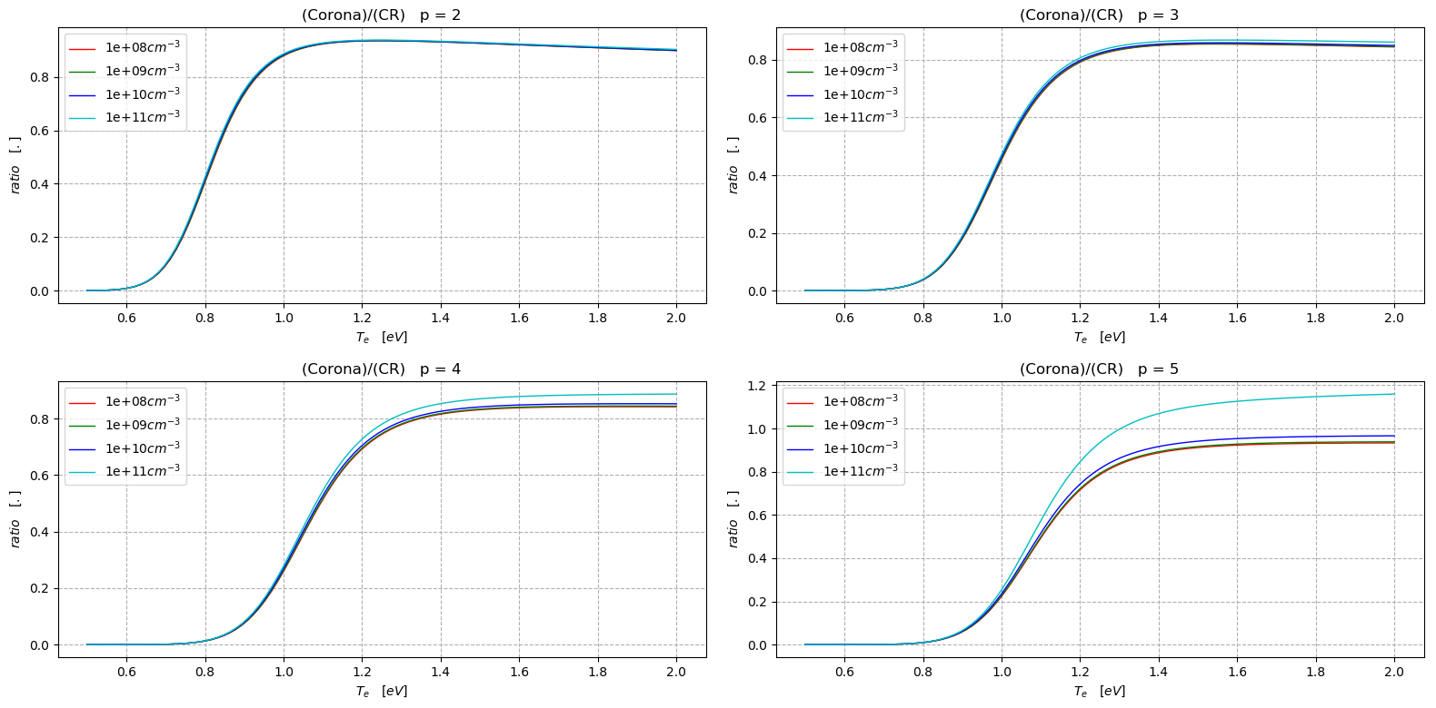}
	\caption[Excited state densities calculated by the CoronaM over those by the CRM ($T_{e}=0.5 \,-\, 2eV$)
	]{Excited state densities calculated by the CoronaM over those by the CRM ($T_{e}=0.5 \,-\, 2eV$)
	}\label{fig:crm_anal_2_2}
\end{figure}

\begin{figure}[htbp]
	\centering
    \includegraphics[width=14cm]{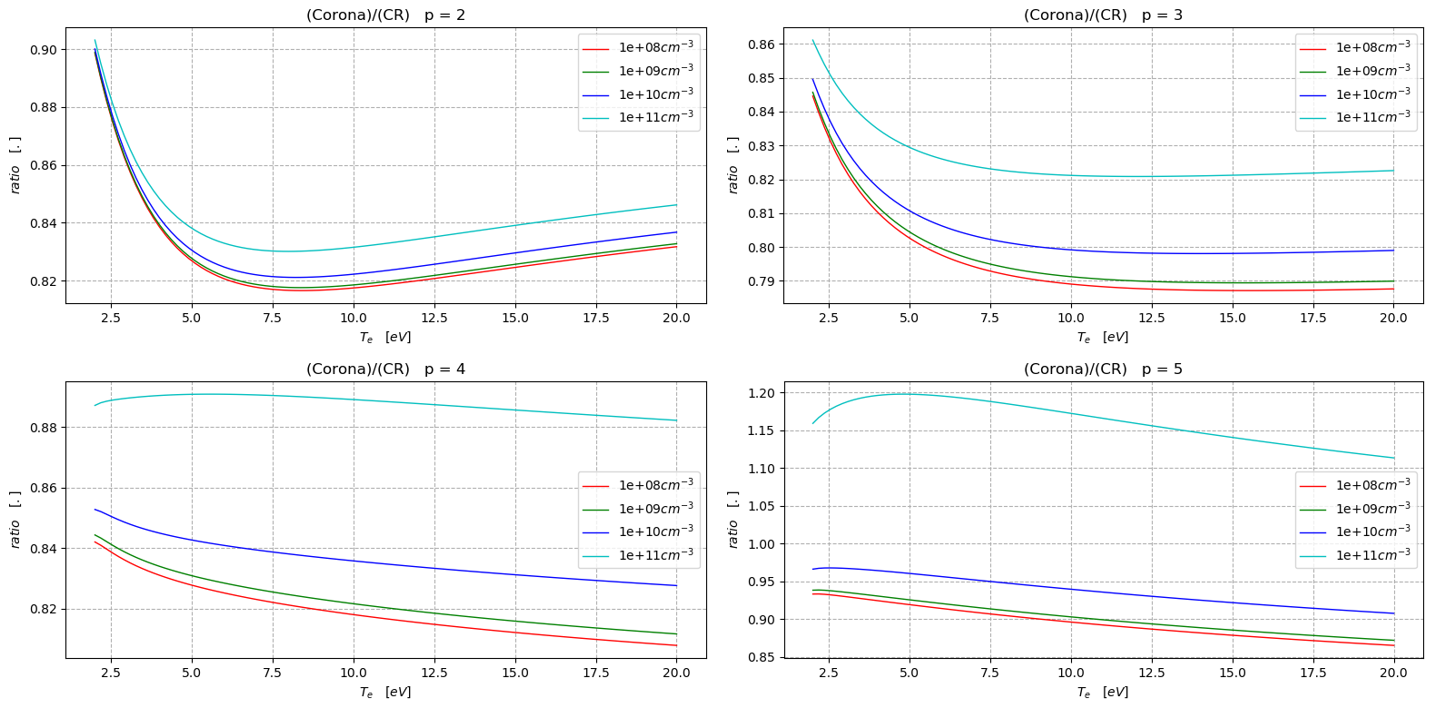}
	\caption[Excited state densities calculated by the CoronaM over those by the CRM ($T_{e}=2 \,-\, 20eV$)
	]{Excited state densities calculated by the CoronaM over those by the CRM ($T_{e}=2 \,-\, 20eV$)
	}\label{fig:crm_anal_2_3}
\end{figure}

The CoronaM only considers the electron impact excitation from the ground to the state of interest and the radiative decay of the state of interest.
Therefore, it is less precise than the CRM.
The following plots (Figure \ref{fig:crm_anal_2_1}, \ref{fig:crm_anal_2_2} and \ref{fig:crm_anal_2_3}) are drawn to analyze the discrepancy between the CoronaM and the CRM.

Figure \ref{fig:crm_anal_2_1} is the plot of excited states calculated by the CRM and CoronaM when $n_{e}=10^{10}cm^{-3}$.
The results from both models are similar except at low $T_{e}$.
This is because the CoronaM does not consider recombination transitions.
This discrepancy is clearer in Figure \ref{fig:crm_anal_2_2}.
Figure \ref{fig:crm_anal_2_2} and \ref{fig:crm_anal_2_3} are the excited state densities calculated by the CoronaM over those calculated by CRM.
Near $T_{e}=1eV$, the CoronaM is severely underestimating the state densities.
Also, mild underestimations are present for high $T_{e}$.
Overestimations occur as well when there should be significant amount of electron impact de-excitation or ionization loss considered, as can be seen in the last plot of Figure \ref{fig:crm_anal_2_3}.
Overall, the calculation results from the CoronaM deviate much from those by the CRM depending on the electron parameters, and therefore, the CoronaM should be used with caution.

\subsection{Number of states included in the CRM}

There are an infinite number of excited state possible for bound electrons.
However, when constructing a CRM, only a finite number of states are included, since highly excited states can be ignored.
20 states are included in the CRM in this work since $E_{H^{+}}-E_{H(20)} < 0.05eV$.
In other words, most of the excited state energy range in $H$ is covered with $H(1 \,-\, 20)$.
In this subsection, the effect of the number of states included in the CRM is investigated.

\begin{figure}[htbp]
	\centering
    \includegraphics[width=14cm]{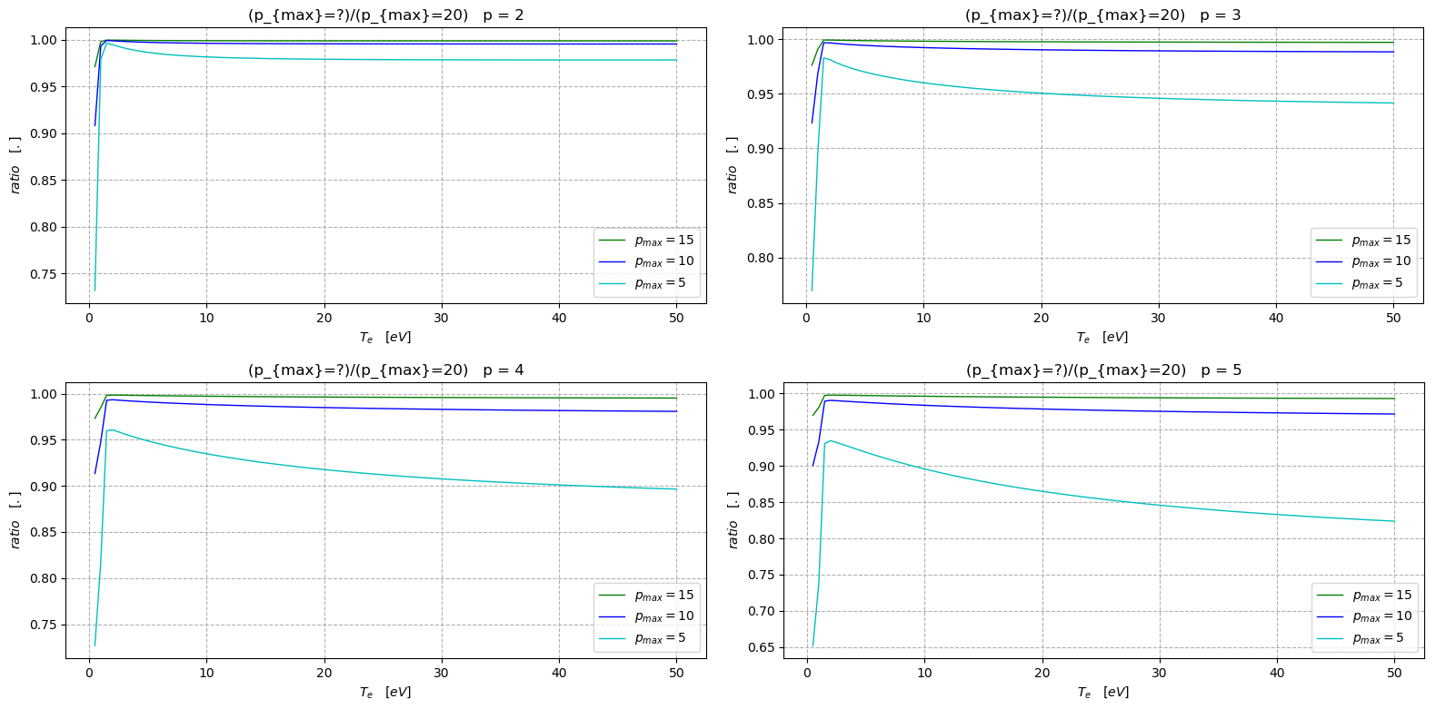}
	\caption[Number of states included and its effect on the CRM calculation
	]{Number of states included and its effect on the CRM calculation
	}\label{fig:crm_anal_3_1}
\end{figure}

Figure \ref{fig:crm_anal_3_1} has plots that shows the excited state densities calculated by other CRM with different numbers of states included divided by those calculated by the CRM with 20 states ($n_{e}=10^{10}cm^{-3}$).
The CRMs with less states underestimate the excited density when $T_{e}$ is either very low or high.
Similar to previous analyses, the recombination is affecting the calculation when $T_{e}$ is low.
Since recombinations most likely produce high energy states and they cascade down to populate lower states, the consideration of the upper higher states is important for determining the lower states.
Similarly, when $T_{e}$ is high, high states are populated by excitation from the lower states and cascade down to populate the lower states.
Therefore, a sufficient number of states must be included in the CRM for precise calculation of state densities.
In addition, the computing time halved for every 5 less states included.

\subsection{Line photon count ratio}

The CRM calculates the excited state densities, and thus, it can calculate the photon emission flux (Equation \ref{eq:spontaneousemission}).
And the ratios between two photon emission fluxes may be sensitive to $T_{e}$ or $n_{e}$.
If the ratios are sensitive to some parameters, those parameters may be inferred by analyzing the photon count ratio of the measured spectra.
So, sensitive ratios are useful in optical diagnostics of plasmas.
Therefore, the sensitivity for each ratio of the $H$ Balmer lines calculated by the CRM is investigated by constructing 3D photon count ratio plots.

\begin{figure}[htbp]
	\centering
    \includegraphics[width=12cm]{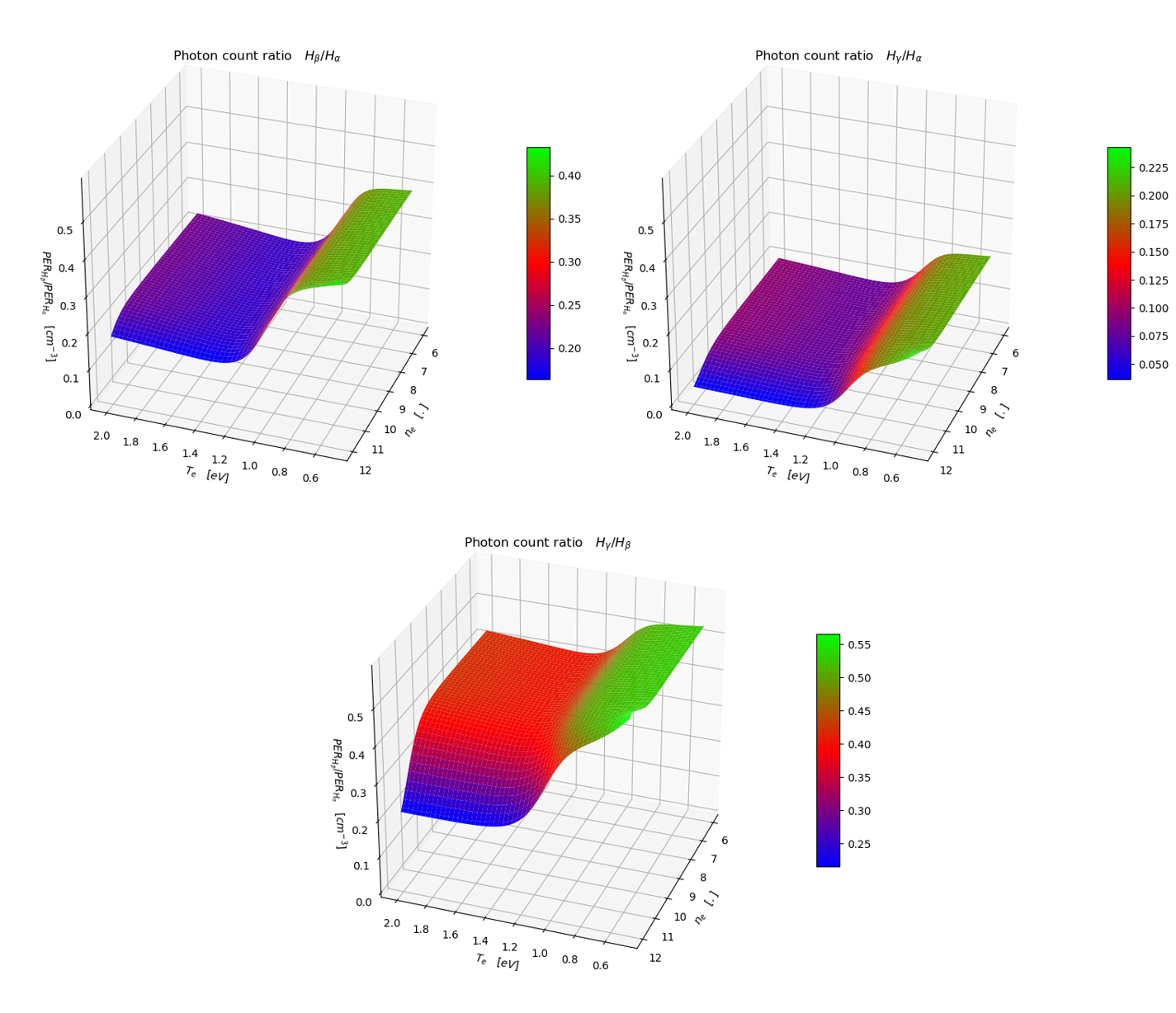}
	\caption[Photon count ratios of Balmer-$\alpha , \beta , \delta$ lines ($T_{e}=0.5 \,-\, 2eV$)
	]{Photon count ratios of Balmer-$\alpha , \beta , \delta$ lines ($T_{e}=0.5 \,-\, 2eV$)
	}\label{fig:crm_anal_4_2}
\end{figure}

\begin{figure}[htbp]
	\centering
    \includegraphics[width=12cm]{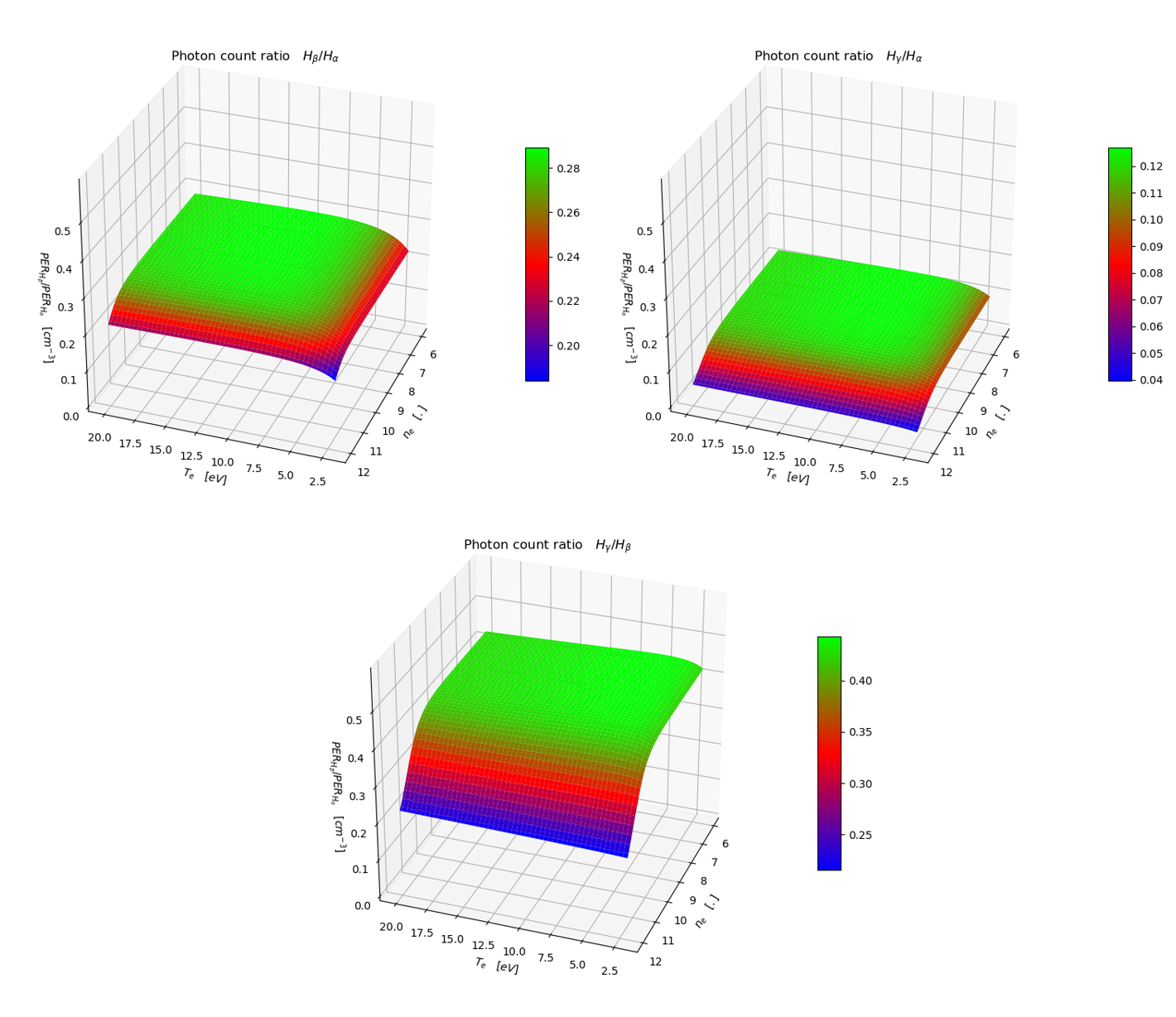}
	\caption[Photon count ratios of Balmer-$\alpha , \beta , \delta$ lines ($T_{e}=2 \,-\, 20eV$)
	]{Photon count ratios of Balmer-$\alpha , \beta , \delta$ lines ($T_{e}=2 \,-\, 20eV$)
	}\label{fig:crm_anal_4_1}
\end{figure}

From Figure \ref{fig:crm_anal_4_2}, it can be seen that the Balmer line ratios are only slightly sensitive to $T_{e}$.
The sensitive region is where $T_{e} \cong 1eV$.
And the ratios are mostly not sensitive to $n_{e}$.
Only the $H_{\gamma}/H_{\beta}$ ratio is sensitive when $n_{e}>10^{10}cm^{-3}$.
Figure \ref{fig:crm_anal_4_1} shows that, mostly, line ratios are only sensitive when $T_{e}$ is low or
$n_{e}$ is high.
The investigation in this subsection only considers two atomic channels, and so, the result may not coincide with actual spectra measured from plasmas generated.

\end{document}